\documentclass[10pt]{article} 
\usepackage[dvips]{graphicx,color}
\usepackage{amssymb,amsmath,latexsym,multicol,array}
\usepackage[ps2pdf]{hyperref}

\newcommand{\bdm}{ \begin{displaymath}}
\newcommand{\edm}{ \end{displaymath}}
\newcommand{\beq}{ \begin{equation}}
\newcommand{\eeq}{ \end{equation}}
\newcommand{\bea}{ \begin{eqnarray}}
\newcommand{\eea}{ \end{eqnarray}}
\newcommand{\lpa}{ \left(}
\newcommand{\rpa}{ \right)}

\newcommand{\prim}{^\prime}
\newcommand{\sgn}{\mathop{\mathrm{sign}}}

\begin{document}

\title{Renormalized charge in a two-dimensional model of colloidal
  suspension from hypernetted chain approach}

\author{Manuel Camargo and Gabriel
T\'ellez\footnote{gtellez@uniandes.edu.co} \\ 
Departamento de F\'{\i}sica \\
Universidad de los
Andes\\ A.A. 4976, Bogot\'a, Colombia.}

\date{}
\maketitle
         
\begin{abstract}
\noindent The renormalized charge of a simple two-dimensional model of
colloidal suspension was determined by solving the hypernetted chain
approximation and Ornstein-Zernike equations. At the infinite dilution
limit, the asymptotic behavior of the correlations functions is used
to define the effective interactions between the components of the
system and these effective interactions were compared to those derived
from the Poisson-Boltzmann theory. The results we obtained show that,
in contrast to the mean-field theory, the renormalized charge does not
saturate, but exhibits a maximum value and then decays monotonically
as the bare charge increases. The results also suggest that beyond the
counterion layer near to the macroion surface, the ionic cloud is not
a diffuse layer which can be handled by means of the linearized
theory, as the two-state model claims, but a more complex structure is
settled by the correlations between microions.
\end{abstract}

                       
\section{Introduction}

The omnipresence of colloidal suspensions in chemistry and biology, 
and the variety and tunability of their particle interactions explains 
their importance both in industrial applications and fundamental 
research. A complete statistical description of this kind of systems 
is an extremely complex task inasmuch as that description must include 
the interaction between all of the species constituting the system: 
colloids, suspended ions and molecules of the solvent. However, as a 
result of the high asymmetry of mass and charge, one hopes that a 
suitable approach is obtained by considering the system, not as a 
mixture, but as a ``new" monodispersed system which involves only 
colloidal particles (macroions) interacting through an effective 
potential \cite{Belloni, Likos}. 

In the framework of the approaches based on linearized
Poisson-Boltzmann theo\-ry, it is found that the effective interaction
between two macro-ions exhibits a Debye-H\"uckel-like (DH) behavior
due to the screening effect produced by ions present in the
surrounding medium. As the charge of macroions ($Q$) increases, the
linear approximation becomes inadequate and it is necessary to use
other approaches. Nevertheless, these approaches show that at
large distances the effective interaction continues to have a DH-like
behavior but it is necessary to replace the charge $Q$ by an effective
or renormalized charge $Q_{ren}$. Then, the concept of charge
renormalization simplifies the description of colloidal interactions
and gives us some insights about the phase behavior of the colloidal
suspensions. However, it remains problematic to find methods that
allow their evaluation for arbitrary conditions of charge and
concentration.
 
In this work we evaluate the renormalized charge for a two-dimensional
(2D) Coulomb system. For this purpose, we solve numerically the
Ornstein-Zernike (OZ) equation in the scope of the hypernetted chain
(HNC) approximation and we use the calculated potential of mean force as
the effective interaction between the components of the system. The
renormalized charge, numerically evaluated, is then compared to the
one predicted by theoretical models based on the Poisson-Boltzmann
(PB) theory.  The results obtained show that, in contrast to the
mean-field theory, the renormalized charge does not saturate, but
exhibits a maximum value and then decays monotonically as the bare
charge increases. The results also suggest that beyond the counterion
layer near to the macroion surface, the ionic cloud is not a diffuse
layer which can be handled by means of the linearized theory, as the
two-state model claims, but a more complex structure is settled by the
correlations between microions.

We use a two-dimensional system due to two main reasons. In the first 
place, classical two-dimensional Coulomb systems with logarithmically 
interacting particles are simplified models which keep the Coulomb 
nature of the particle interaction. Besides, by using field theoretical 
tools, some of these models admit an exact solution, which makes it 
possible  to judge the quality of approximate formulations. On the other
hand, 2D Coulomb systems can be used to model stiff rod-like polyelectrolytes, 
which interact with the ions present in the solvent via a logarithmic 
potential.

This article is organized as follows. In section 2 we describe briefly
the system considered. Section 3 is devoted to explain the numerical
method used to solve the OZ and HNC equations.  The central part of
this work is presented in section 4, where the numerical results of
the HNC calculations are displayed and discussed for two models of 2D
Coulomb systems: the guest particle system and a simple model of
colloidal suspensions at the infinite dilution limit. Finally, some
conclusions are summarized in the last section.

\section{Guest charge in an electrolyte}

In order to take the first steps toward a solution of the renormalized
charge problem in colloidal suspensions at the infinite dilution
limit, we consider the properties of a ``guest" charge immersed in an
electrolyte, which is modeled as a two-dimensional two-component
plasma (2D-TCP). We consider here a system composed by a single
central guest particle with charge $Q=Ze$ and radius $\sigma_0$
immersed in a 2D-TCP formed by ne\-gatively and positively hard-disks
ions of radius $\sigma_{\pm}=\sigma$ and charges $q_\pm=\pm z_\pm e$
(with $z_{\pm}>0)$. The bulk densities of positive and negative ions
are denoted by $n_{+}$ and $n_{-}$, and the total density is
$n=n_{+}+n_{-}$. The electrolyte is globally neutral, $q_{+} n_{+} +
q_{-} n_{-} = 0$. The interaction potential between disks $i$ and $j$,
separated by a distance $r_{ij}$, is given by
\begin{equation}
\beta u_{ij}(r_{ij}) =
\begin{cases} \infty &  r_{ij}<\sigma_i+\sigma_j \\
 \beta v_{ij}(r_{ij}) = - z_i z_j\sgn(q_i q_j) \Gamma\ln \lpa
 \frac{r_{ij}}{L}\rpa & r_{ij} >\sigma_i+\sigma_j
\end{cases}  \label{Coul_HDpot}
\end{equation}
where $\beta$, $z_i$, and $\Gamma = \beta e^2$ are, respectively, the
inverse temperature, the valence of the charge $i$ and the coupling
(strength) parameter. $L$ is an irrelevant length scale used to fix
the zero of the Coulomb potential. At the limit of point-like charges,
$\sigma_0=0$ and $\sigma=0$, the systems is stable against the
collapse of positive and negative particles provided
$-2/z_{+}<Z\Gamma<2/z_{-}$ and $z_{+}z_{-}\Gamma<2$.

From field theoretical arguments in the limit $\sigma_0\to0$ and
$\sigma=0$, the asymptotic large-distance behavior of the effective
interaction between the guest particle and the TCP charges is related
to the large-distance beha\-vior of the corresponding two-points
correlation function of the exponential fields associated to the 2D
sine-Gordon theory \cite{Samaj_Guest, Samaj_HGuest}. According to this
correspondence, the asymptotic beha\-vior of the effective interaction
can be exactly evaluated and this result indicates that the
electrostatic potential produced by the guest charge has the same 
asymptotic behavior as the one evaluated through the DH approach
provided that both the inverse decay length and the charge are
adequately renormalized.

On the other hand, the short-distance asymptotics of the density
profiles of the coions and counterions near the guest charge has been
studied for the symmetric case, i.e. $z_\pm=1$~\cite{Tellez_Guest,
Tellez_small-r}. In this case, it has been shown that the coion
density profile exhibit a change of behavior if the guest charge
becomes large enough $(Z\Gamma \ge 2-\Gamma)$. This behavior has been
interpreted as a first step of the counterion condensation (for large
coulombic coupling); the second step taking place at the usual
Manning-Oosawa threshold $(Z\Gamma = 2)$.

Naturally, the latter effect is related to the settlement of the
renormalized charge. In this way, it can be stated that this model
encloses the most important facts involved in the non-linear screening
phenomenon. In order to compare these exact results with the numerical
ones in more general conditions, we have calculated the correlation
functions between the different components of the system by solving
the OZ and HNC equations as described below.
 
\section{Numerical method}
 
As usual, we denote by $g_{ij}(r)$, $h_{ij}(r)=g_{ij}(r)-1$ and
$c_{ij}(r)$ the radial distribution function, the total correlation
function and the direct correlation function, respectively, between
two components $i$ and $j$ of the system separated by a distance
$r$. The subscripts $i$ and $j$ can be $+$, $-$, or $0$, denoting
respectively a positive ion, a negative ion, or the guest charge. The
OZ equations for a multicomponent system reads
\begin{equation}
  h_{ij}(|\mathbf{r}_1-\mathbf{r}_2|)
  =c_{ij}(|\mathbf{r}_1-\mathbf{r}_2|)+
  \sum_{k} n_{k} \int h_{ik}(|\mathbf{r}_1-\mathbf{r}_3|)
  c_{kj}(|\mathbf{r}_3-\mathbf{r}_2|) \,d\mathbf{r}_3
\end{equation}
where $n_k$ the density of the species $k$. In the infinite dilution
limit considered here, $n_0=0$, the OZ equations describing the
correlation functions between ions of the TCP do not involve the guest
charge. By defining $\gamma_{ij}(r)=h_{ij}(r)-c_{ij}(r)$, these
equations can be written in the Fourier space as
\begin{eqnarray} \hat \gamma_{++}(k) &=& \frac{
\lpa 1-D(k)\rpa \hat c_{++}(k) + n_-\lpa \hat c_{+-}^2(k)-\hat
c_{++}(k)\hat c_{--}(k) \rpa}{D(k)} \label{gk_pp}\\ \hat
\gamma_{+-}(k) &=& \frac{\hat c_{+-}(k)(1-D(k))}{D(k)} \label{gk_pm}\\
\hat \gamma_{--}(k) &=& \frac{ \lpa 1-D(k)\rpa \hat c_{--}(k) +
n_+\lpa \hat c_{+-}^2(k)-\hat c_{++}(k)\hat c_{--}(k) \rpa
}{D(k)}\label{gk_mm} 
\end{eqnarray}
where the hat denotes the Fourier transform of a function, and $D(k) =
(1-n_+\hat c_{++}(k)) (1-n_-\hat c_{--}(k)) - n_+n_-\hat
c_{+-}^2(k)$. The guest charge-ion correlation functions, in the
infinite dilution limit, satisfy
\begin{eqnarray}
\hat \gamma_{0+}(k)
&=& \frac{\hat c_{0+}(k) \lpa 1-n_-\hat c_{--}(k) -D(k) \rpa +n_-\hat
c_{0-}(k) \hat c_{+-}(k)}{D(k)}
\label{gk_op}\\
\hat \gamma_{0-}(k) &=& \frac{ \hat c_{0-}(k) \lpa 1-n_+\hat c_{++}(k) -D(k) \rpa +n_+\hat c_{0+}(k)\hat c_{+-}(k)
}{D(k)}.\label{gk_on}
\end{eqnarray}
In order to solve equations (\ref{gk_pp})--(\ref{gk_on}) it is necessary 
to split  functions $\gamma(r)$ and $c(r)$ in   short- and  large-distance
 components. Since  $h(r)\rightarrow 0$ y $c(r)\rightarrow  -\beta v(r)$ 
  when $r \rightarrow \infty$, it may be written 
\beq \nonumber
v(r) = v^{(s)}(r) +  v^{(l)}(r),\hspace{2ex}   c(r) = c^{(s)}(r) - \beta v^{(l)}(r),\hspace{2ex} 
\gamma(r)  = \gamma^{(s)}(r) + \beta v^{(l)}(r) 
\eeq
where the large-distance component of the potential is chosen to be
$v^{(l)}(r)=(1-e^{-\alpha r})v(r)$ with $\alpha$ a cutoff parameter. In
this way, the corresponding HNC closure relations are defined through 
\beq \nonumber
g_{ij}(r) = \begin{cases} 0 & r<\sigma_i+\sigma_{j} \\
 \exp\lpa \gamma^{(s)}_{ij}(r) - \beta v^{(s)}_{ij}(r) \rpa & r\ge\sigma_i+\sigma_{j} \end{cases} 
\eeq
\beq
c^{(s)}_{ij}(r) = g_{ij}(r)  - 1 - \gamma^{(s)}_{ij}(r) \label{csr_closure}
\eeq
with $v^{(s)}_{ij}(r)$ the short-range part of the Coulomb potential.

Equations (\ref{gk_pp})--(\ref{csr_closure}) were solved by using a
direct iterative method. In order to improve convergence, the new
input function for iteration $i+1$, $c^{(i+1)}_{\text{in}}(r)$, was
evaluated by using a linear relaxation (Broyles' mixing of the old
input function $c^{(i)}_{\text{in}}(r)$ and the output function
$c^{(i)}_{\text{out}}(r)$) 
\beq 
c^{(i+1)}_{\text{in}}(r) =
c^{(i)}_{\text{in}}(r) + \mu \lpa c^{(i)}_{\text{out}}(r) -
c^{(i)}_{\text{in}}(r)\rpa 
\eeq 
where $\mu$ $(0\le \mu \le 1)$ is the
relaxation parameter. The numerical evaluation of the two-dimensional
Fourier transform was carried out through the quasi-discrete Hankel
transform method \cite{Guizar}, by using $N=6000$ grid points and
maximum radius $10<\kappa R<16$, with $\kappa$ the inverse Debye
length $\left(\kappa^2 = 2\pi z_+z_-n\Gamma\right)$. The iterations
were stopped when the larger difference between successive $c(r)$'s
was less than $10^{-6}$, i.e.  
\beq \tau = \max |c^{(i+1)}_{\text{out}}(r) -
c^{(i)}_{\text{out}}(r)| \le 10^{-6}. \label{tolerance} 
\eeq 
Finally, the
numerical accuracy was monitored by checking both the
electroneutrality condition and the Stillinger-Lovett screening sum
rule~\cite{SL}. In all the considered cases, the numeric results for
these conditions differ at most by $0.01\%$ with regard to the
theoretical values.
 
\section{Results}

In order to calculate the renormalized charge $Z_{ren}$ we follow a
similar procedure to the described one in \cite{LegerLevesque}. Let us
recall that under the DH approximation the potential of mean force
$w_{0s}^{\text{DH}}(r)$ between the guest charge and an ion of type
$s=\pm$ is given by
\begin{equation}
  -\beta w_{0s}^{\text{DH}}(r)=\frac{\Gamma Z z_{s}}{\kappa \sigma_{c}
   K_1(\kappa \sigma_{c}) } K_0(\kappa r)
\end{equation}
with $K_0$ and $K_1$ the modified Bessel functions of order 0 and 1,
and $\sigma_c = \sigma_0+\sigma$.

Once the correlation functions $g_{0\pm}(r)$ are iteratively evaluated
from the OZ equations, the potential of mean force $w_{0\pm}(r)$ is
obtained from the definition $g_{0\pm}(r) = e^{-\beta w_{0\pm}(r)}$. We assume that, at
large distances $\kappa r\gg1$, the potential of mean form has a
DH-like functional form, i.e.~$w(r)\propto K_0(\kappa_{ren} r)\sim
\sqrt{\pi} \exp(-\kappa_{ren} r)/\sqrt{(2\kappa_{ren} r)}$. Besides,
we can also assume that all correlation functions fall off in the same
way at large distances. These assumptions are supported in the exact
solution for point-like particles, which indicates that in the
stability region the correlation functions exhibit a DH
behavior. However, it is necessary to renormalize not only the charge
$Z_{ren}$ but also the decay length $\kappa_{ren}^{-1}$ \cite{Samaj_Guest}.

In this way,  we may suppose that 
\beq
-\beta w_{0s}(r) = \ln g_{0s}(r) \sim A_{0s} \frac{\exp{\lpa-\kappa_{ren} r\rpa}}{\sqrt{r}}
\eeq
where 
\beq
A_{0s} = \sqrt{\frac{\pi}{2\kappa_{ren}}}\frac{\Gamma Z_{ren} z_s}{\kappa_{ren} \sigma_c K_1(\kappa_{ren} \sigma_c)}.
\eeq
 Then, the parameters  $A_{0s}$ and $\kappa_{ren}$ can be
obtained from a linear fit of the functions
\beq
L_{0s}(r) = \ln\lpa \sqrt{r} |\ln g_{0s}(r)|\rpa = \ln\lpa |A_{0s}|\rpa - \kappa_{ren} r  \label{linear_fit}
\eeq
for  $r$ values large enough. We select the range of $r$ to be used in
the fit through the condition 
$10^{-4} \le |\log g_{0s}(r)| \le 10^{-2}$.  We choose this  range because 
the values of the fitted functions are two orders of magnitude greater than 
the tolerance and the numerical uncertainties (see eq.~(\ref{tolerance})). 
In some cases, the correlation functions exhibit  an oscillation and then 
it also becomes necessary to impose empirically a minimum value for $\kappa r$.  

\subsection{Guest charge}
\subsubsection{Symmetric TCP}

\begin{figure}[!htbp] 
\begin{center}
\includegraphics[width=0.48\textwidth]{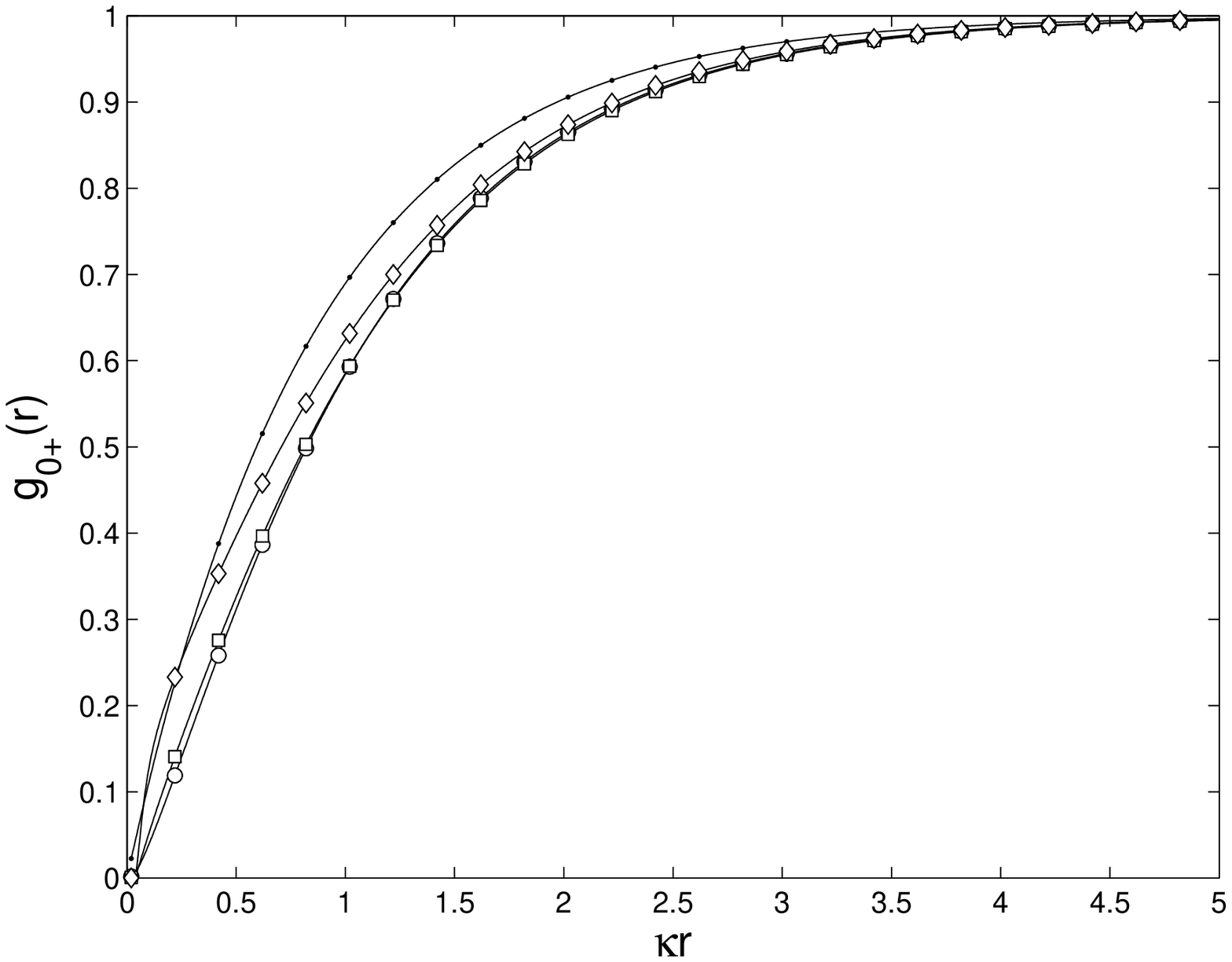}
\includegraphics[width=0.48\textwidth]{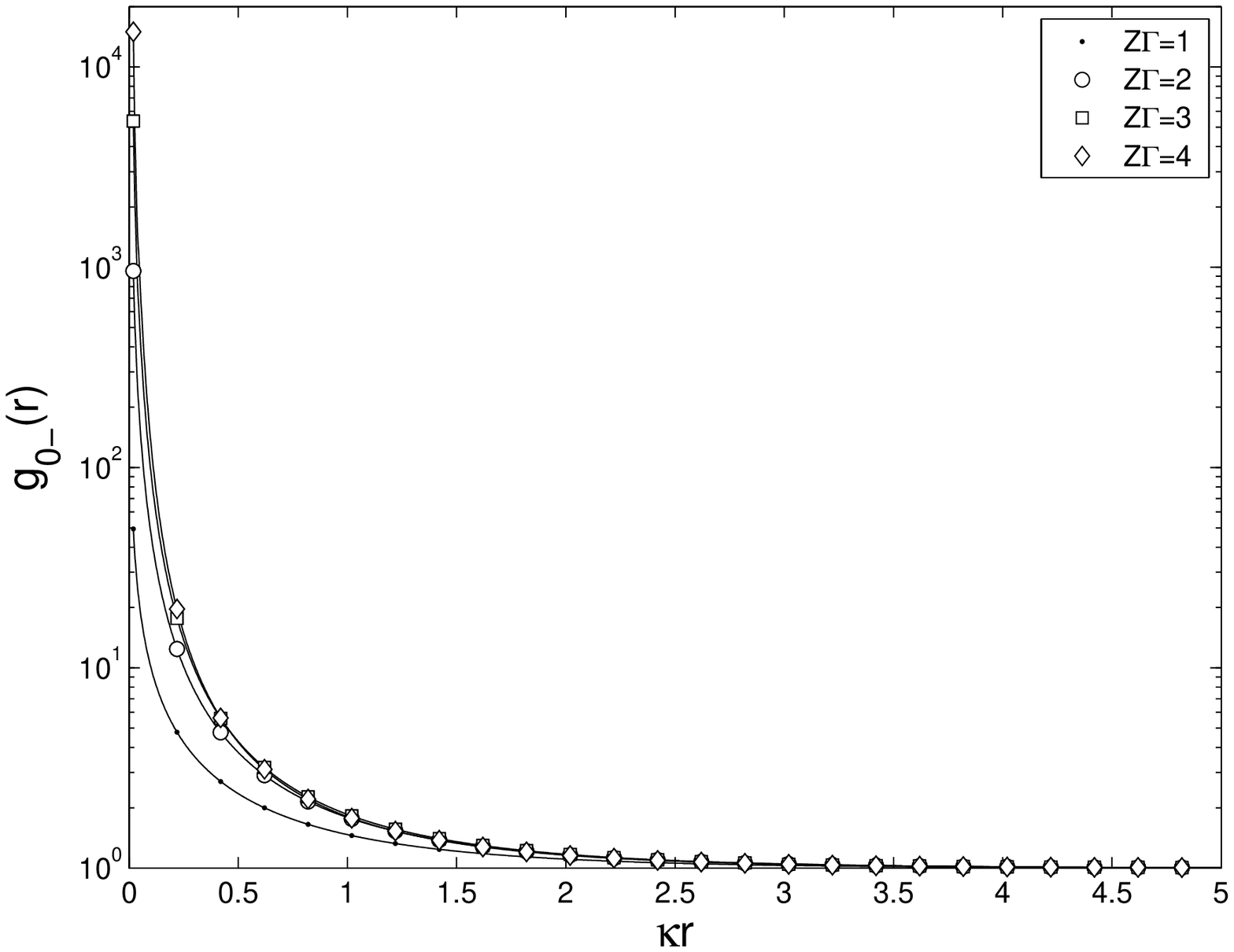}
\includegraphics[width=0.48\textwidth]{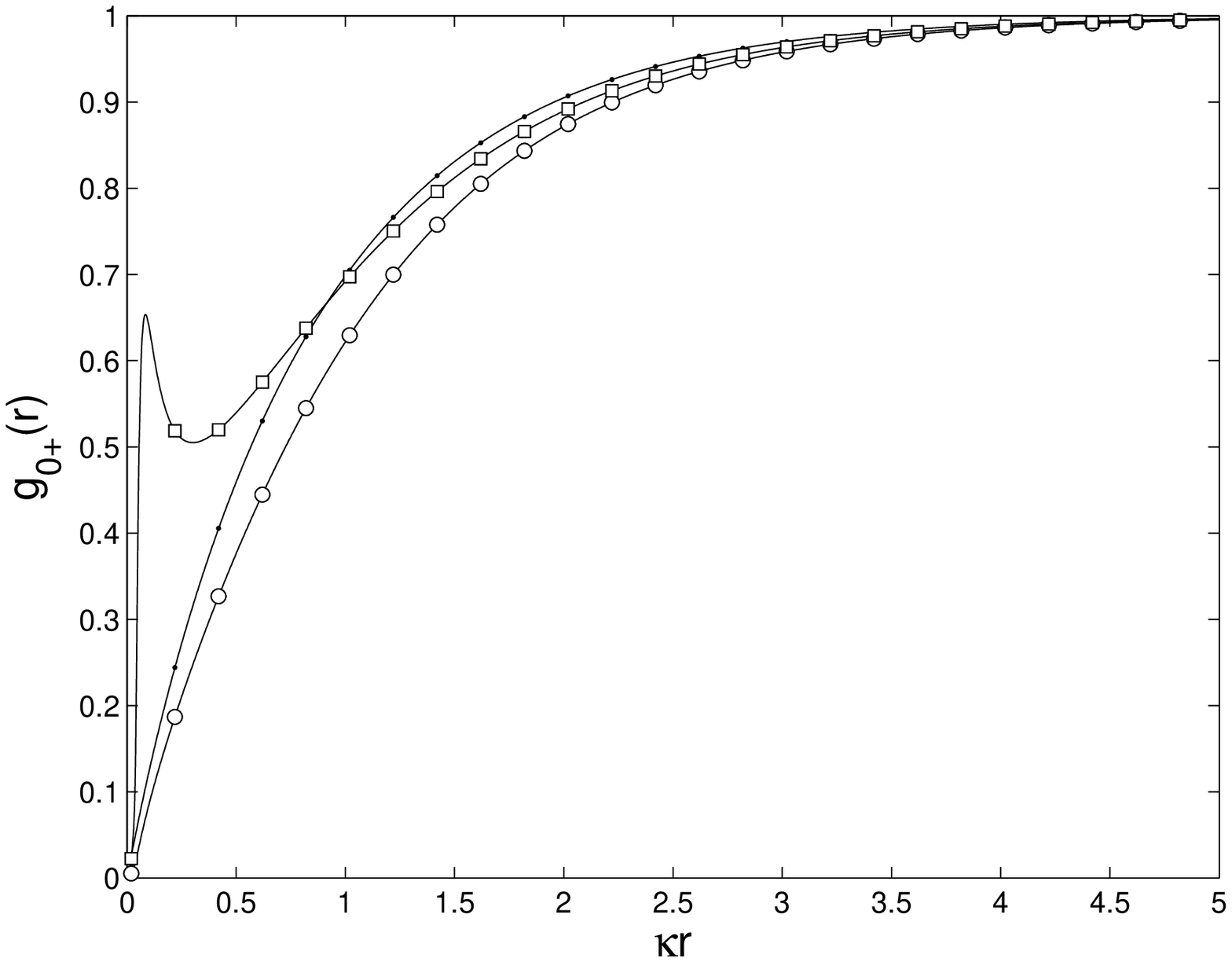}
\includegraphics[width=0.48\textwidth]{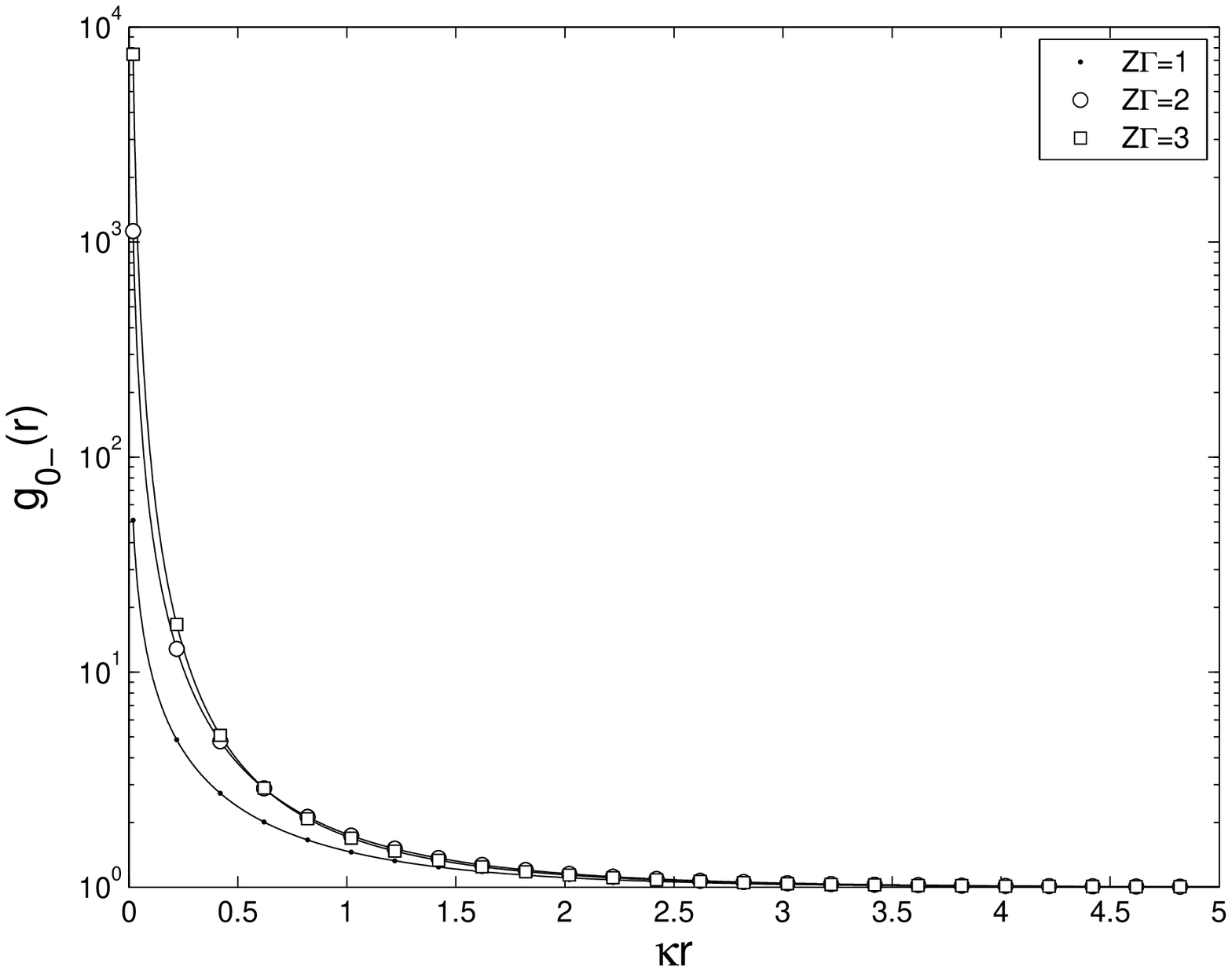} 
\begin{quotation}
\caption{Correlation functions $g_{0\pm}(r)$ at different values of $Z\Gamma$ for a guest charge immersed in a symmetric TCP with coupling para\-meters  $\Gamma=0.2$ (top) and $\Gamma=0.4$ (bottom).}  
\label{Guest_groi}
\end{quotation}
\end{center}
\end{figure}

\begin{figure}[!htbp] 
\begin{center}
\includegraphics[width=0.7\textwidth]{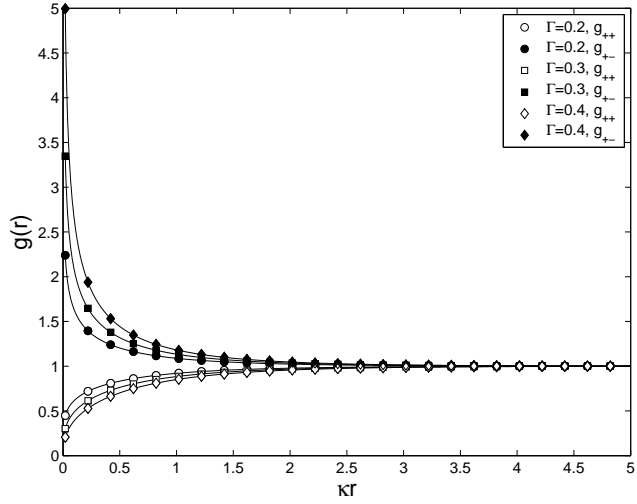} 
\begin{quotation}
\caption{Correlation functions between microions of the symmetric TCP at different values of coupling parameter $\Gamma$. 
The plotted functions correspond to the ones evaluated  when $Z\Gamma=3$; however these functions are 
independent of the magnitude of $Z\Gamma$, as expected from the equations (\ref{gk_pp})--(\ref{csr_closure}).} 
\label{Guest_grpm}
\end{quotation}
\end{center}
\end{figure}

We consider all particles as equally sized hard-disks satisfying
$\kappa\sigma_0=\kappa\sigma=10^{-2}$, and the ions have charges
$z_{\pm}= 1$. In figure~\ref{Guest_groi} are shown the
correlation functions between the guest charge and the ions
of a symmetric TCP for different values of $Z>0$ and the coupling
parameter $\Gamma$.  Qualitatively, these profiles are consistent with
the common image of an ionic cloud (mainly formed by counterions)
surrounding the central charge.  From the figure, it is clear that for
$\Gamma=0.2$ the correlation functions $g_{0\pm}(r)$ behave
monotonically for all displayed values of $Z\Gamma$. However, when
$\Gamma=0.4$ and $Z\Gamma=3$ the correlation function $g_{0+}(r)$
exhibit a peak at $\kappa r\approx 0.1$, which rise up as a
consequence of the increase of $g_{0-}$ jointly with more marked
microions correlations, as it can be seen in figure~\ref{Guest_grpm}.

\begin{figure}[!htbp] 
\begin{center}
\includegraphics[width=0.48\textwidth]{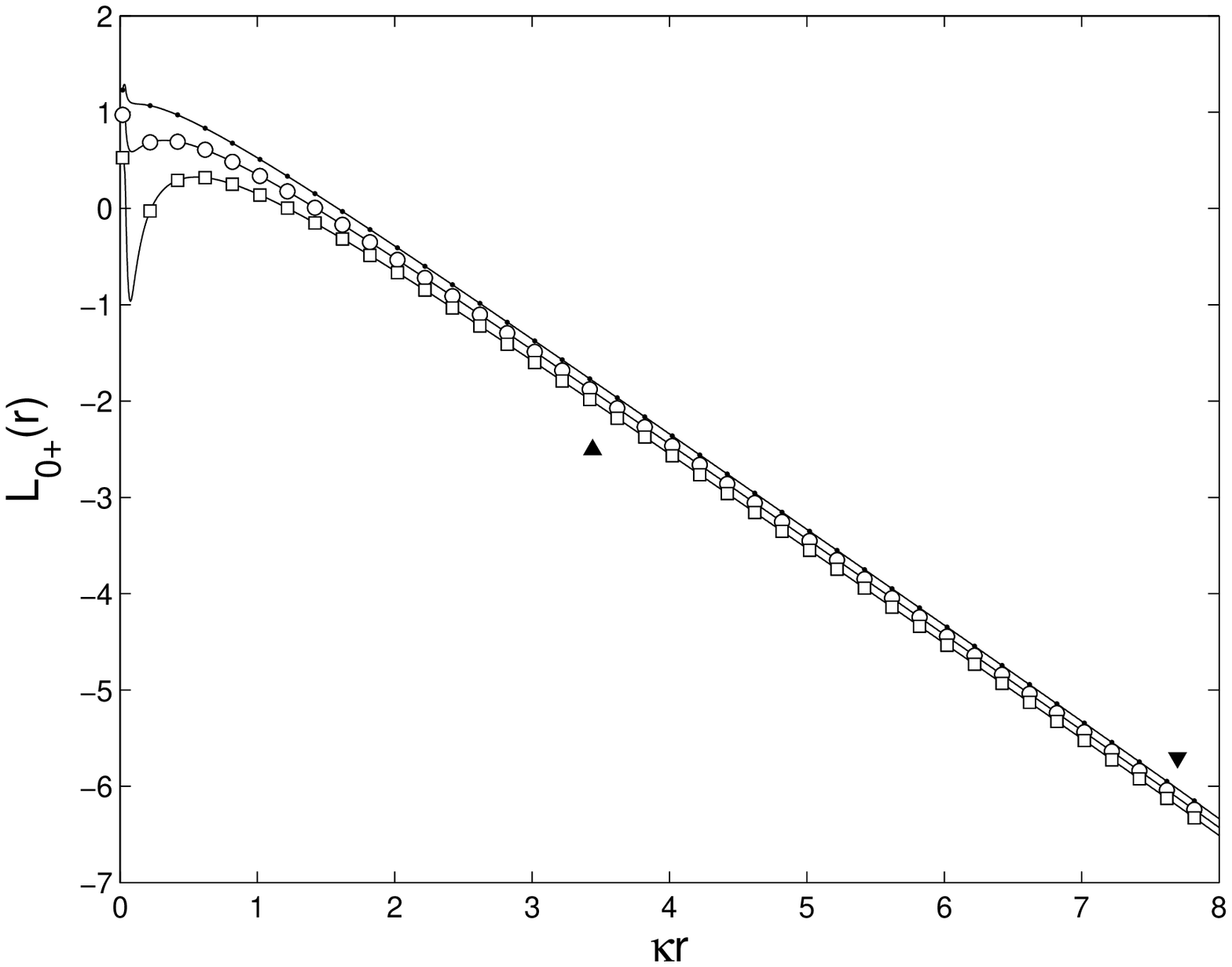}
\includegraphics[width=0.48\textwidth]{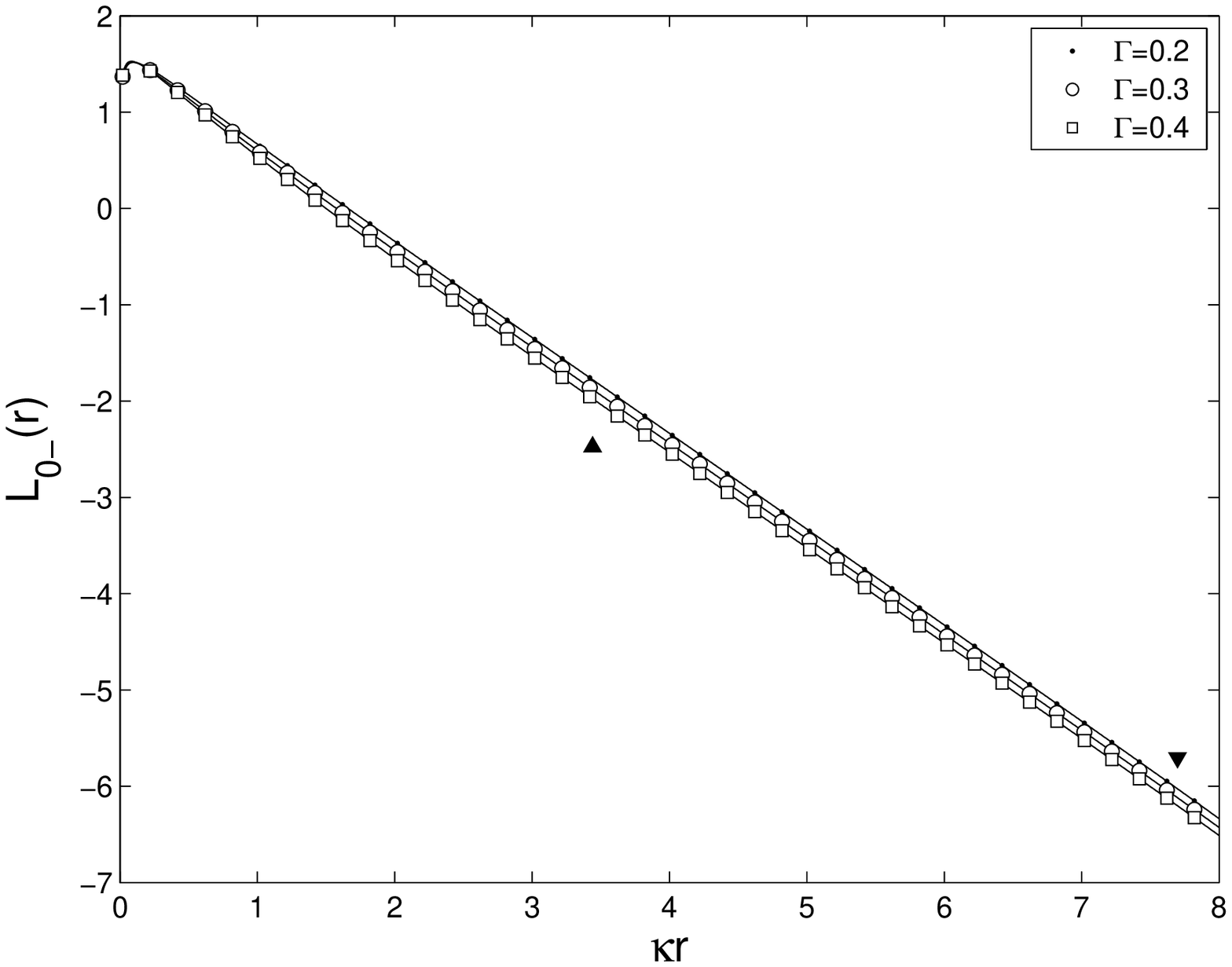}
\begin{quotation}
\caption{Functions $L_{0\pm}(r)$ used to evaluate renormalized
  parameters. The curves were calculated from the correlation functions at  $Z\Gamma=3$. Triangles delimit the $r$ range used for the linear fit.}  
\label{Guest_Lroi}
\end{quotation}
\end{center}
\end{figure}

As stated before, the renormalized parameters were evaluated from a
linear fit (in a least squares sense) of the functions $L_{0\pm}(r)$
similar to the ones shown in figure~\ref{Guest_Lroi}, whose non-linear
behavior extends roughly from $\kappa r=0$ to $\kappa r\approx
1$. From now on, a computed renormalized parameter is equivalent to
the mean value of the two ones obtained from both $L_{0+}(r)$ and
$L_{0-}(r)$ taken independently.  Note that the curves were calculated
at $Z\Gamma=3$ which correspond to strong coupling regime between the
guest charge and the microions.

We explored a range of couplings $0.1\leq\Gamma\leq 0.5$. Within this range,
the renormalized inverse screening length $\kappa_{ren}$ is found to
be systematically smaller than $\kappa$, and it is a decreasing
function of the coupling $\Gamma$, in agreement with the exact
results~\cite{Samaj_Guest}. However $\kappa_{ren}/\kappa$ only differs
from 1 by at most 2\% for the largest coupling $\Gamma=0.5$.

\begin{figure}[!htbp] 
\begin{center}
\includegraphics[width=0.7\textwidth]{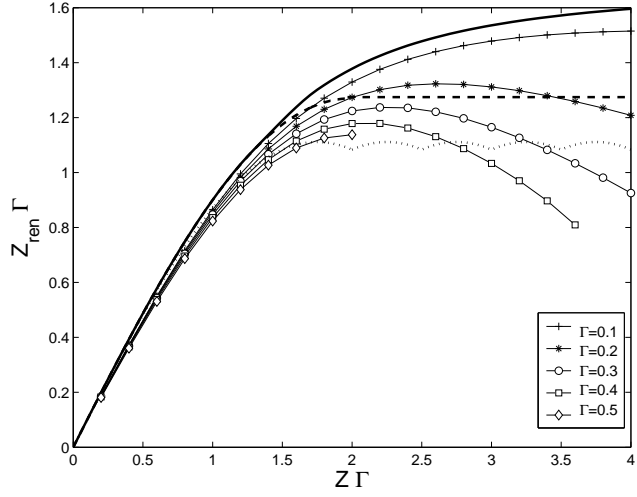}
\begin{quotation}
\caption{Renormalized charge as a function of the bare one for different coupling parameters. Dashed  and thick continuous lines represent the mean-field PB analytical result for $\sigma_0 = \sigma = 0$ and the numerical one when $\kappa\sigma_0 = 10^{-2} , \sigma = 0$, respectively \cite{Tellez_Guest}. Dotted line indicates the exact result from field theoretical arguments ($\kappa\sigma_0 \to 0 , \sigma = 0$, $\Gamma=0.5$) \cite{Samaj_HGuest}.} 
\label{Guest_Zren}
\end{quotation}
\end{center}
\end{figure}

Figure~\ref{Guest_Zren} displays $Z_{ren}$ at different coupling
parameters. In contrast to the PB prescription, $Z_{ren}$ is not a
monotonically increasing function of $Z$, but it displays a maximal
value after which it drops to a point where the implemented iterative
method fails to converge. This behavior qualitatively resembles the
one predicted by the exact solution when the microions and the guest
charge are considered as point-like particles $(\sigma=0)$ and a small
hard disk $(\kappa\sigma_0 \to 0 )$, respectively. According to this
result, the renormalized charge exhibit a maximum at $Z_m\Gamma = 2 -
\Gamma/2$ and decrease until the collapse point $Z_c\Gamma=2$ at which
$Z_{ren}= Z_c-1$, as indicated by the dotted line in
figure~\ref{Guest_Zren}. This behavior reflects the effect of the
counterion condensation, i.e. the renormalized charge associated with
the collapse value $Z_c$ of the bare charge is identical to that of
the bare charge $Z_c - 1$ because of the condensation of one
counterion from the TCP onto the guest particle \cite{Samaj_Guest,
Samaj_HGuest}. By increasing $Z$ further more counterions become
condensed and the previous description predicts a renormalized charge
oscillating between two extremes.

As it can be noted from figure~\ref{Guest_Zren}, as $\Gamma$
decreases, $Z_{ren}$ moves closer to the mean-field PB theory
predicted values; however, the renormalized charge predicted by PB
theory is always greater than the one computed from HNC
approach. Naturally, this behavior can be ascribed to the correlations
between the components of the system and in a minor grade, to the
finite radius of the ions, as it will become clearer below. When the
coupling becomes stronger, the HNC collapse point is closer to the
one predicted for point-like particles, i.e.~$Z\Gamma=2$. Although,
in the other cases the hard-disk potential inclusion allows us to go
beyond this point, at $\Gamma=0.5$ we found that the coupling between
the guest particle and the counterions is strong enough to provoke the
divergence of the HNC method at smaller values of $Z\Gamma$. This
effect can be related to the short-range structure of the correlations
between the guest particle and the microions.
 
By using the sine-Gordon model for the TCP, it has been argued that a
change in the short-range behavior of the coion density profile can be
interpreted as a ``precursor" of the (Manning) counterion condensation
\cite{Tellez_Guest}. When $Z\Gamma$ increases above $2-\Gamma$
($Z\Gamma<2$, $\Gamma<2$), the coion cloud $n_+(r) = n_+ g_{0+}(r)$
shows a change of behavior near the guest particle. This fact is
reflected in the parameter $\alpha_{0\pm}$ describing the potential of
mean force for short-distances defined below. Near the guest charge
this latter must correspond to the Coulomb potential, i.e. 
\beq
-\beta w_{0s}(r) = \ln g_{0s}(r) \underset{r\to\sigma_0}{\sim}
\alpha_{0s}\ln r  \label{meanforce_short} 
\eeq 
where $\alpha_{0s}$ is
a constant depending on the bare charge and temperature. Ana\-lo\-gous
arguments, involving the electrostatic potential $\psi(r)$ calculated
from the exact PB equation, establish the change in the short-range
behavior of $\psi(r)$ as a fingerprint of the counterion condensation
phenomenon \cite{TellezTrizac_PB}.

\begin{figure}[!htbp] 
\begin{center}
\includegraphics[width=0.7\textwidth]{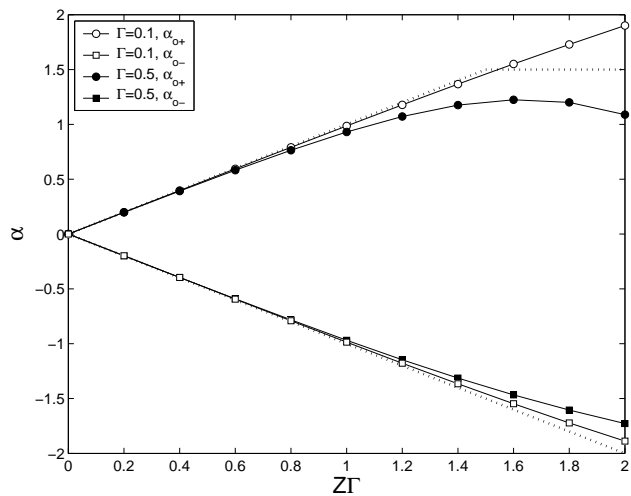}
\begin{quotation}
\caption{$\alpha_{0s}$ factors characterizing the short-distance
  potential of mean force at $\Gamma=0.1$ and $\Gamma=0.5$. In order
  to exclude the hard-disk effect, the data shown was calculated from
  a linear fit of $\ln g(r)$ at $6\sigma<r < 10\sigma \approx
  0.1\kappa^{-1}$. Dotted lines represent theoretical limits
  (\ref{slope_short}) at $\Gamma=0.5$.}
\label{Guest_slope}
\end{quotation}
\end{center}
\end{figure}

From the field theoretical
approach~\cite{Tellez_Guest,Tellez_small-r}, it is expected that 
\beq
\alpha_{0+} = \begin{cases} Z\Gamma  & \hspace{0.5cm}Z\Gamma<2-\Gamma \\ 2-\Gamma & \hspace{0.5cm}Z\Gamma>2-\Gamma\end{cases}
\hspace{2cm} \alpha_{0-} = - Z\Gamma, \label{slope_short} 
\eeq 
i.e.~at $Z\Gamma>2-\Gamma$ the coions interact with the guest charge as if the
latter one carries a charge $\hat Z=2/\Gamma-1=Z_c-1$ independent of
$Z$, whereas the counterions always interact with the bare charge
$Z$~\cite{Tellez_Guest}.  Figure~\ref{Guest_slope} shows the values of
$\alpha_{0s}$ evaluated from the short-range values of
$g_{0s}(r)$. In order to evaluate them, we carried out linear fits of
$\ln g_{0s}(r)$ according to (\ref{meanforce_short}). The comparison
between the numerical results at $\Gamma=0.5$ to the ones given by
(\ref{slope_short}) indicates that the short-range behavior of
$g_{0+}(r)$ when $\kappa\sigma=10^{-2}$ is qualitatively consistent
with that of a system of point-like charges.  However, from the figure
it can be noted that as $Z\Gamma\to 2$ a change in $g_{0-}(r)$ is also
found, as it was suggested in \cite{Tellez_Guest}. This fact indicates
that as the coupling increases, the counterions ``see" the guest
charge with a smaller charge than $Z$. At this point, the effect
produced by the correlation between the counterions jointly to the
hard-disk potential provoke the increase of distance between the guest
particle and the counterions. Also, as a consequence of this change at
large enough coupling, the maximum value of $Z_{ren}$ do not reach
the saturation value predicted by PB theory, which is indicated in
figure~\ref{Guest_Zren} through the dashed line at $Z\Gamma>2$.

\begin{figure}[!htbp] 
\begin{center}
\includegraphics[width=0.7\textwidth]{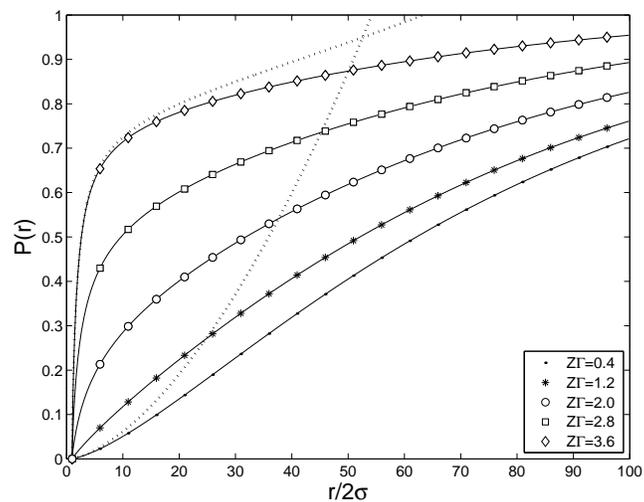}
\begin{quotation}
\caption{Integrated charge distribution  as a function of the radial distance at $\Gamma=0.4$ and different values of $Z\Gamma$.  Dashed lines represent the distribution of counterions  when  $Z\Gamma=0.4$ and $Z\Gamma=3.6$. Radial range corresponds to two Debye lengths. } 
\label{Guest_Qfrac}
\end{quotation}
\end{center}
\end{figure}

On the other hand, the charge accumulation around $Z$ can be also measured
  through the integrated charge distribution 
\beq
P(r) =  -\frac{2\pi}{Z}\int_0^r \rho(r\prim)r\prim dr\prim \label{Charge_r}
\eeq
where $\rho(r)=(nz_+z_-)\lpa g_{0+}(r)-g_{0-}(r)\rpa/(z_++z_-)$. $P(r)$  
indicates the overall electric charge found within a disk of radius $r$ 
and can be used to define  the cutoff distance, the so-called Manning radius 
$r_m$,  which separates the condensed counterions from the uncondensed ones.  
According to a geometrical construction,  the cutoff distance  is identified
 by an inflexion point in $P(r)$ when it is displayed as a function of 
 $\ln r$ \cite{TellezTrizac_PB, Belloni_Cond, Deserno}, i.e.
\beq
P_{yy}(r)\Big|_{r=r_m} = \frac{d^2P(r)}{d(\ln r)^2}\Big|_{r=r_m} = 0. \label{inflection_crit}
\eeq
Figure~\ref{Guest_Qfrac} shows  $P(r)$. As expected, at larger
$Z\Gamma$, the short distance behavior of $P(r)$ is dominated by the
counterion density.  Besides, as demonstrated before in the third
panel of the figure~\ref{Guest_groi} for which $Z\Gamma=3$ and
$\Gamma=0.4$, the onset of a peak in coion profile at high enough
$Z\Gamma$ and $\Gamma$  does mark the extension of condensed
counterions layer and give rise to inflection points in the integrated
charge distribution.   

\begin{figure}[!htbp] 
\begin{center}
\includegraphics[width=0.48\textwidth]{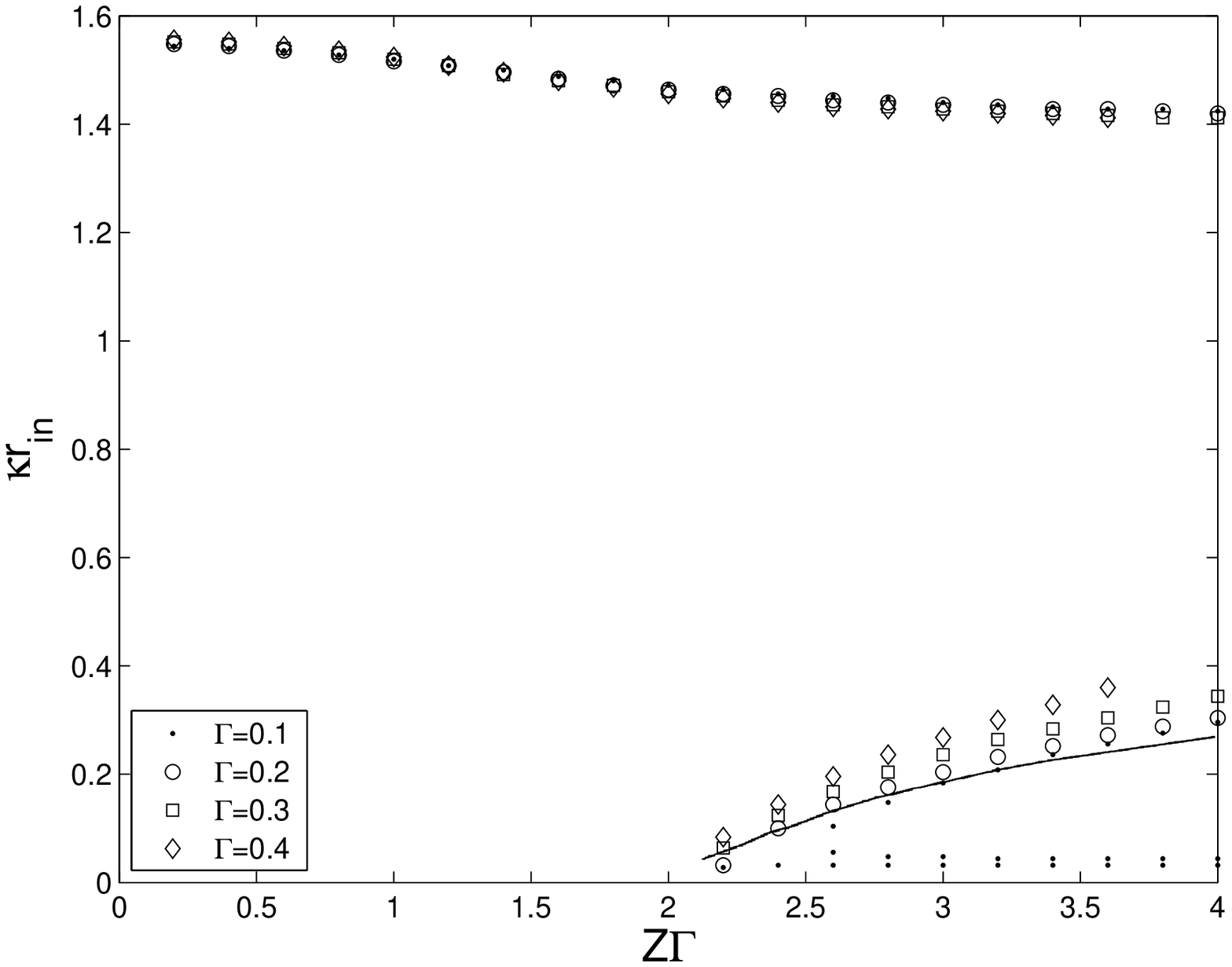}
\includegraphics[width=0.485\textwidth]{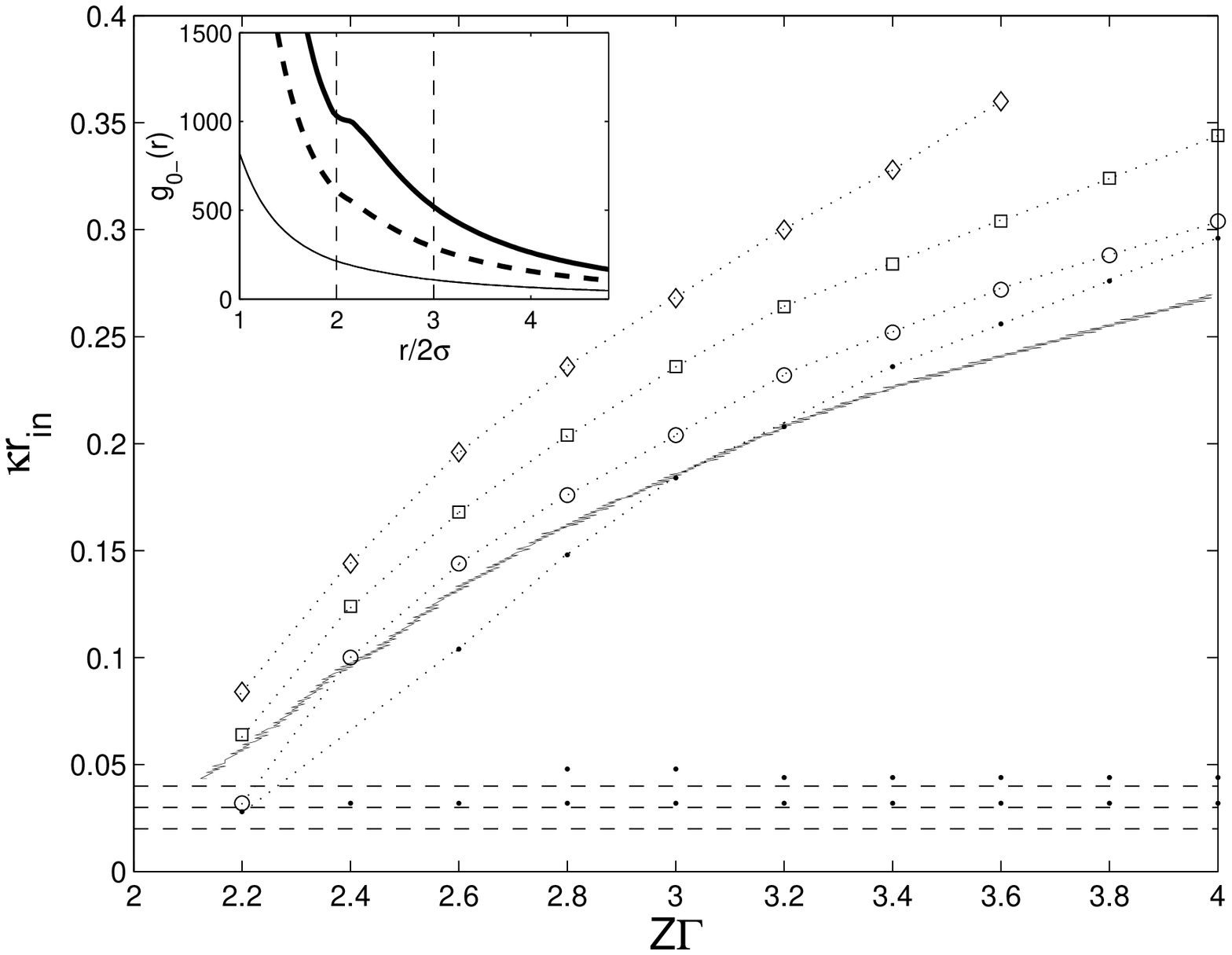}
\begin{quotation}
\caption{Inflection points of the charge distribution as a function of $Z\Gamma$ for different values of $\Gamma$ (left). Continuous line correspond to the numerical solution of PB equation (data taken from \cite{TellezTrizac_PB}).  Detailed plot  showing the Manning radius $r_m$ (right). In the inset are shown the correlation functions $g_{0-}(r)$ at $\Gamma=0.1$ for  $Z\Gamma=2,3,4$, which suggest that the onset of a well-defined second layer of counterions provoking  smaller values of $r_{in}$. Thin dashed lines indicate the position of one, two and three particle diameters. Dotted lines are a guide to the eye.} 
\label{Guest_Rmann}
\end{quotation}
\end{center}
\end{figure}

The locus of points satisfying $P_{yy}(r_{in})=0$ is plotted in
figure~\ref{Guest_Rmann}.  This figure shows that for all values of
$Z\Gamma$ an inflection point beyond the Debye length is found. This
point represents a characteristic change of convexity of $P(r)$ (as a
function of $\ln r$) induced by screening not by the condensed
layer~\cite{Deserno}; the inflection points associated with this layer
appear when $Z\Gamma\gtrsim 2$. As it was mentioned above, 
for  point-like particles it is expected that counterion condensation
begins at $2-\Gamma$; however, as consequence of finite radius of
particles, the onset of inflection points for different values of
$\Gamma$ takes place at larger values, as can be inferred from
figure~\ref{Guest_Rmann}. It is important to note that smallest values
of $r_{in}$ arise as an effect of the hard-disk radius, as becomes
clearer from the inset.

As figure~\ref{Guest_Rmann} shows, increasing $\Gamma$ produces a
thicker layer of condensed counterions, that is, the counterion
condensation is enhanced as a consequence of increasing microions
correlations.  If the renormalized charge is considered as a result of
the condensation phenomenon, then a smaller value of $Z_{ren}\Gamma$
than the one predicted by PB theory is expected. It is indeed the
case, as the figure~\ref{Guest_Zren} displays: at high enough
$Z\Gamma$, the renormalized charge decreases as the bare charge
increases because $r_m$ becomes larger.

\subsubsection{Asymmetric TCP}

In order to study the effects of the charge asymmetry on $Z_{ren}$, we
carried out a similar analysis to the one presented in the previous
section when particles forming the TCP are characterized by $z_+=1$
and $z_-=1/2$. If all particles were point-like, the stability region
would be restricted to $-2<Z\Gamma<4$. We choose to work with these
``valences" because, some exact results exist concerning the pair
correlation functions of this kind of TCP
\cite{Samaj_aTCP,Tellez_Inversion}. In this section, we consider both
negative and positive values of $Z$, so that, in the cases when $Z>0$
and $Z<0$ we may refer to them as 1:$\frac{1}{2}$ and $\frac{1}{2}$:1
cases, respectively.

\begin{figure}[!htbp] 
\begin{center}
\includegraphics[width=0.48\textwidth]{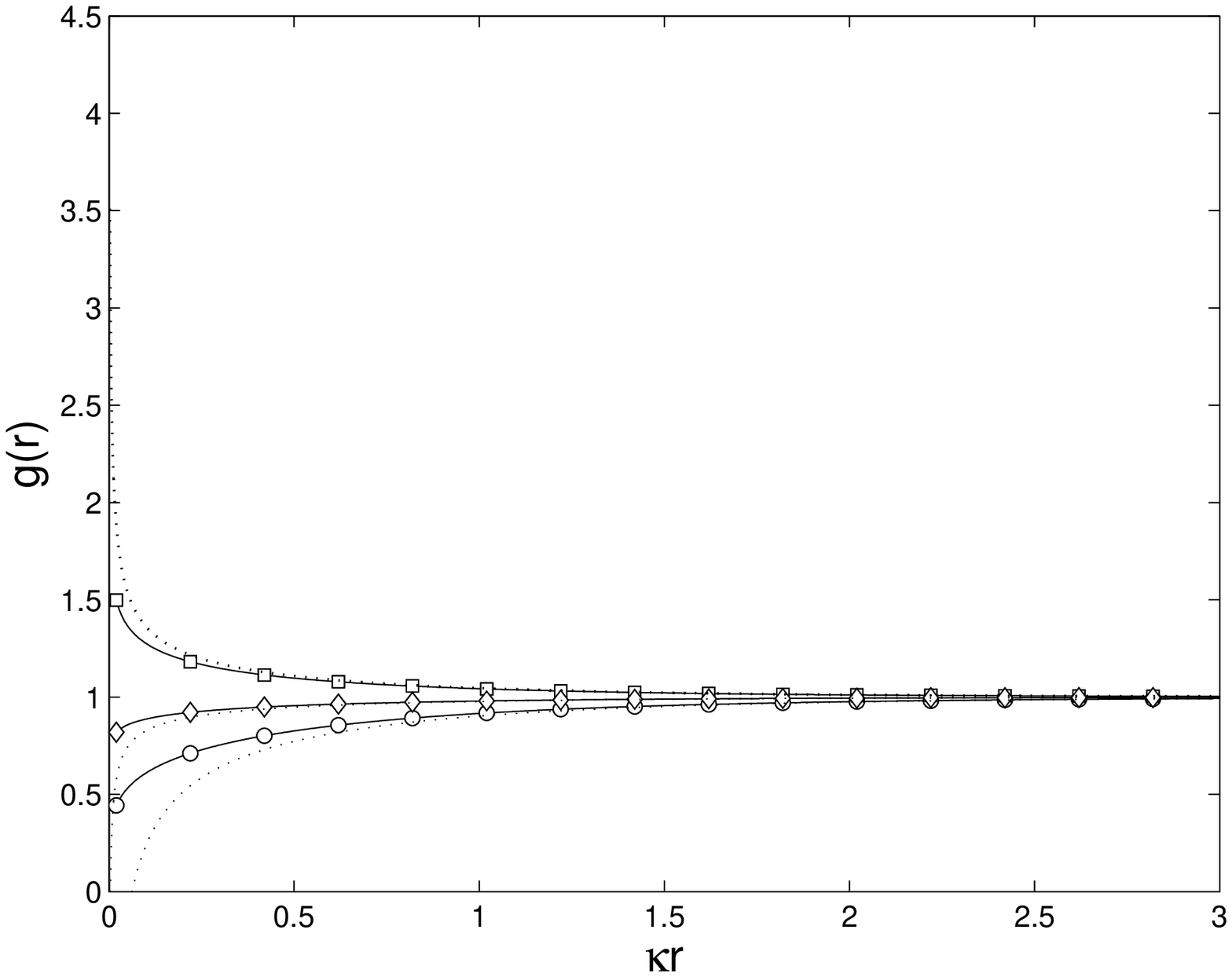}
\includegraphics[width=0.48\textwidth]{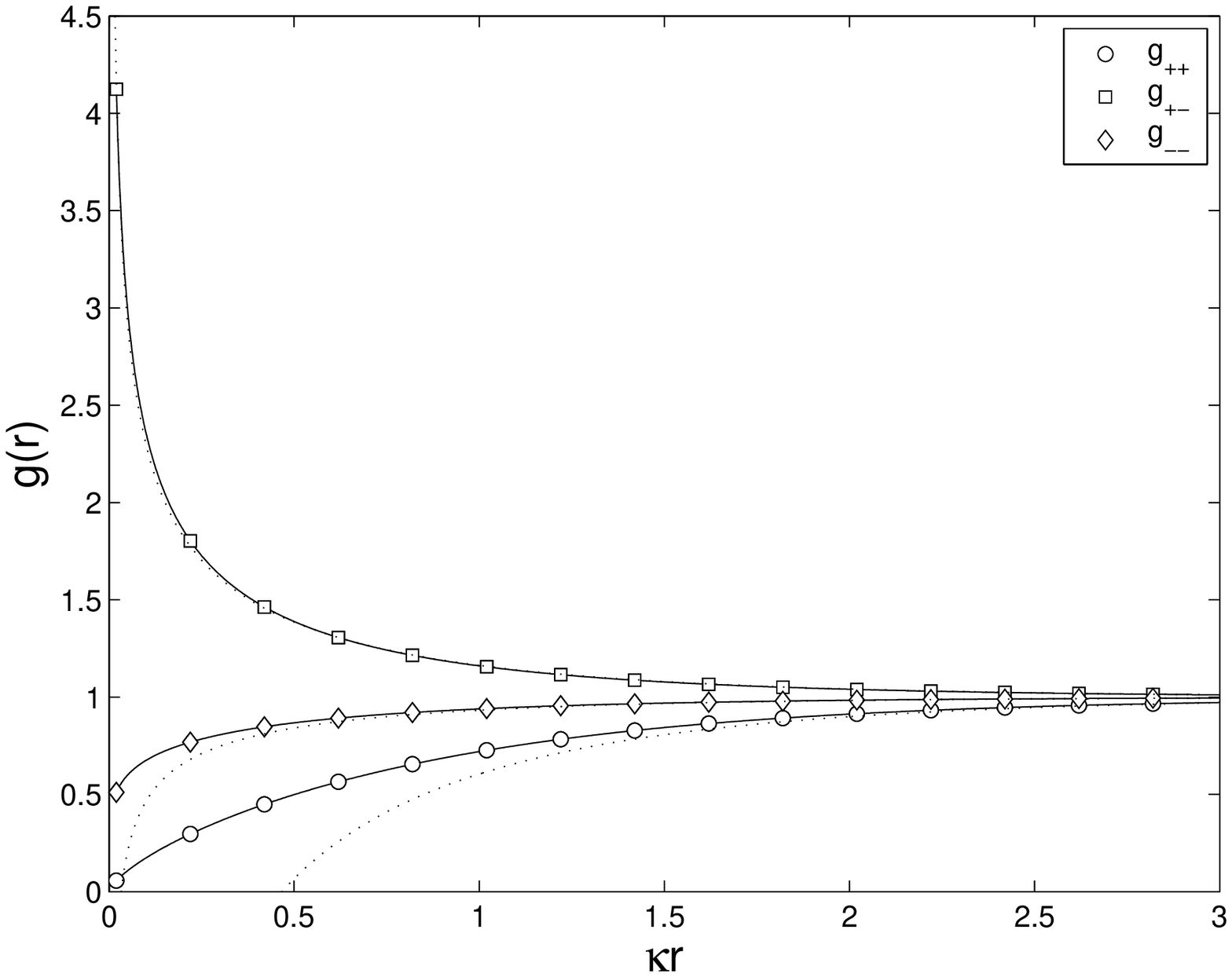}
\begin{quotation}
\caption{ Correlation functions between microions at $\Gamma=0.2$ (left) and 
$\Gamma =0.7$ (right) for the asymmetric TCP. Dotted lines denote theoretical 
asymptotic limits given by field theoretical arguments \cite{Samaj_aTCP}.} 
\label{Guesta_grpm}
\end{quotation}
\end{center}
\end{figure}

Figure~\ref{Guesta_grpm} shows the correlation functions between the
microions at two different values of $\Gamma$. Due to our
approximation, in the infinite dilution limit the correlation
functions $g_{ss\prim}(r)$ are independent on $Z$. As it can be seen, at
$\Gamma=0.2$ and $\Gamma=0.7$ the agreement between the numerically
evaluated correlation functions $g_{ss\prim}(r)$ and those predicted
by field theoretical arguments in~\cite{Samaj_aTCP} (valid at large
distances) is fairly good at large enough distance ($\kappa r\gtrsim
0.5$ and $\kappa r\gtrsim 1.5$, respectively); this is particularly
the case for $g_{+-}(r)$ and $g_{--}(r)$. Additionally, the
electroneutrality condition is fulfilled for this TCP. It was found
that the zero moment defect was less than $2\times10^{-3}$ for all
values of $Z$ and $\Gamma$ reported in this section. 

\begin{figure}[!htbp] 
\begin{center}
\includegraphics[width=0.48\textwidth]{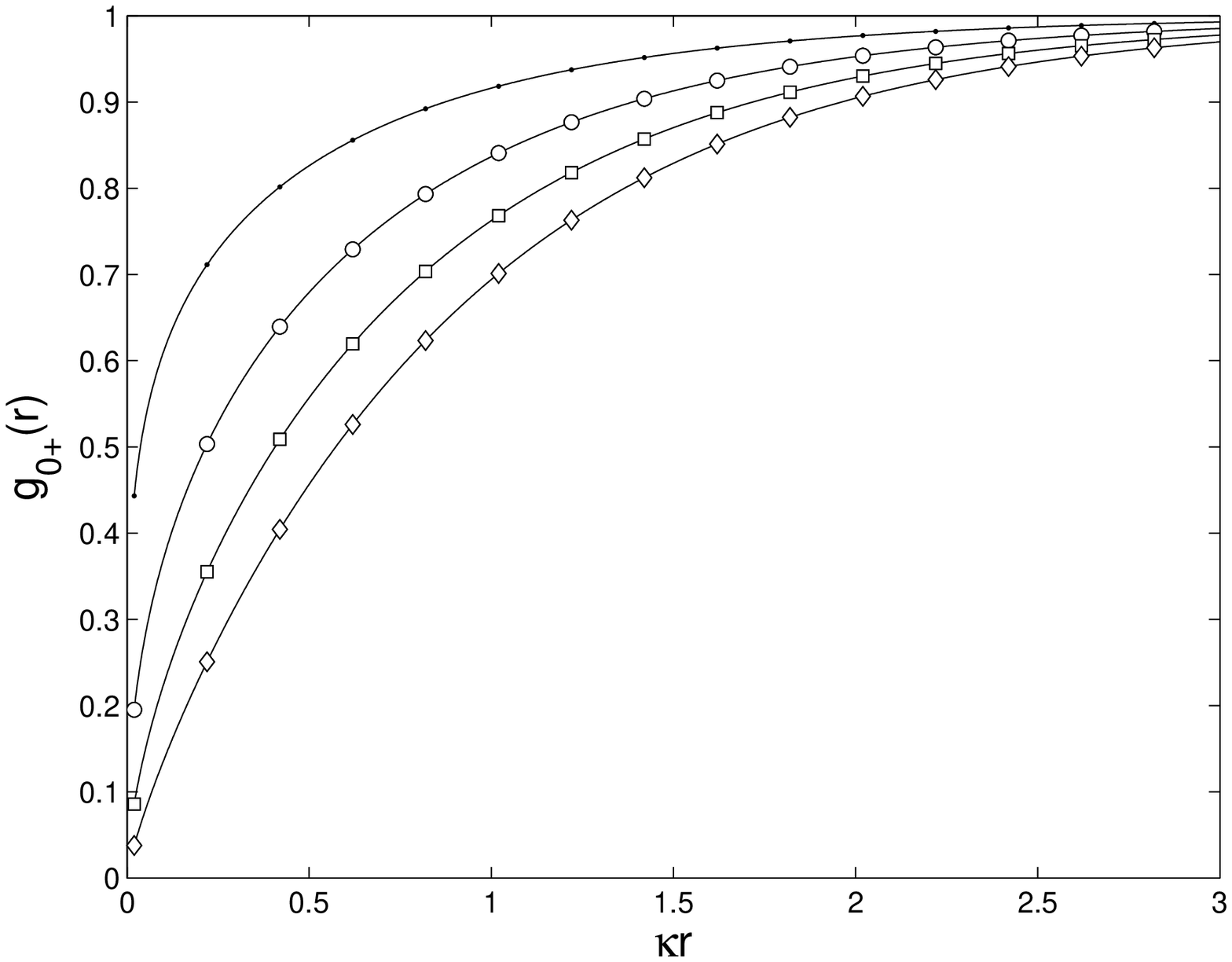}
\includegraphics[width=0.48\textwidth]{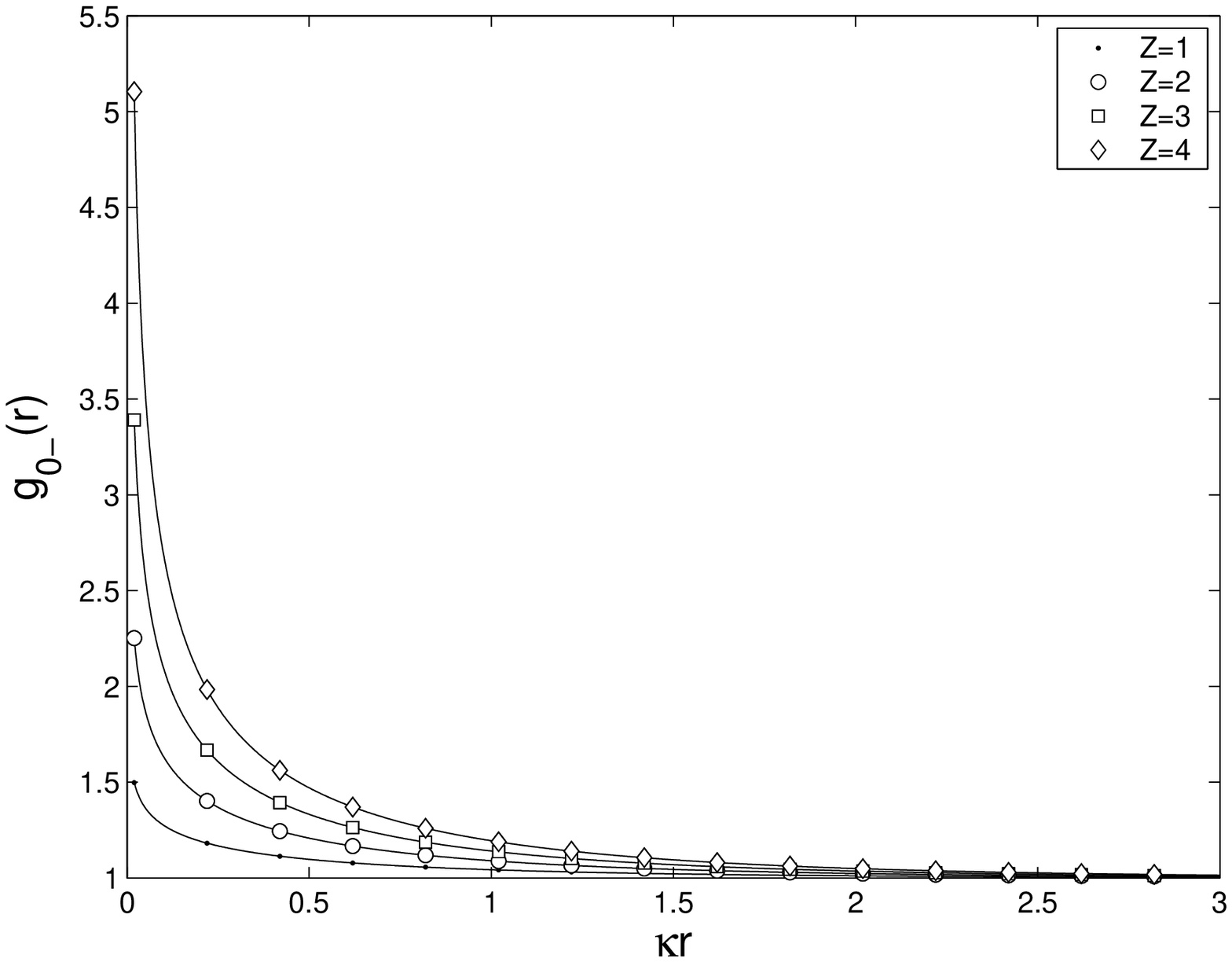}
\includegraphics[width=0.48\textwidth]{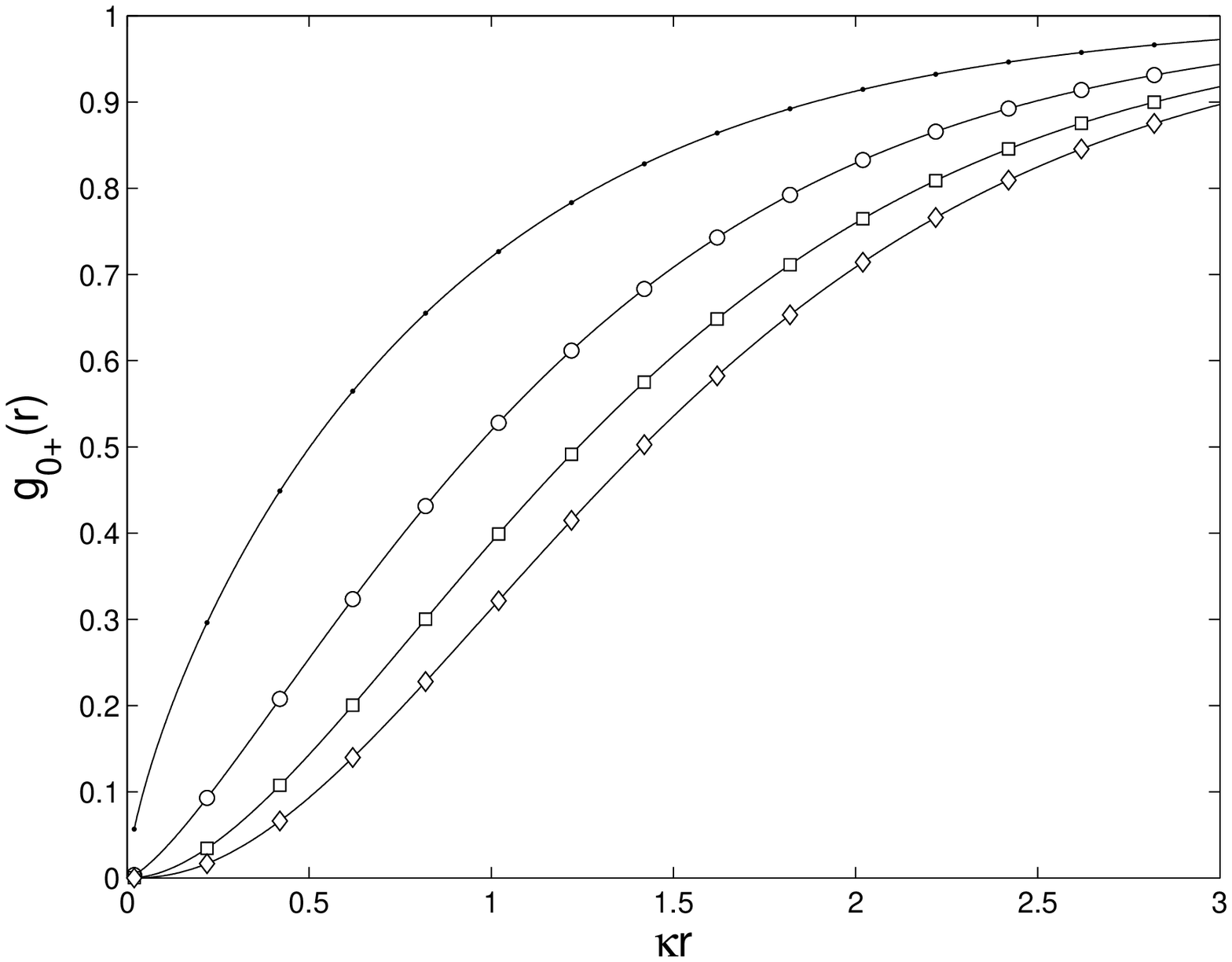}
\includegraphics[width=0.48\textwidth]{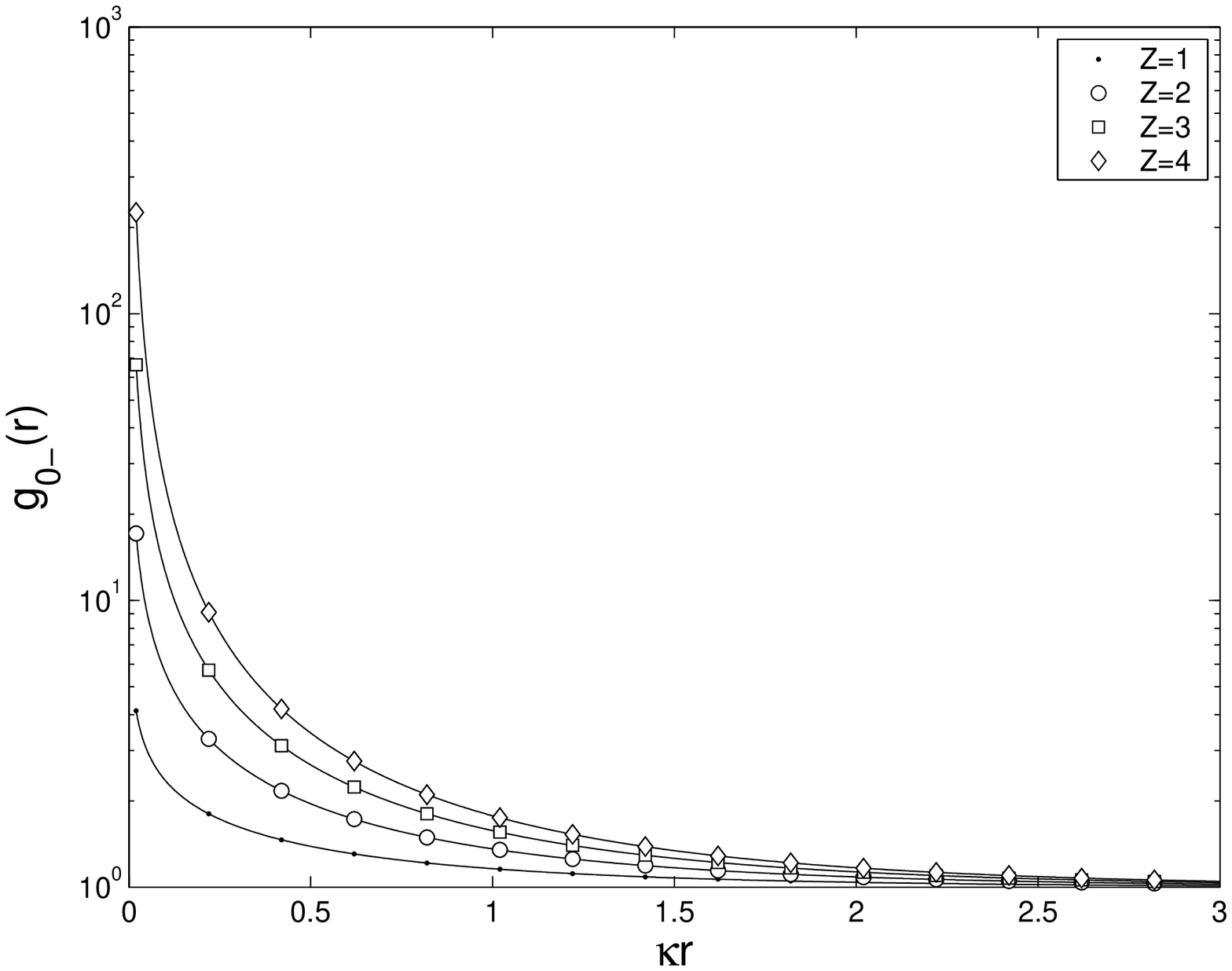} 
\begin{quotation}
\caption{Correlation functions $g_{0\pm}(r)$ for a guest charge immersed in an asymmetric 
TCP for different values of $Z>0$ (1:$\frac{1}{2}$ case). The curves were evaluated 
at $\Gamma=0.2$  (top) and $\Gamma=0.7$ (bottom).}  
\label{Guesta_groi1}
\end{quotation}
\end{center}
\end{figure}

\begin{figure}[!htbp] 	
\begin{center}
\includegraphics[width=0.48\textwidth]{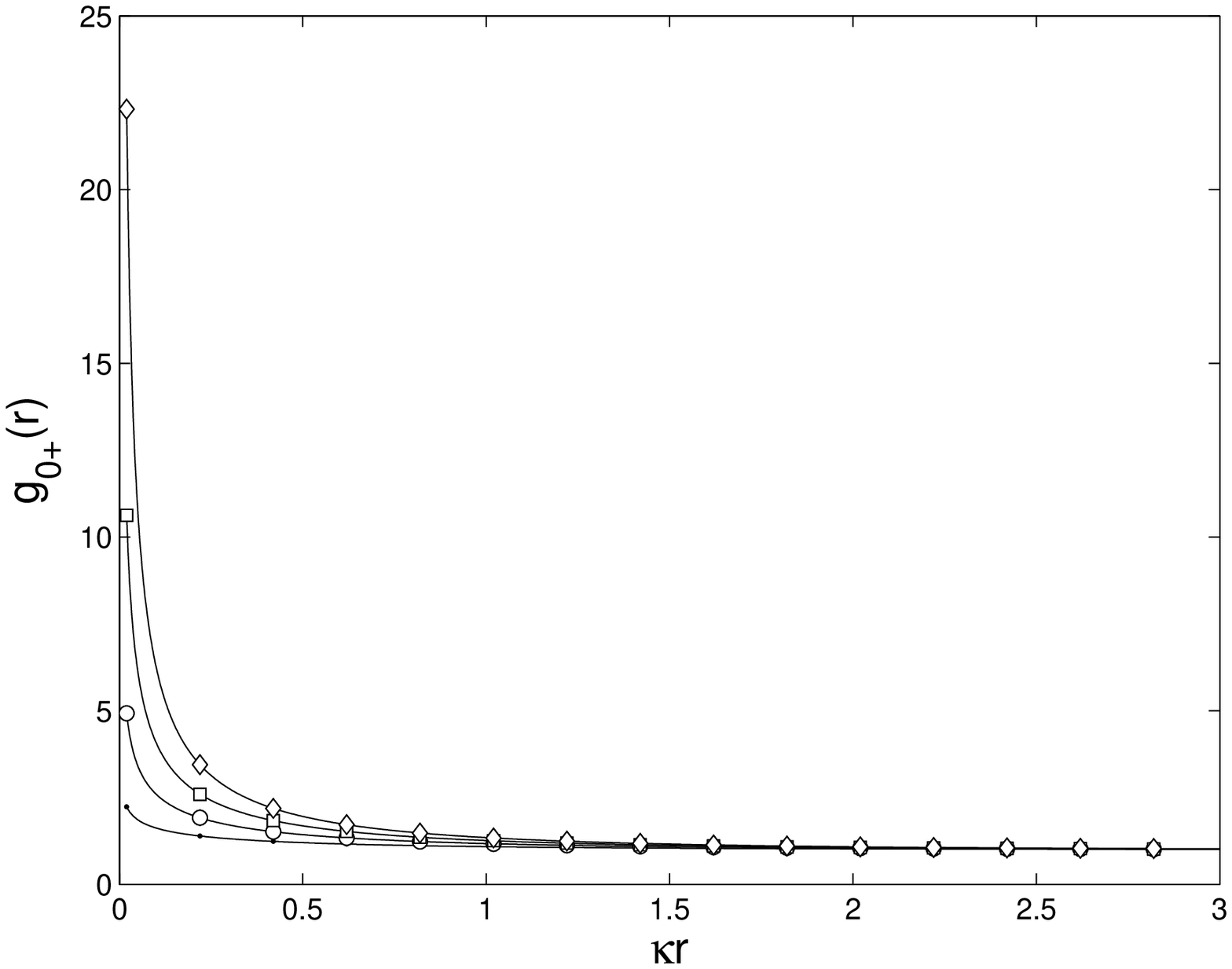}
\includegraphics[width=0.48\textwidth]{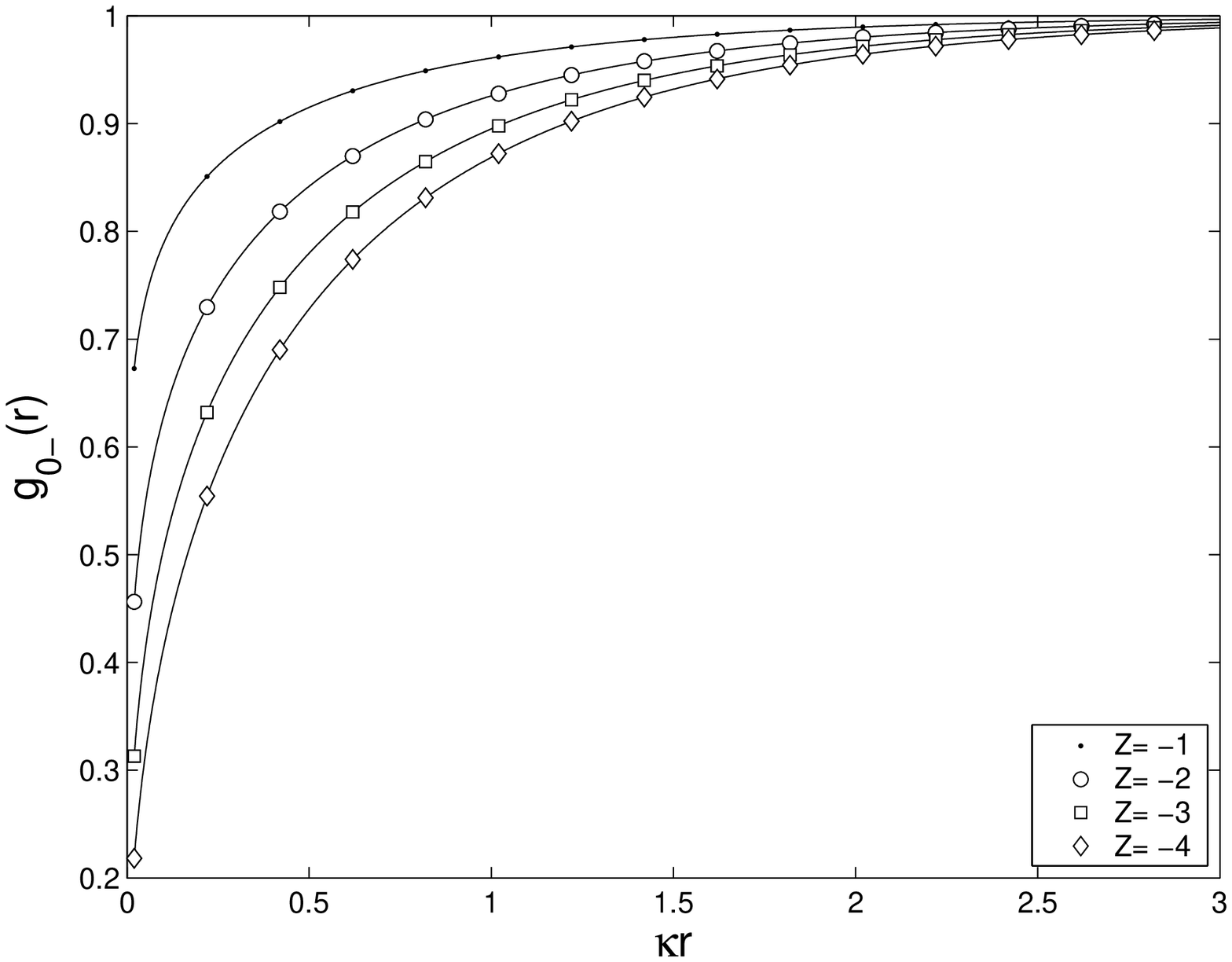}
\includegraphics[width=0.48\textwidth]{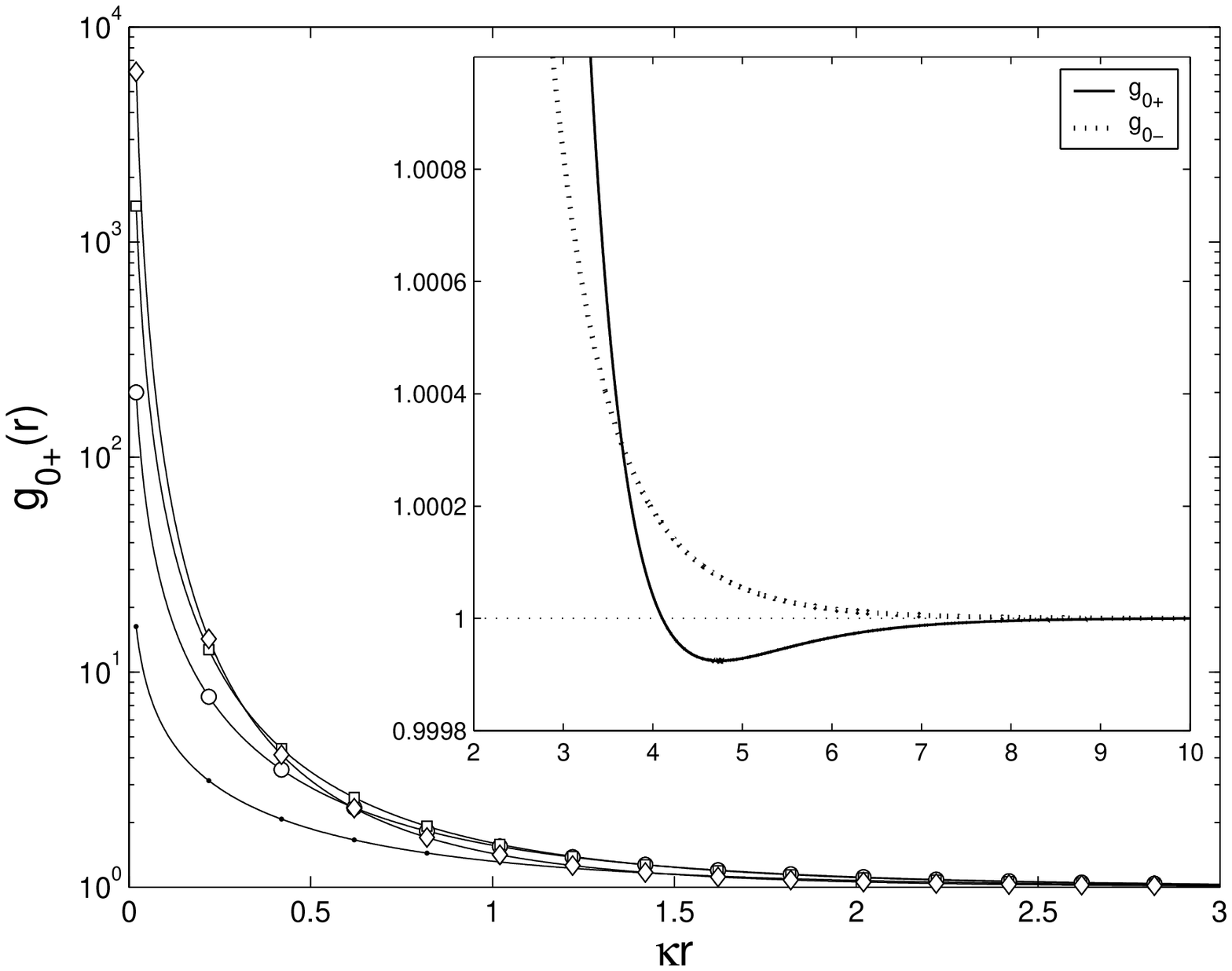}
\includegraphics[width=0.48\textwidth]{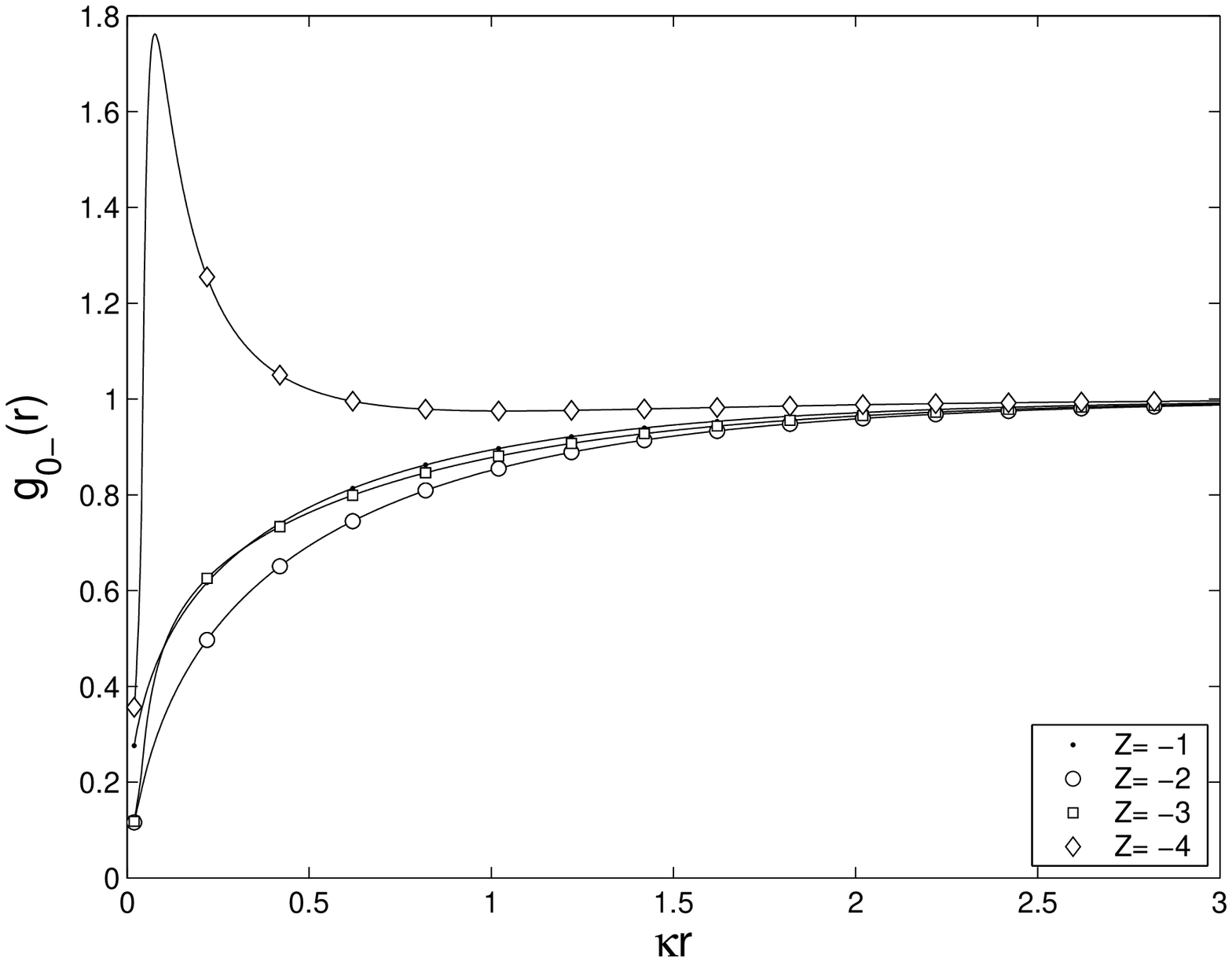}
\begin{quotation}
\caption{Same as the figure (\ref{Guesta_groi1}) but $Z<0$ ($\frac{1}{2}$:1 case). The inset shows a detail of $g_{0\pm}(r)$ for $Z=-3.6$ and $\Gamma=0.8$.}  
\label{Guesta_groi2}
\end{quotation}
\end{center}
\end{figure}

Figures~\ref{Guesta_groi1} and~\ref{Guesta_groi2} show plots of the
correlation functions between the guest particle and the microions at
different values of the bare charge for 1:$\frac{1}{2}$ and
$\frac{1}{2}$:1 cases, respectively. These curves were calculated for
$\Gamma=0.2$ and $\Gamma=0.7$. As it can be expected, the correlations
between the guest particle and the counterions exhibit larger values
when the latter ones have higher valence. It should be noted that the
curves behave monotonically and do not cross for any coupling in case
1:$\frac{1}{2}$; the same happens at $\Gamma=0.2$ in $\frac{1}{2}$:1
case. However, for the latter case at $\Gamma=0.7$ it is observed at
larger $Z$, that the counterion profile near the guest particle (which
is associated with $g_{0+}(r)$) falls off much faster; also, the
coions profile shows a peak at $\kappa r\approx 0.1$. As in the
symmetric case, these facts suggest the presence of a counterion shell
strongly bounded to the guest particle, which considerably screens the
repulsion between the guest charge and the coions, and consequently
plays an important role in the renormalized charge $Z_{ren}$ setting.

\begin{figure}[!htbp] 
\begin{center}
\includegraphics[width=0.48\textwidth]{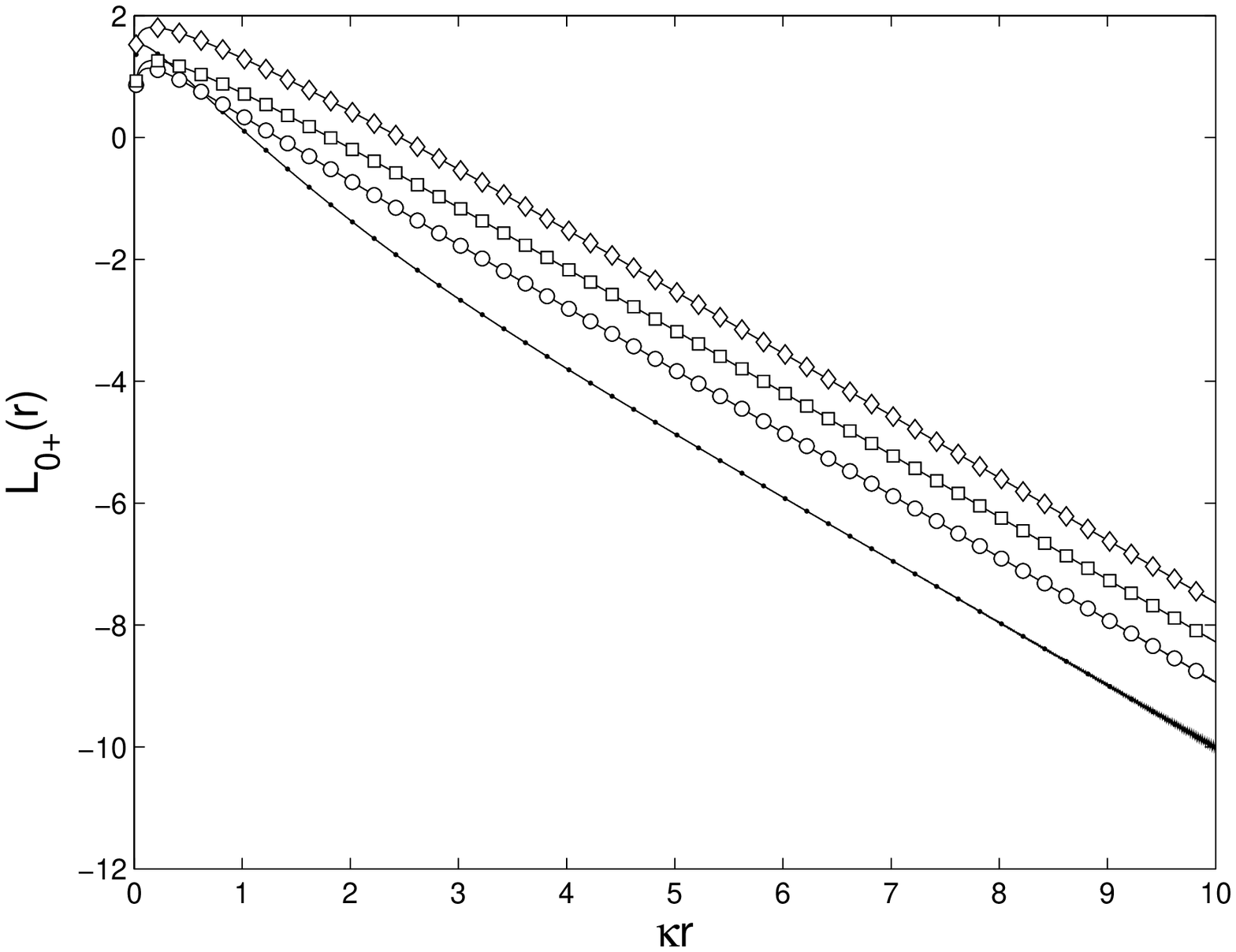}
\includegraphics[width=0.48\textwidth]{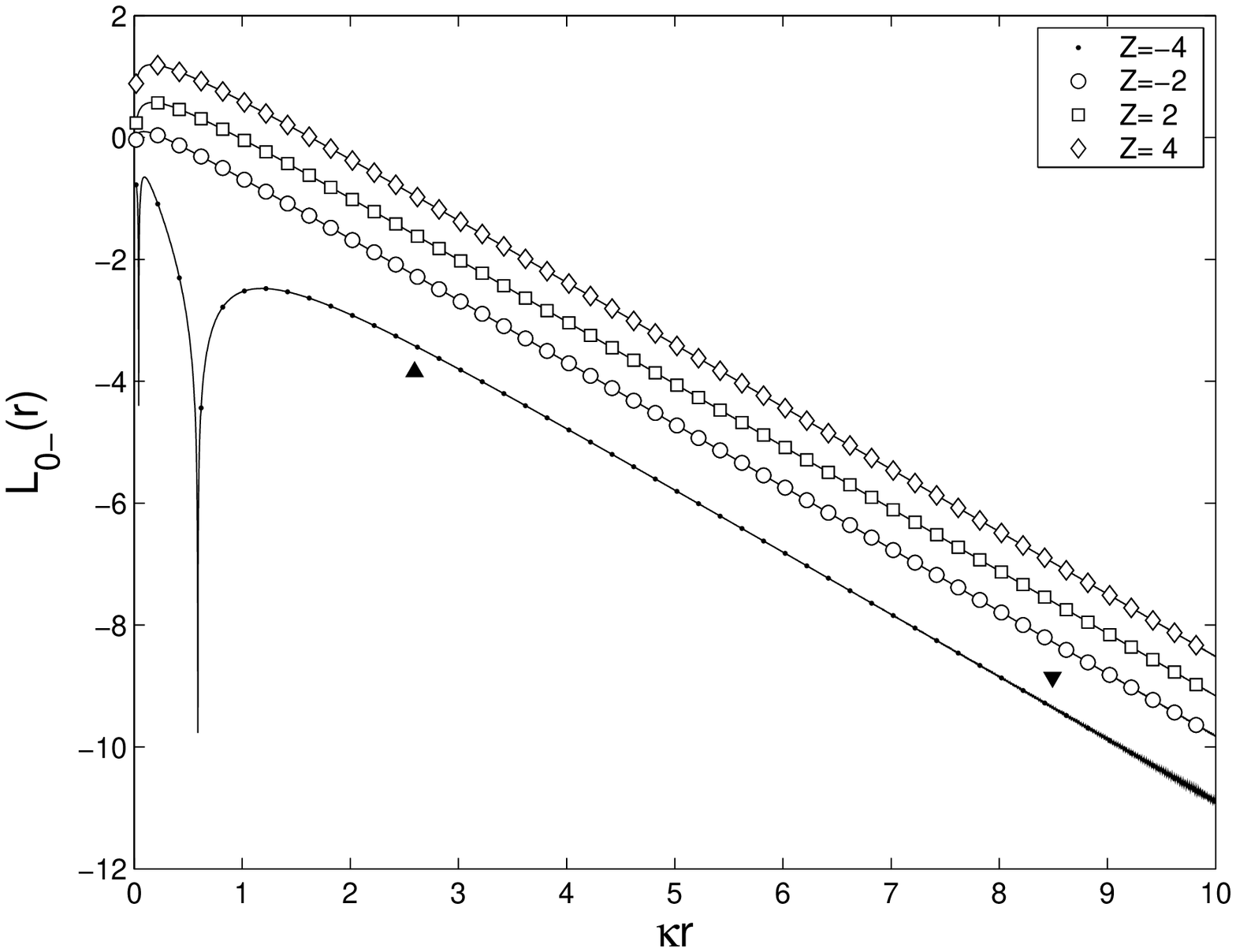}
\begin{quotation}
\caption{Functions $L_{0\pm}(r)$ used to evaluate renormalized
  parameters. The curves were calculated from the correlation functions at  $\Gamma=0.7$. Triangles delimited the range of $r$ according to $10^{-4} \le |\log g_{0s}(r)| \le 10^{-2}$ for $Z=-4$.}  
\label{Guesta_Lroi}
\end{quotation}
\end{center}
\end{figure}

As in the symmetric case, the renormalized parameters were evaluated
by a linear fit of the functions $L_{0\pm}(r)$ as those shown in
figure~\ref{Guesta_Lroi}, where the curves displayed were evaluated at
$\Gamma=0.7$ and for different values of bare charge $Z$. In addition
to the condition $10^{-4} \le |\log g_{0s}(r)| \le 10^{-2}$, in some
cases it was mandatory to introduce empirically a minimum value of $r$
in order to select an adequate range to be fitted. Such a situation
was found particularly at large coupling and for $Z<0$, where the
correlation function between the guest charge and the coions exhibits
a pronounced oscillation as it is shown in figure~\ref{Guesta_Lroi} 
for $Z=-4$.

Within the range of couplings explored, $0.1 \leq\Gamma\leq 0.9$, the
renormalized inverse screening length $\kappa_{ren}$ is found to be
systematically larger than $\kappa$, and it is an increasing function
of the coupling $\Gamma$, in agreement with the exact
results~\cite{Tellez_Guest}. Notice that this is the opposite behavior
than the one found in the symmetric electrolyte. The ratio
$\kappa_{ren}/\kappa$ only differs from 1 by at most 3\% for the
largest coupling $\Gamma=0.9$.

\begin{figure}[!htbp] 
\begin{center}
\includegraphics[width=0.7\textwidth]{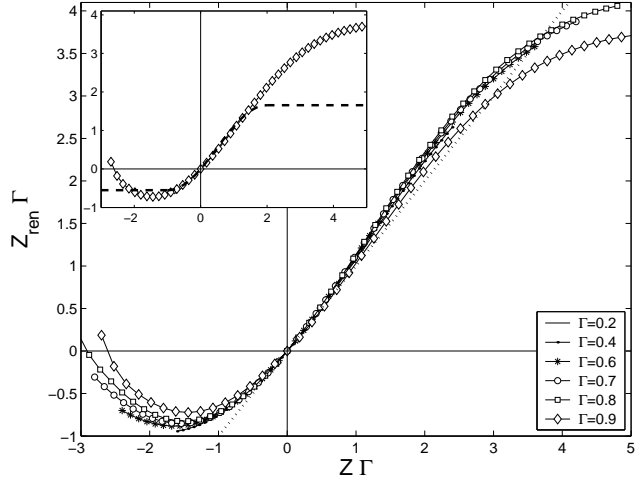}
\begin{quotation}
\caption{Renormalized charge for a guest charge immersed in an
  asymmetric TCP at different coupling parameters. Dotted line
  indicates the condition $Z_{ren}=Z$. In the inset the HNC result for
  $\Gamma=0.9$ is compared to the one predicted by PB theory  (dashed line). } 
\label{Guesta_ZKap_ren}
\end{quotation}
\end{center}
\end{figure}

In figure~\ref{Guesta_ZKap_ren}, the renormalized charge resulting
from the fitting procedure is illustrated as a function of $Z$.  The
global behavior of $Z_{ren}$ shares much of the features described in
\cite{Tellez_Inversion}, where an analysis based on the integrable
complex Bullough-Dodd (cBD) model was used.  The most remarkable
outcome corresponds to $\frac{1}{2}$:1 case at $\Gamma\ge 0.8$. In
such a situation, we observe a very strong neutralization of $Z_{ren}$
as $|Z|\Gamma$ increases beyond 2, which gives rise to the charge
inversion phenomenon, i.e.~the renormalized and bare charges have
opposite signs.

Because of the asymmetry, the correlations between microions provoke
an extra attraction of the counterions to, and repulsion of the coions
from, the guest particle.  This causes the accumulation of a large
number of counterions around of the guest particle, as shown in
figure~\ref{Guesta_groi2}, whose charge exceeds the amount necessary
to neutralize $Z$. As it was mentioned, charge inversion is predicted
by an exactly solved 2D model based on cBD
theory~\cite{Tellez_Inversion}. However, it has also been reported in
systems having other symmetries by using the HNC approximation
\cite{LegerLevesque, Greberg}. The decisive characteristic of the
systems in which this process takes place is the asymmetry (either of
size or charge) among coions and counterions. Particularly, when the
ions are equally sized particles, it seems necessary that the
counterions have larger valence (in absolute value) than the coions,
as in the present situation.

The charge inversion phenomenon involves microions correlations and
therefore the mean field PB treatment is not able to predict it. In
addition to the charge inversion, in the case 1:$\frac{1}{2}$ the
``overshooting'' phenomenon predicted in~\cite{TellezTrizac},
i.e.~effective charge becomes larger than the bare charge for
intermediate values of the latter one, is also observed. In such
situation, coions are repelled further away in the double layer than
they are in the symmetric case.  The inset of
figure~\ref{Guesta_ZKap_ren} illustrates the comparison between HNC
with PB results; the latter ones were obtained from
($\sigma_0=\sigma=0$)~\cite{TellezTrizac_PB} 
\beq Z_{ren}\Gamma =
\frac{\sqrt{3}}{\pi}\lpa 2\sin\lpa\frac{\pi
Z\Gamma}{3}-\frac{\pi}{6}\rpa+1\rpa.
\label{ZGren_asim}
\eeq 
Equation (\ref{ZGren_asim}) is strictly valid when
$-1<Z\Gamma<2$, i.e.~if $Z\Gamma$ does not exceed the Manning
thresholds for counterion condensation, otherwise $Z_{ren}\Gamma$
saturates to $\frac{-\sqrt{3}}{\pi}$ and $\frac{3\sqrt{3}}{\pi}$,
respectively \cite{Tellez_Inversion}.

\begin{figure}[!htbp] 
\begin{center}
\includegraphics[width=0.48\textwidth]{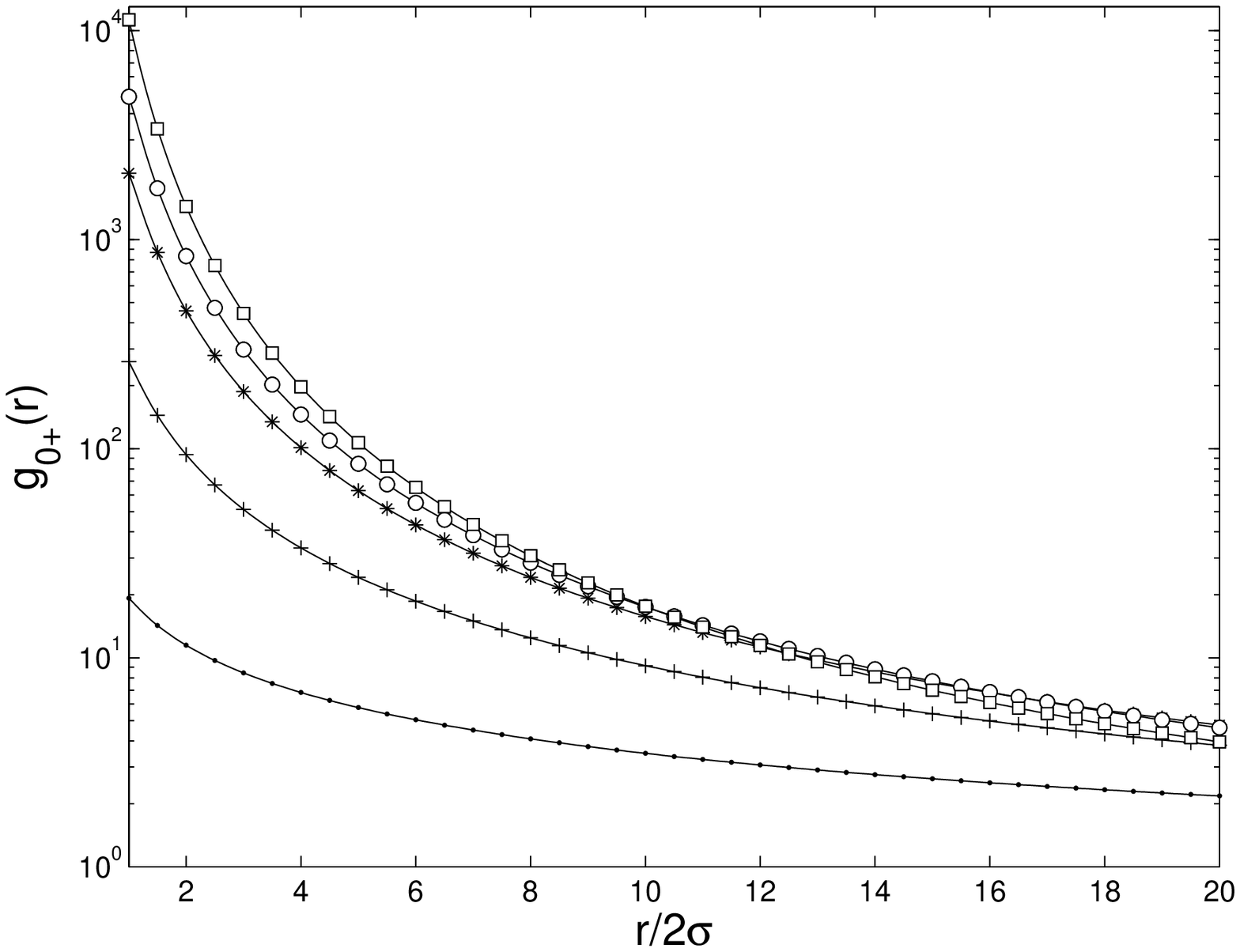}
\includegraphics[width=0.48\textwidth]{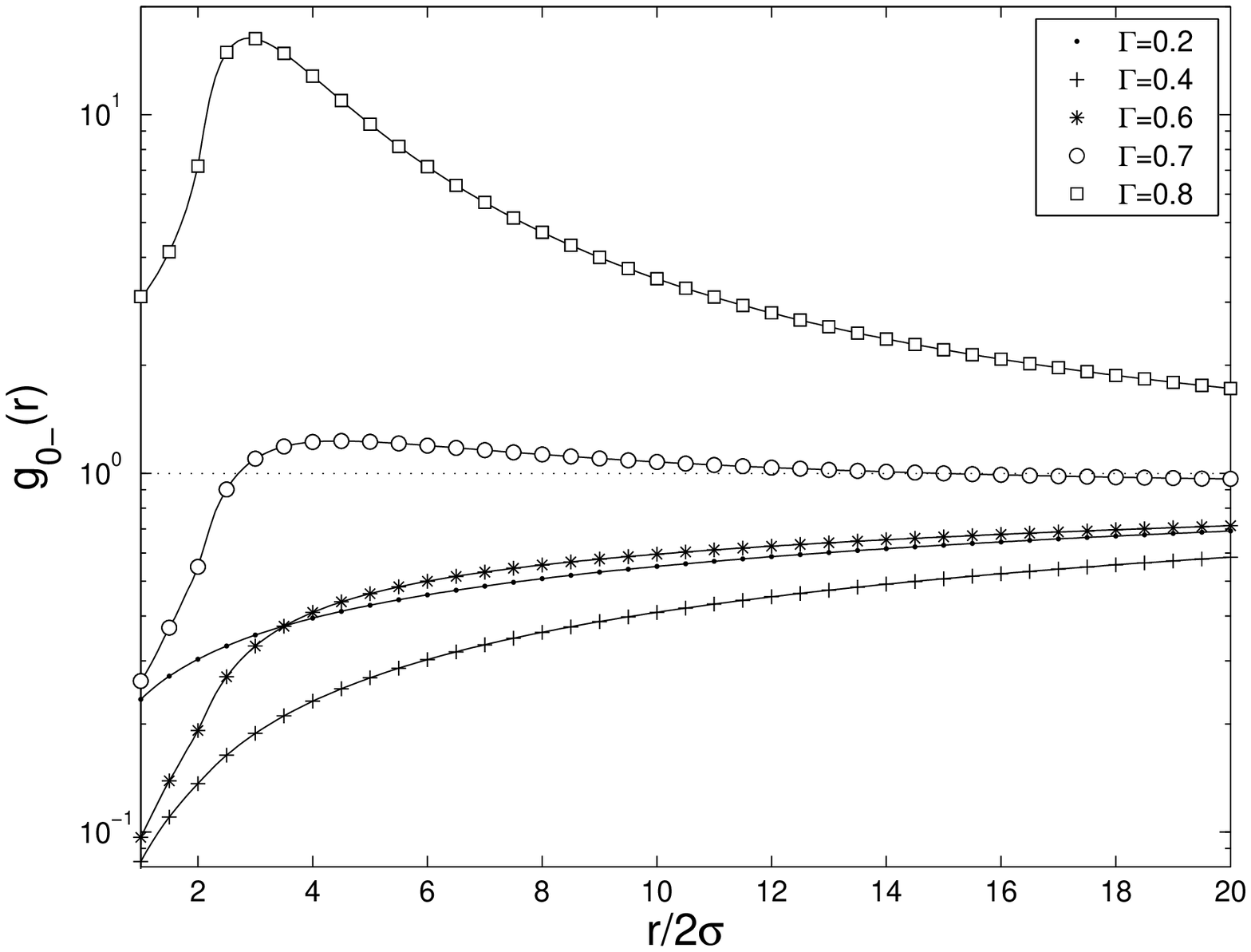}
\begin{quotation}
\caption{Short distance behavior of correlation functions $g_{0\pm}(r)$ for $Z=-3.6$ and  different coupling parameters.} \label{Guesta_inversion}
\end{quotation}
\end{center}
\end{figure}

\begin{figure}[!htbp] 
\begin{center}
\includegraphics[width=0.7\textwidth]{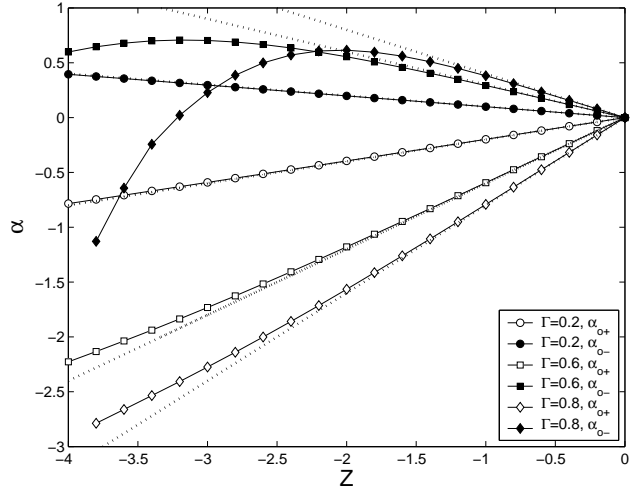}
\begin{quotation}
\caption{$\alpha_{0s}$ factors characterizing the short-distance behavior of the potential of mean force for different values of $\Gamma$. Dotted lines indicate  factors related to the Coulomb potential  ($|\alpha_{0\pm}^{(coul)}|=|Zz_\pm\Gamma  |$).} 
\label{Guesta_slope}
\end{quotation}
\end{center}
\end{figure}

As in the symmetric case, condensation thresholds mark the change of
behavior of the correlation functions at short
distance~\cite{Tellez_Inversion}. In figures~\ref{Guesta_inversion}
and~\ref{Guesta_slope} the short-range correlation functions and the
$\alpha_{0\pm}$ factors are plotted (see equation
(\ref{meanforce_short})). As before, the $\alpha_{0\pm}$ factors
determine the strength of the potential of mean force $w_{0\pm}(r)$
between the guest particle and the microions at short distance. The
curves were evaluated at $Z=-3.6$ and for different values of the
coupling parameter $\Gamma$. From the figure we can observe that there
exists a point from which the interaction between the guest particle
and the coions (i.e. negative ions) becomes less and less
repulsive. As before, that change can be interpreted as a first step
in the counterion condensation \cite{Tellez_Inversion}. At
$\Gamma=0.8$ that change is much more dramatic to the extent that the
interaction begins to be attractive. Subsequent to the sign change in
the short-range of $w_{0-}(r)$, the inversion of the renormalized
charge (which is evaluated at large distance) takes place. On the
contrary, the interaction between the guest charge and the counterions
is hardly modified as $|Z|$ increases, i.e.~at all coupling
parameters, the counterions seem to ``see'' just the bare charge at
short distance.

\begin{figure}[!htbp] 
\begin{center}
\includegraphics[width=0.48\textwidth]{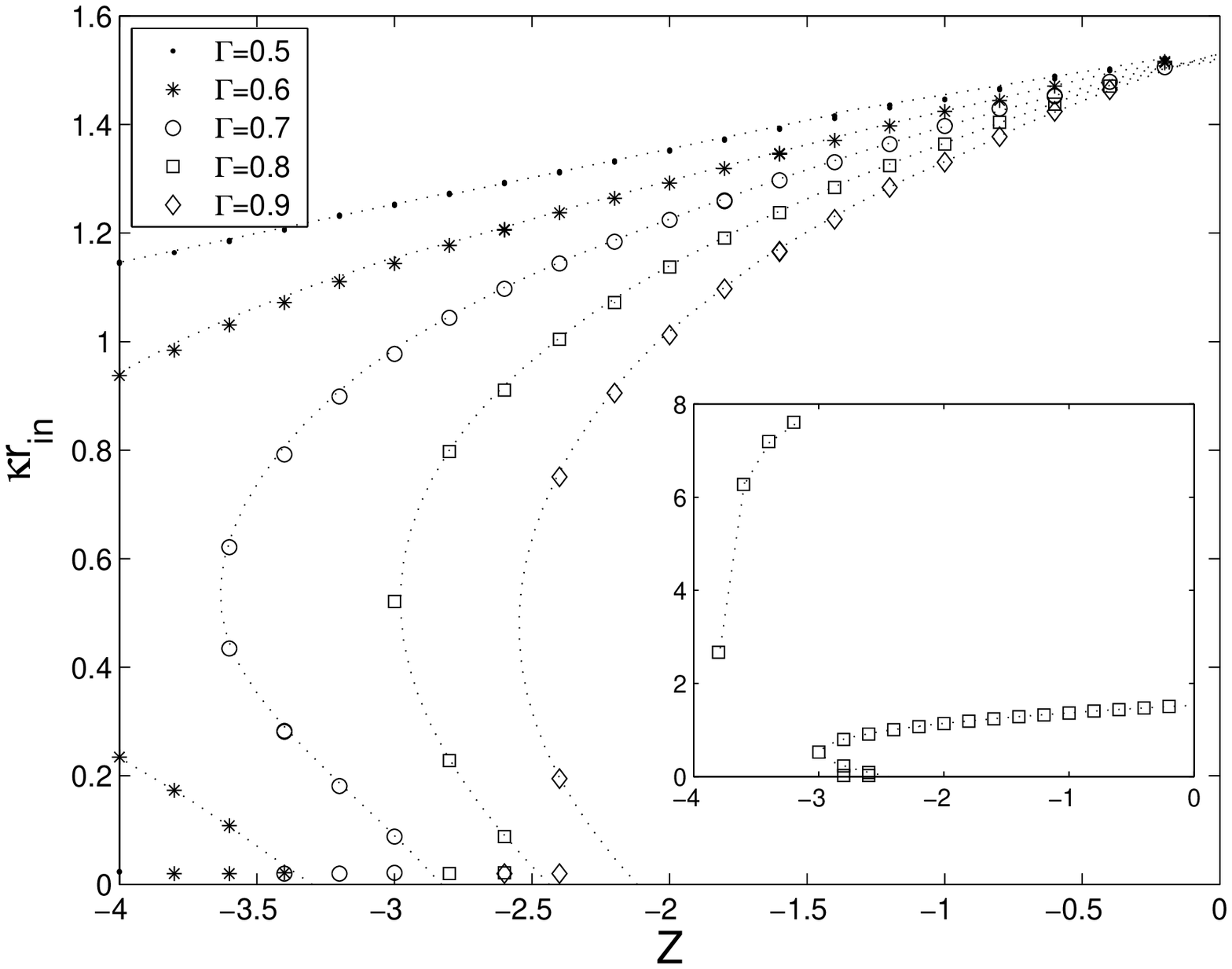}
\includegraphics[width=0.48\textwidth]{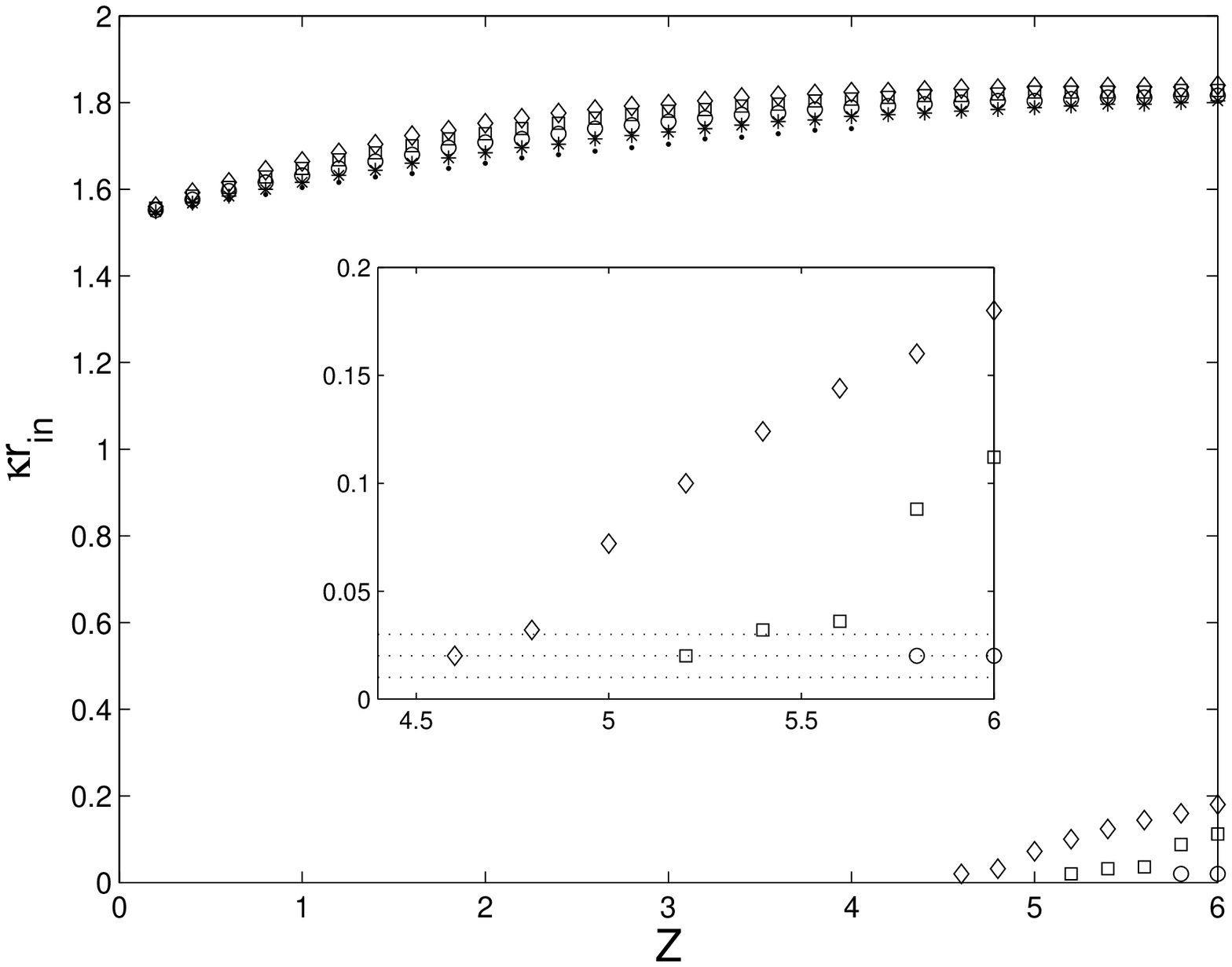}
\begin{quotation}
\caption{Inflection points of charge distribution as a function of $Z$
  and $\Gamma$ for $\frac{1}{2}$:1 (left) and 1:$\frac{1}{2}$ (right)
  cases. At the left, the inset shows an additional  inflection point at
  larger distance when $\Gamma=0.8$, which is related to the charge
  inversion. Dotted lines were obtained by  interpolation but are
  intended  as a guide to eye. As in the symmetric case, the smallest
  values of $r_{in}$ are related to the finite radius of particles, as
  dotted lines indicate in the inset at the right.} 
\label{Guesta_Rmann}
\end{quotation}
\end{center}
\end{figure}

In order to study the thickness of the condensed layer we use again
the inflection point
criterion~(\ref{inflection_crit}) for the integrated charge
distribution $P(r)$. Figure~\ref{Guesta_Rmann} shows the
locus of the inflection point for both $\frac{1}{2}$:1 and
1:$\frac{1}{2}$ cases. In the first place, for the case
1:$\frac{1}{2}$ it is found that inflection points are located beyond
the Debye length when $\Gamma<0.8$. In this way, for the considered
range of $Z>0$ the renormalized charge is well described by the PB
approach at coupling strength below 0.8. On the other hand, when
$\Gamma\gtrsim 0.8$ and $Z\gtrsim4.5$ a new inflection point is
detected at $\kappa r_{m}\sim 0.1$, i.e.~at these bare charge and
coupling parameters values the correlations give rise to the
counterion condensation. Note that at such conditions the renormalized
charge is close to reach its maximum value, as
figure~\ref{Guesta_ZKap_ren} suggests.

With regard to the $\frac{1}{2}$:1 case, a richer scenario is found.
For a large enough coupling, two inflection points are found, the
smallest of them corresponding to $r_m$. As in the other cases, the
onset of $r_m$ is associated with coupling conditions at which the
maximum (absolute) value of $Z_{ren}$ is reached. Upon a further
increase of $|Z|$ these inflection points shift one toward the other
and finally coalesce and annihilate, as it can be seen in
figure~\ref{Guesta_Rmann}. In addition, the bare charge at which such
annihilation takes place, $Z^*$, is related to the onset of a peak in
the coions density, i.e. $g_{0-}(r)$.

Within the cell model of polyelectrolytes, that annihilation  occurs
when the typical salt  screening length (Debye length) interferes with
the size of the condensed counterion layer and therefore it  becomes 
unnecessary to distinguish between condensed and uncondensed counterions 
\cite{Deserno}.  Moreover, it has been argued about  the merit of  the 
inflection point criterion, which  unifies  the  phenomenology of  an 
infinite dilution/finite salt system (as the studied one here) and that
of the a finite density/vanishing salt cell model \cite{TellezTrizac_PB, 
Shaughnessy}. However, our results suggest a  different situation: 
because of the microions correlations, for  $Z>Z^*$ a second layer 
mainly constituted by coions  starts  to be formed at some distance 
of the guest particle. 

On the other hand, according to the two-state model, the distribution
of microions around the macroion can be divided into a condensed
counterion region and a free population; the latter one being
considered within the linearized PB theory. So, the role of a
condensed layer is to impose a new boundary condition to the electric
potential and consequently fixing the value of the renormalized charge
\cite{Shaughnessy}.  However, since there is no inflection point of
$P(r)$ for $Z>Z^*$, we may not differentiate such populations at large
enough coupling (i.e. strong microions correlations). In fact, the
renormalized charge and subsequently the charge inversion are
determined by a ``layered'' ionic cloud, as the insets in
figures~\ref{Guesta_groi2} and~\ref{Guesta_Rmann} suggest. An
additional discussion about that ``layered'' cloud will be carried out
later.

\begin{figure}[!htbp] 
\begin{center}
\includegraphics[width=0.7\textwidth]{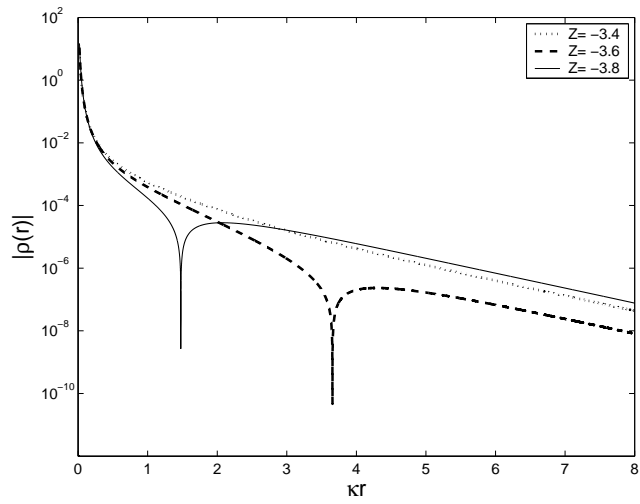}
\begin{quotation}
\caption{Absolute value of the charge density around guest particle for the $\frac{1}{2}$:1 case at $\Gamma=0.8$. Peaks indicate a change of sign of $\rho(r)$ and therefore a change of sign in the renormalized charge. Note that the onset of such peaks imposes a restriction on the  $r$-range  used to the fitting procedure.}  \label{Guesta_rhor}
\end{quotation}
\end{center}
\end{figure}

In this way, the process behind  charge inversion  appears to be  a  similar one to the secondary cause of the ``Mechanism I'' proposed  in \cite{Greberg}. Since the net charge around the guest particle has opposite sign to that of  $Z$,  coions at some distance from the guest charge are attracted electrostatically towards the guest particle, while the counterions are repelled electrostatically from it, leading to the existence of a range of $r$-values  at which the mobile ion system  is locally opposedly charged, as figure~\ref{Guesta_rhor} corroborates for $\Gamma=0.8$.

\subsection{Colloidal suspension at infinite dilution limit }
\newcommand{\An}{\overset{\text{\tiny{o}}}{\text{\footnotesize{A}}}}

As it was previously mentioned, 2D Coulomb systems can be used as a first approach to model some rod-like colloids and polyelectrolytes. Many synthetic as well as biological examples   can be classified within this category. Depending  on the number of charges per monomer unit bound to the backbone, the linear charge density $\lambda$ can be tuned, so that typical dimensionless values comprise $|\xi|= |\lambda| l_B = 3-6.6$, where $l_B$ is the Bjerrum length. 

In order to study the effective charge in a little more realistic model 
of a cylindrical macroion immersed in an electrolytic solution,  we follow
  similar assumptions to those given in \cite{LegerLevesque} and 
  \cite{Tellez_Inversion}. At first, we consider a long charged 
cylindrical macroion with diameter $a_0=20\An$ immersed in an electrolytic 
solution composed by ions with diameter $a = 5\An$, all of  them interacting 
through a logarithmic potential. Temperature is fixed through the Bjerrum 
length $l_B=7.2\An$, which corresponds to aqueous suspensions at $T =$ 293 K.
 In order to study the effect  of salinity we consider both symmetric  (1:1) 
 and asymmetric  (1:2 and 2:1)  electrolytes at different concentrations $n$. 

\begin{figure}[!htbp] 
\begin{center}
\includegraphics[width=0.7\textwidth]{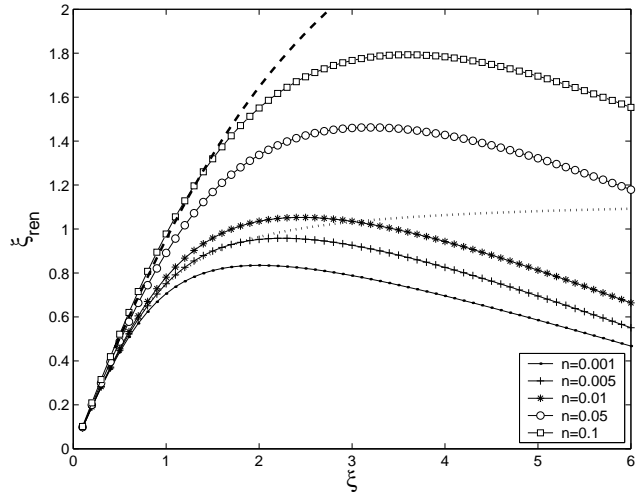}
\begin{quotation}
\caption{Dependence on the salinity of the renormalized charge for a charged cylinder immersed in a 1:1 electrolyte ($[n]$ = M). The renormalized linear charge density is compared to the exact PB results for $n=0.001$M ($\kappa\sigma_c\approx 0.074$, dotted line) and $n=0.1$M ($\kappa\sigma_c\approx 0.74$, dashed line).}  \label{colloid_Zren}
\end{quotation}
\end{center}
\end{figure}

By assuming the DH-like decay of the correlation functions
$g_{0\pm}(r)$ at large distances, we have obtained for the symmetric
case the renormalized charge exhibited in
figure~\ref{colloid_Zren}. We found a qualitative behavior that
resembles the one shown by the guest charge system.  Again, the most
significant difference between the HNC and analytical PB results
\cite{TellezTrizac_PB} is the presence of a maximum value of
$\xi_{ren}$ followed by a monotonic decay in the HNC outcomes.

A similar dependence has been described for the zeta potential  of a 
charged cylinder \cite{Gonzales}   and for the renormalized charge 
within a spherical cell model of ionic condensation \cite{Groot, Hsin}. 
By using  the density-functional theory based on a  weighted (i.e.~non-local) 
density approximation, the monotonic decrease of $\xi_{ren}$ as $\xi$ 
increases can be explained in terms of the excess chemical potential 
$\mu_{exc}(r)$. As  $\xi$  increase, the number of counterions is also 
increased  due to the electroneutrality condition and therefore 
$\mu_{exc}(r)$ near the particle surface decreases, resulting in   
an additional  number of counterions  attracted towards the surface  
\cite{Groot}. In this way, $\xi_{ren}$ must decrease when $\xi$ surpasses 
the value where the maximum $\xi_{ren}$, which is as well related to the 
plateau predicted by PB theory, is reached.

\begin{figure}[!htbp] 
\begin{center}
\includegraphics[width=0.48\textwidth]{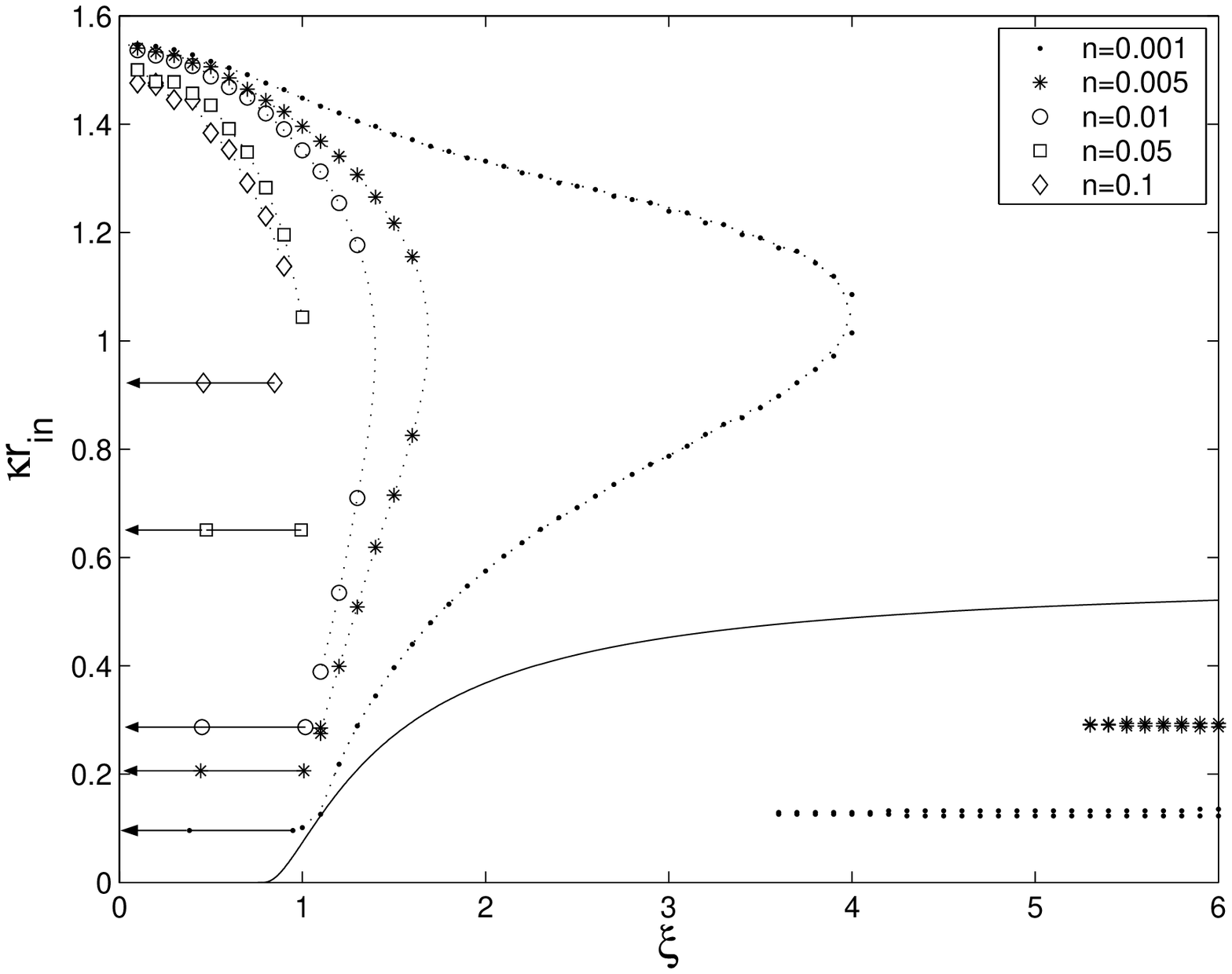}
\includegraphics[width=0.48\textwidth]{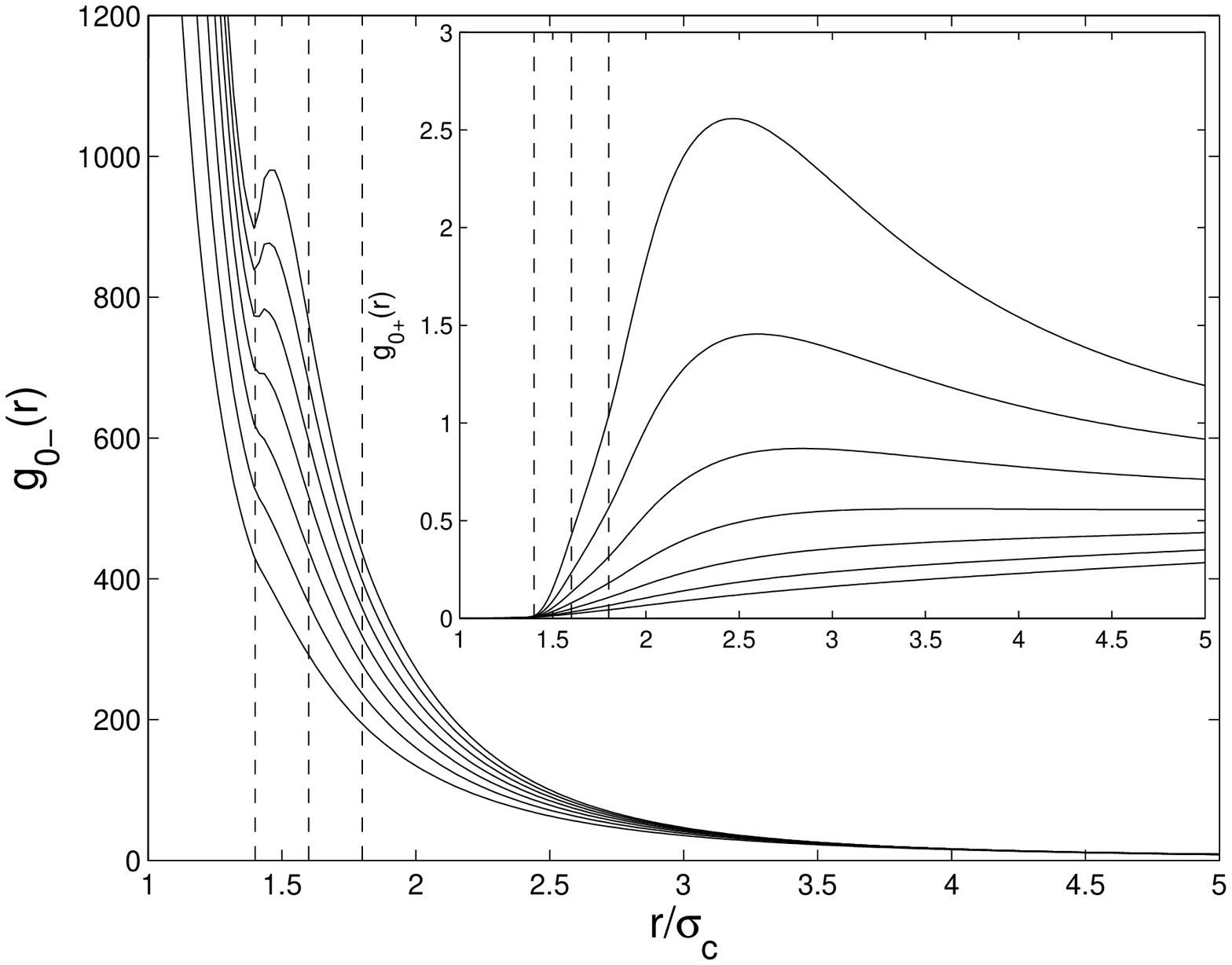}
\begin{quotation}
\caption{Inflection points in the integrated charge distribution
  (left). Arrows indicate values of $\kappa\sigma_c$ for each density
  ($[n]$ = M). Dotted lines are intended as a guide to the eye; the
  continuous line indicates the exact PB result for $n=0.001$
  ($\kappa\sigma_c\approx 0.074$) \cite{TellezTrizac_PB}. At the right
  are shown the correlation functions $g_{0\pm}$ for $\xi=3\rightarrow
  6$ at $n=0.001$M. At higher values of $\xi$, the correlation functions
  exhibit inflection points as a consequence of finite radius of
  microions, which suggest the formation of a second layer of
  counterions. }  \label{colloid_manning_a20}
\end{quotation}
\end{center}
\end{figure}

\begin{figure}[!htbp] 
\begin{center}
\includegraphics[width=0.485\textwidth]{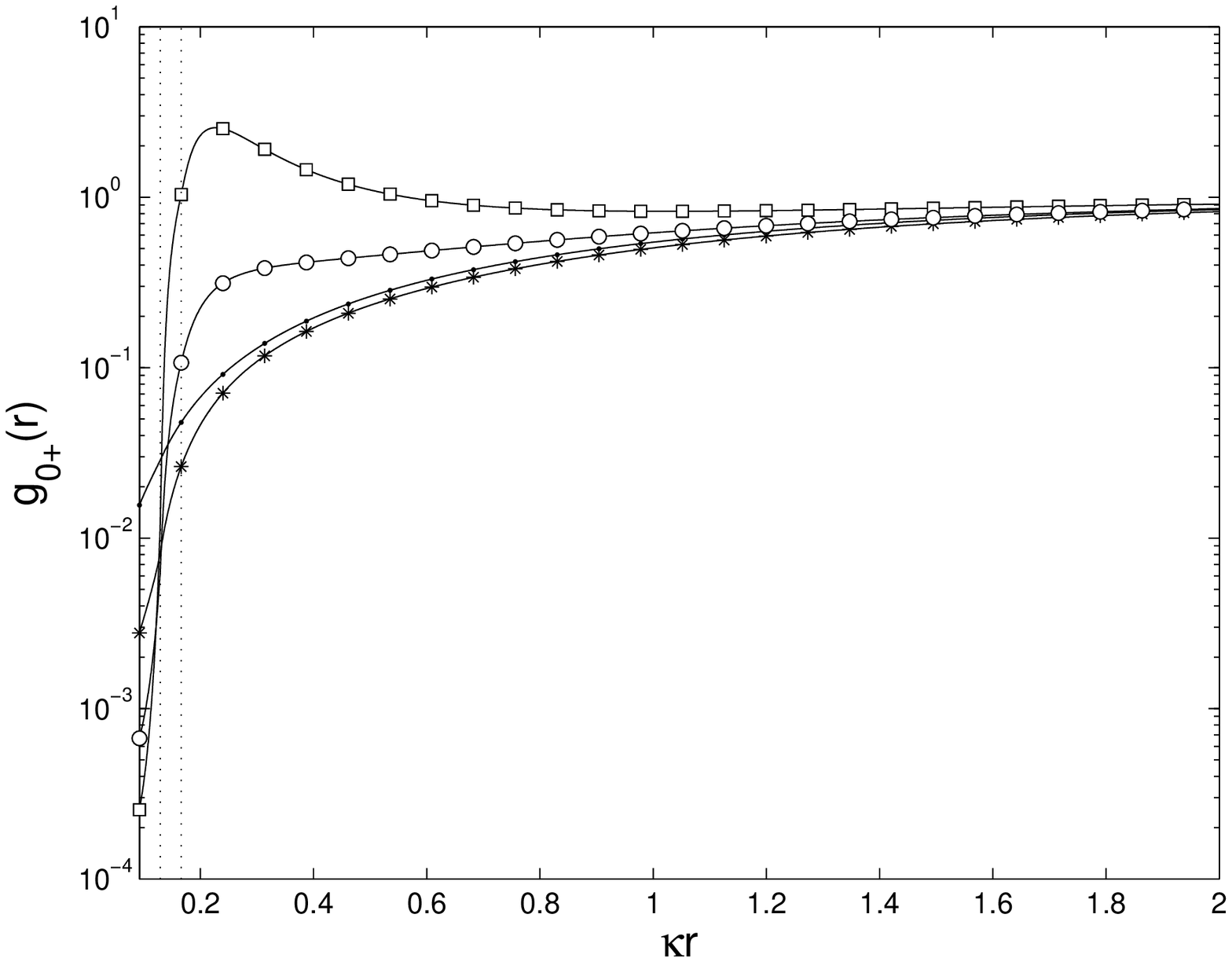}
\includegraphics[width=0.48\textwidth]{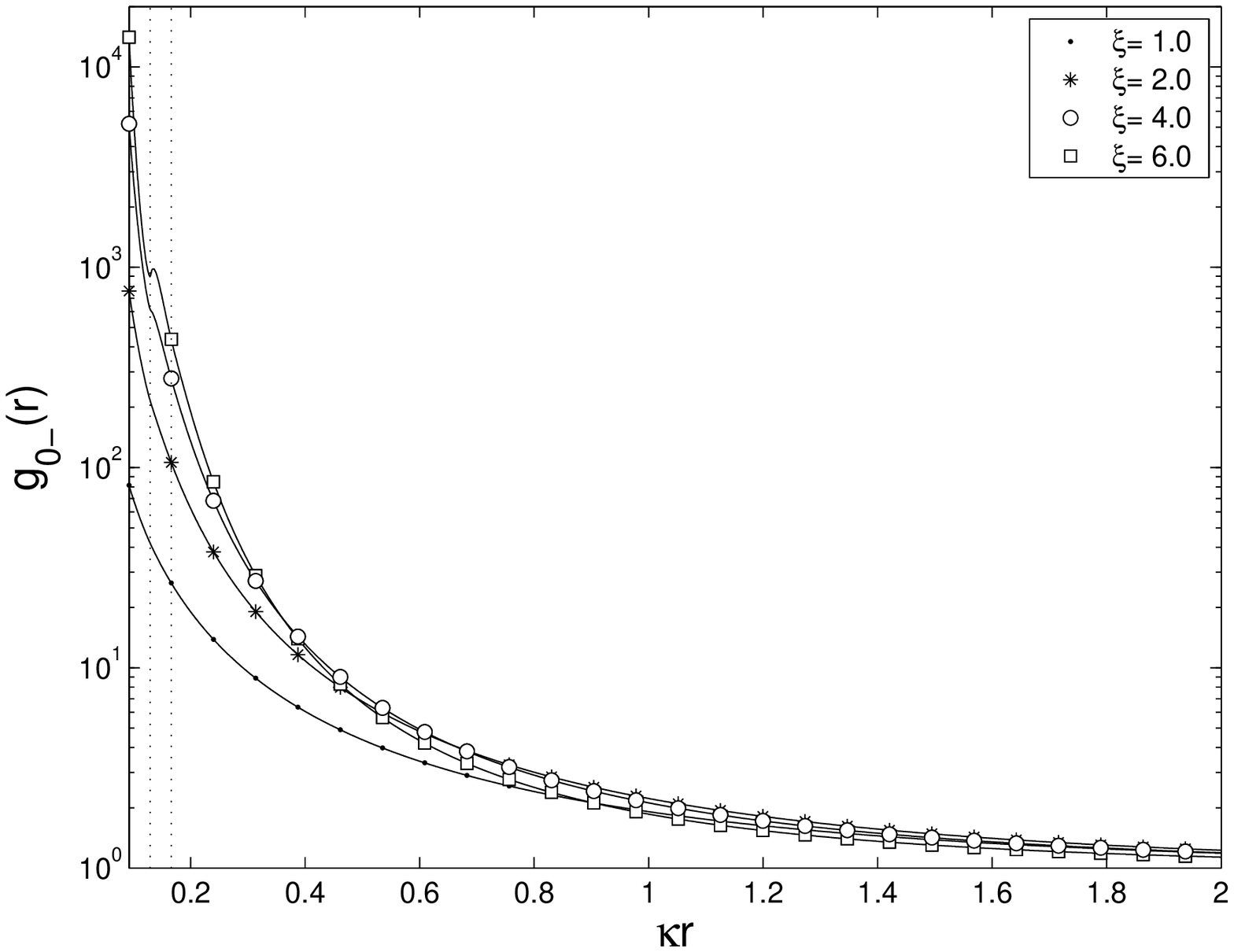}
\includegraphics[width=0.48\textwidth]{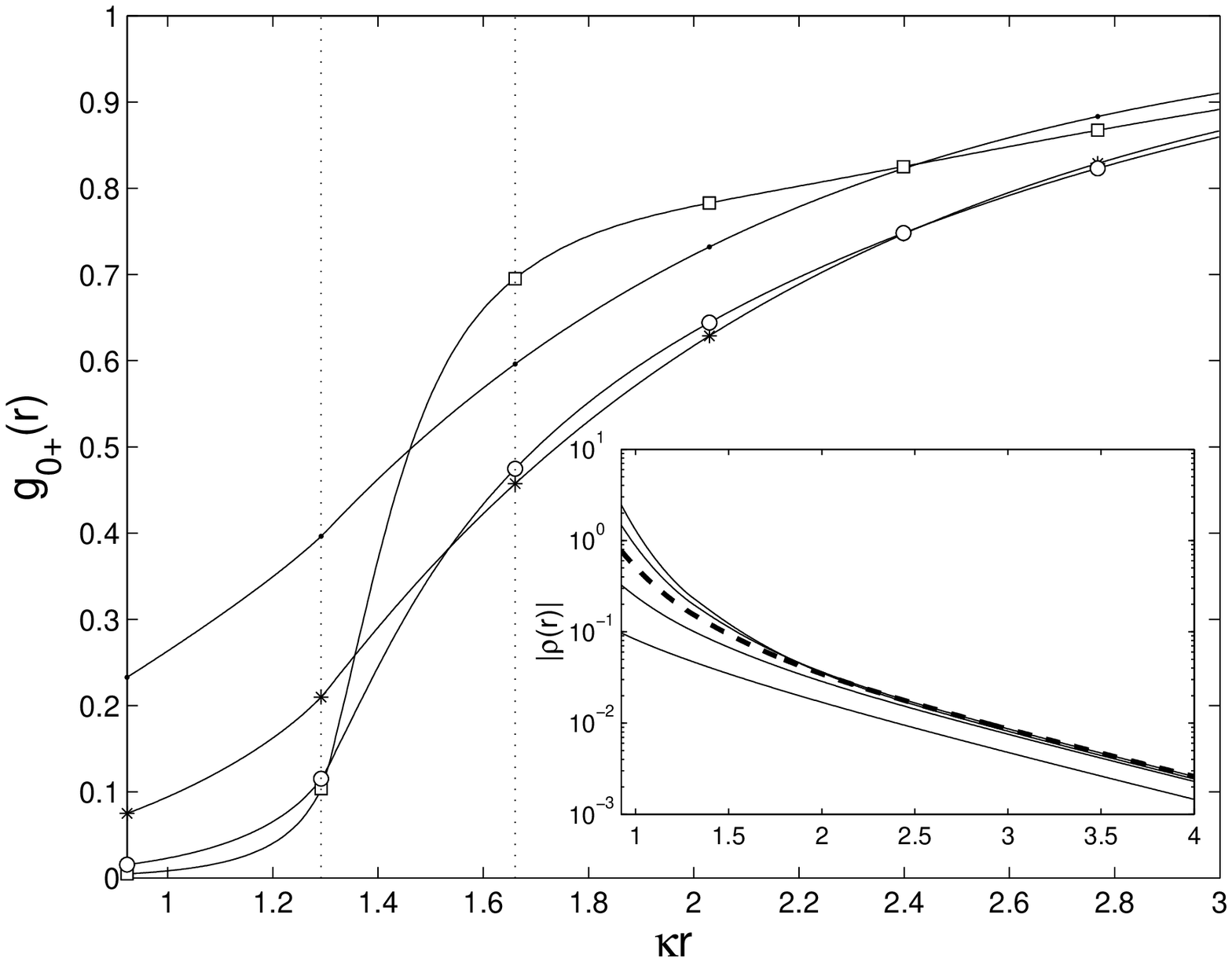}
\includegraphics[width=0.48\textwidth]{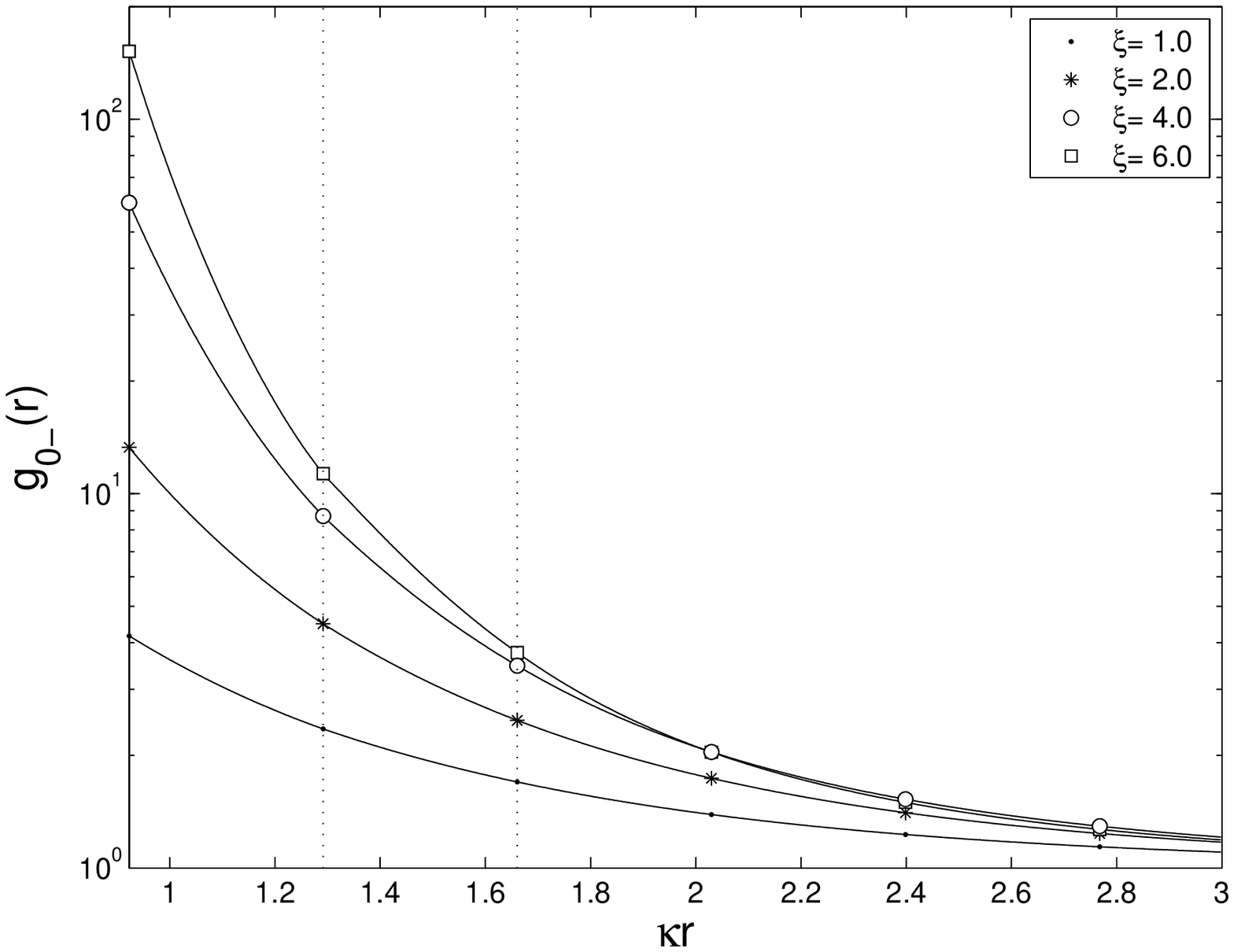}
\begin{quotation}
\caption{Correlation functions $g_{0\pm}(r)$ at $n=0.001$M (top) and $n=0.1$M (bottom) for a cylindrical macroion. Dotted lines delimit the region $r=\sigma_c+2\sigma$ and $r=\sigma_c+4\sigma$. When $n=0.001$M and $\xi=4$, the onset of a peak in $g_{0+}(r)$ takes place. This point is associated with the annihilation of inflections points (see the figure (\ref{colloid_manning_a20})). In the inset is  shown the charge density at $n=0.1$M for $\xi=\{1,2,3,4,5\}$ (dashed line indicates $\xi=3$).  }  \label{colloida_groi1}
\end{quotation}
\end{center}
\end{figure}

As before, the extent of the condensed counterion layer can be
measured by using the integrated charge distribution
(\ref{Charge_r}). In figure~\ref{colloid_manning_a20} inflection
points of $P(r)$ are plotted as a function of the bare linear charge
density for different salt concentrations. At high enough density
($n>0.05$M), the counterion condensation appears to occur at the
macroion surface and involves a single layer of counterions. This
layer is moderately bound to the charged cylinder, as it can be concluded
by comparing the order of magnitude of $g_{0-}(\sigma_c)$ at
$n=0.001$M to that at $n=0.1$M. Beyond this region, the charge
distribution is dominated by screening and falls off in a DH-like
manner after approximately one Debye length, as it is shown in the
inset of the figure~\ref{colloida_groi1}.

At low density ($n<0.01$M), $r_{in}$ depends on $\xi$ in a similar way
to what is shown in figure~\ref{Guesta_Rmann}.  For $n=0.001$M, we can
associate the smallest value of $r_{in}$ to the Manning radius within
the range $1\lesssim\xi\lesssim 3$. Besides, that range roughly
discriminates the region where $\xi_{ren}$ passes from an increasing
to a decreasing function, as it can be seen in figure~\ref{colloid_Zren}.
When $\xi> 3.4$, inflection points related to the finite size of
microions enter to scene, indicating the accumulation of a second
layer of counterions and consequently, the additional decrease in
$\xi_{ren}$. Further increasing $\xi$ produces annihilation of the
inflexion points at $\xi\approx 4$ and simultaneously $g_{0+}(r)$
starts exhibiting a peak, as figure~\ref{colloida_groi1}
illustrates.

Previously, when the charge inversion phenomenon was discussed, we
associated the onset of a peak in the coions density profile to the
settling on of a ``layered'' ionic cloud and we argued that the
renormalized charge is determined by the net charge enclosed in that
cloud (see figure~\ref{Guesta_rhor}). A similar conclusion has
been given from the comparative study at very low ionic strength of
the electrokinetic and the effective charges, which were obtained from
electrophoresis and light scattering data of suspensions of latex
particles, respectively~\cite{Quesada}.

According to the study carried out in \cite{Quesada}, it may be stated
that, interacting at large distance, colloidal particles act as if
they carry an effective charge equal to the one enclosed by the
so-called outer Helmholtz plane (OHP), from which the diffuse double
layer begins. To a certain extent, the definition of OHP can be
associated to the choice of the $r$ value, from which $g_{0+}(r)$ and
$g_{0-}(r)$ fall off symmetrically with respect to 1 in a DH-like
form, i.e. the choice of an adequate range to fit correlation
functions within the linearized theory.  On the other hand, we would
like to point out that the determination of the electrokinetic charge
depends on the previous evaluation of the zeta potential and, as it
was aforementioned, the behavior of $\xi_{ren}$ shown in
figure~\ref{colloid_Zren} and that of the zeta potential for a charged
cylinder found in \cite{Gonzales} are qualitatively similar. In this
way, our results suggest that the HNC approach captures appropriately,
at least qualitatively, the short-range correlations determining the
double layer structure.

\begin{figure}[!htbp] 
\begin{center}
\includegraphics[width=0.65\textwidth]{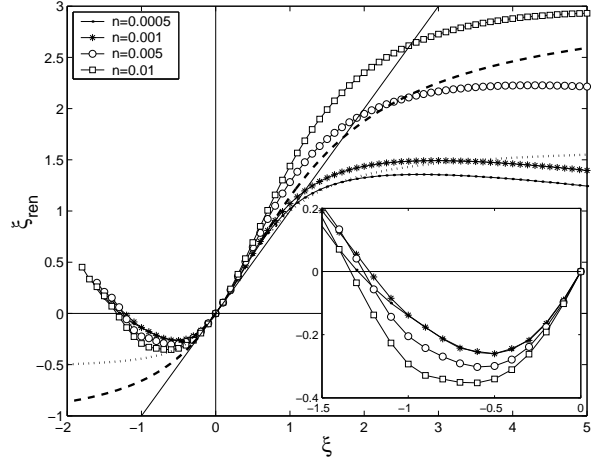}
\begin{quotation}
\caption{Renormalized charge for a cylindrical macroion  immersed in an asymmetric electrolyte at different salinities. Dotted and dashed lines  indicate  analytical PB solutions at $n=0.0005$M ($\kappa\sigma_c\approx 0.9$) and $n=0.01$M ($\kappa\sigma_c\approx 0.41$), respectively. }  \label{colloida_Zren_a20}
\end{quotation}
\end{center}
\end{figure}

\begin{figure}[!htbp] 
\begin{center}
\includegraphics[width=0.48\textwidth]{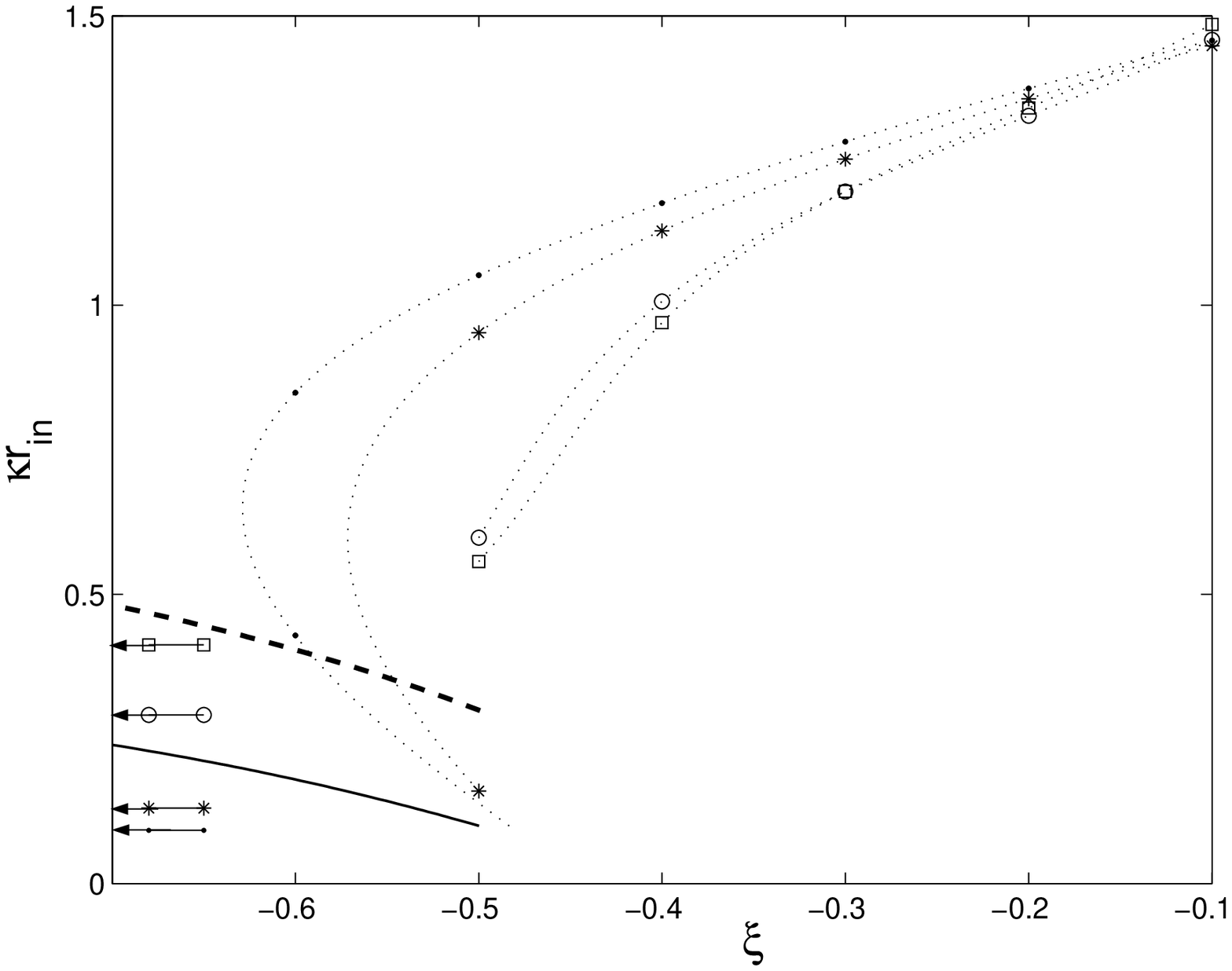}
\includegraphics[width=0.48\textwidth]{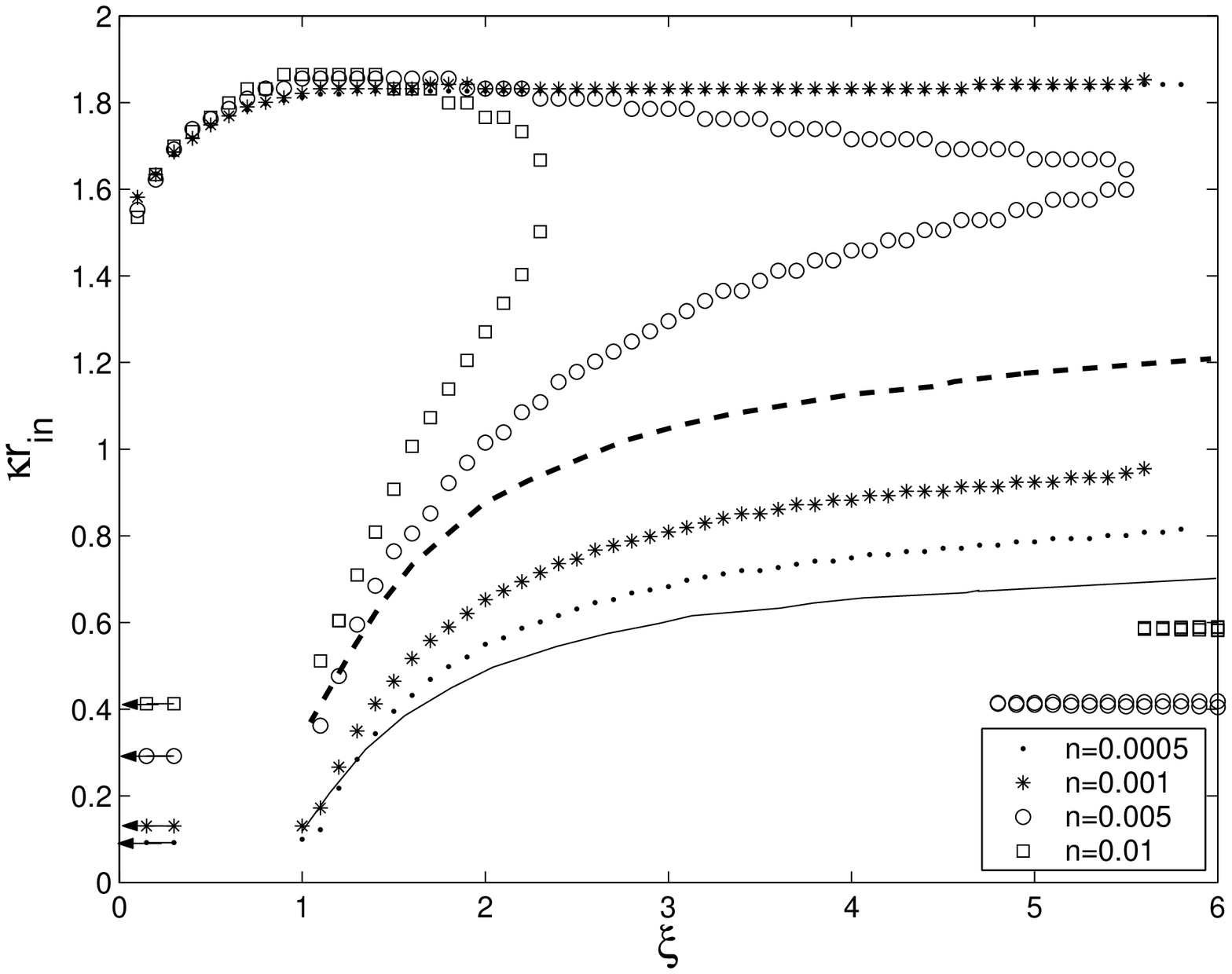}
\begin{quotation}
\caption{Locus of the inflection points of $P(r)$ for a charged cylinder of radius $\sigma_0 = 10\An$ immersed in an 1:2 (left) and 2:1 (right) electrolyte. Continuous and dashed lines correspond to calculation of $\kappa r_m$ by solving the PB equation  for $\kappa\sigma_c=0.1$ and $\kappa\sigma_c=0.3$,  respectively  \cite{TellezTrizac_PB}. Note that for $n=0.0005$M and $n=0.005$M correspond $\kappa\sigma_c\approx0.09$ and $\kappa\sigma_c\approx0.29$, respectively. Arrows indicate  the $\kappa\sigma_c$ limits.}  \label{colloida_Rmann}
\end{quotation}
\end{center}
\end{figure}

On the other hand, when the macroion is immersed in an asymmetric
electrolyte composed by divalent and monovalent ions, the
renormalized charge eva\-lua\-ted from HNC calculation behaves as
shown in figure~\ref{colloida_Zren_a20}. Again, the overshooting
and charge inversion phenomena are observed. In the 2:1 case at
$n=5\times 10^{-4}$M ($\kappa\sigma_c\approx 0.09$) and $n=0.001$M
($\kappa\sigma_c\approx 0.13$), figure~\ref{colloida_Rmann} suggests
the presence of a well defined layer of condensed counterions. At
these densities, the Manning radius dependence on $\xi$ is
qualitatively similar to the one predicted by PB theory; however, it
appears to underestimate the thickness of the condensed layer. If we
would consider the renormalized charge as the net charge enclosed by
the Manning radius (two-state model), then the re\-nor\-ma\-lized
charge should be smaller than the bare one, and therefore, the
overshooting phenomenon would not be present.

In spite of a thicker condensed layer, depending mainly on short-range
correlations between the macroion and the counterions, the HNC
approach also predicts overshooting. This phenomenon lies in the fact
that divalent coions are expelled further away in the double layer
than the monovalent ones in the 1:1 situation, which results in a
stronger electrostatic potential at larger
distance~\cite{TellezTrizac}. Although the PB theory accounts for this
phenomenon, figure~\ref{colloida_Zren_a20} shows that the mean field
solution underestimates the magnitude of the overshooting effect as
well as the range of $\xi$ where it is present, which is particularly
notorious at high density.
 
\begin{figure}[!htbp] 
\begin{center}
\includegraphics[width=0.48\textwidth]{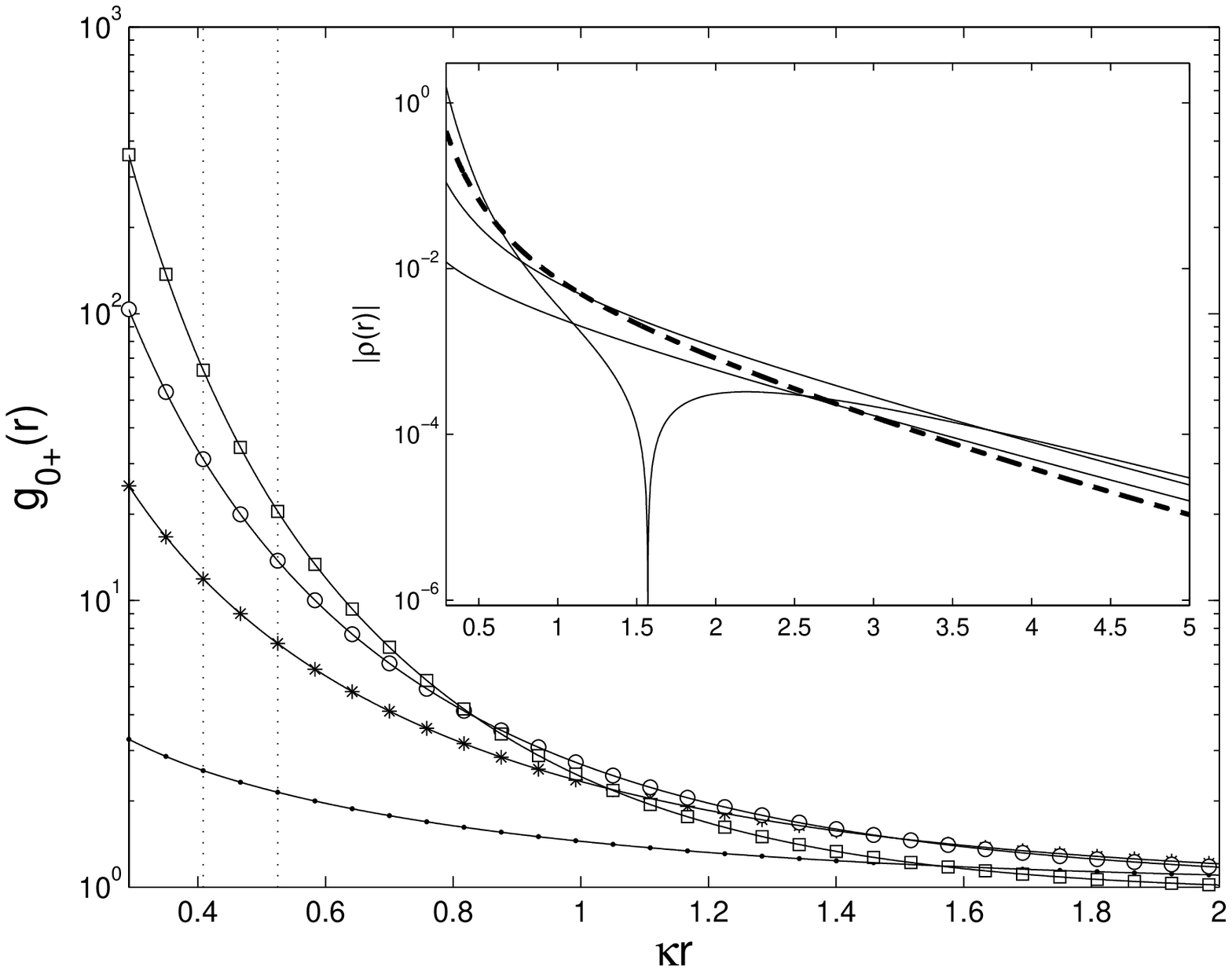}
\includegraphics[width=0.48\textwidth]{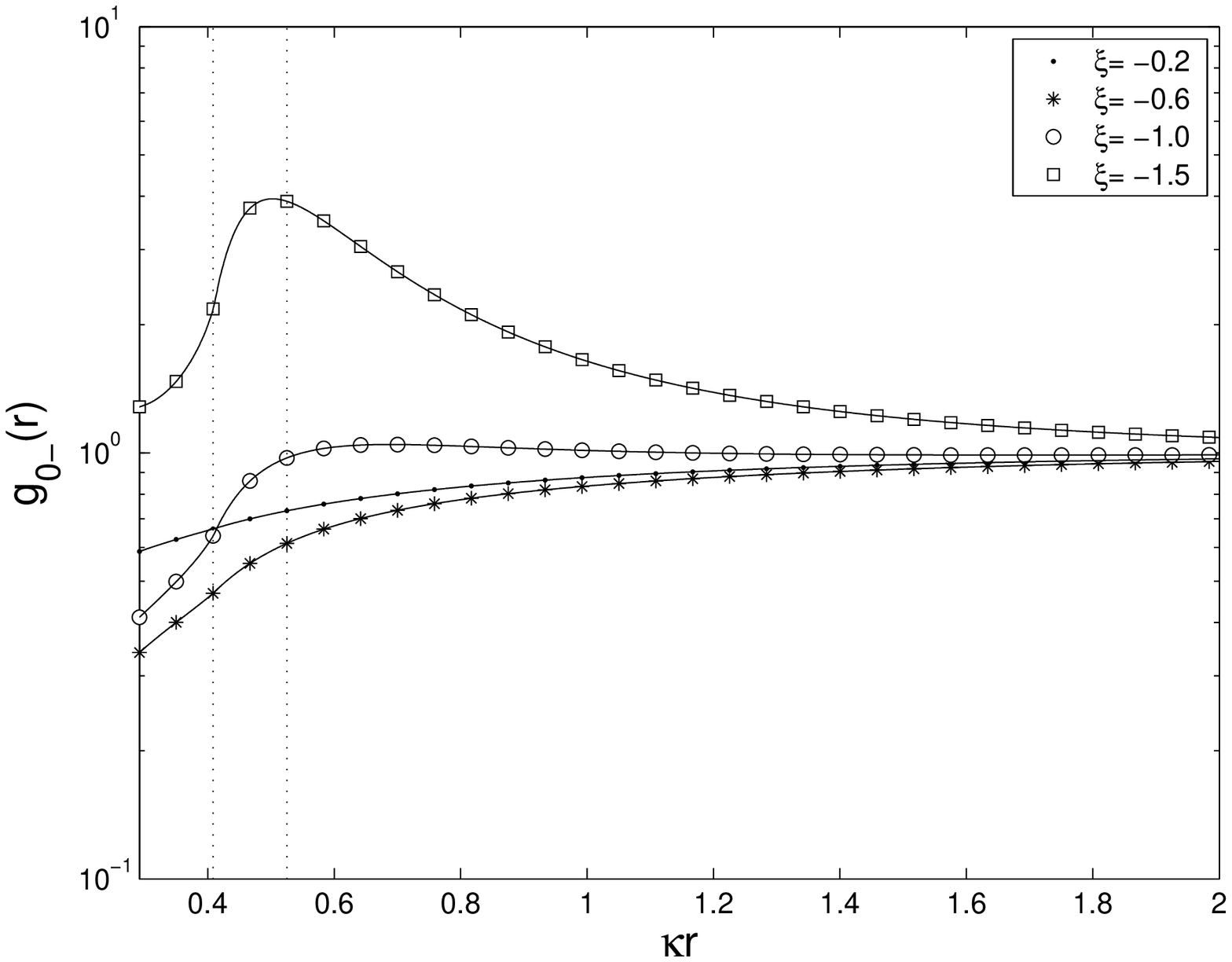}
\includegraphics[width=0.48\textwidth]{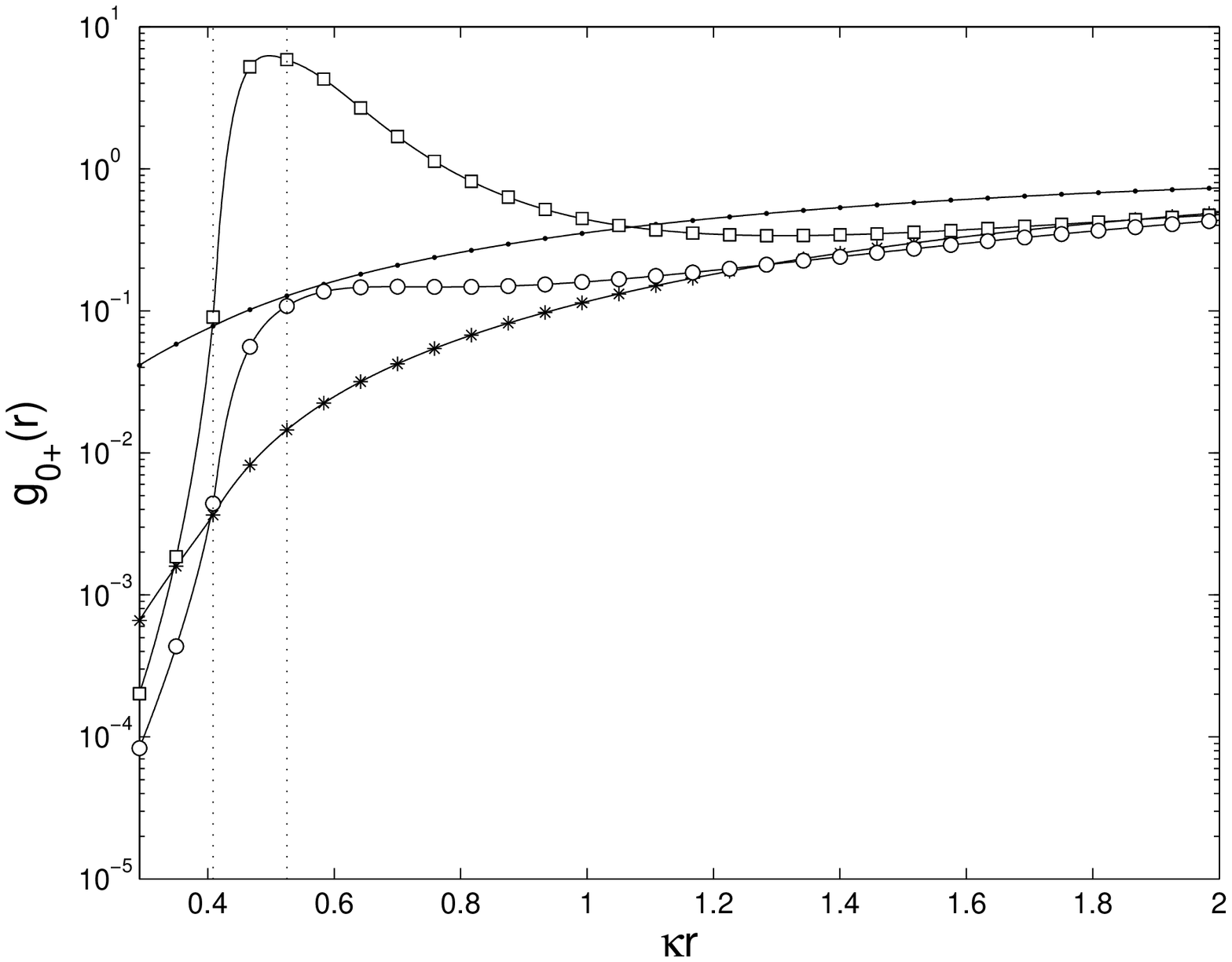}
\includegraphics[width=0.48\textwidth]{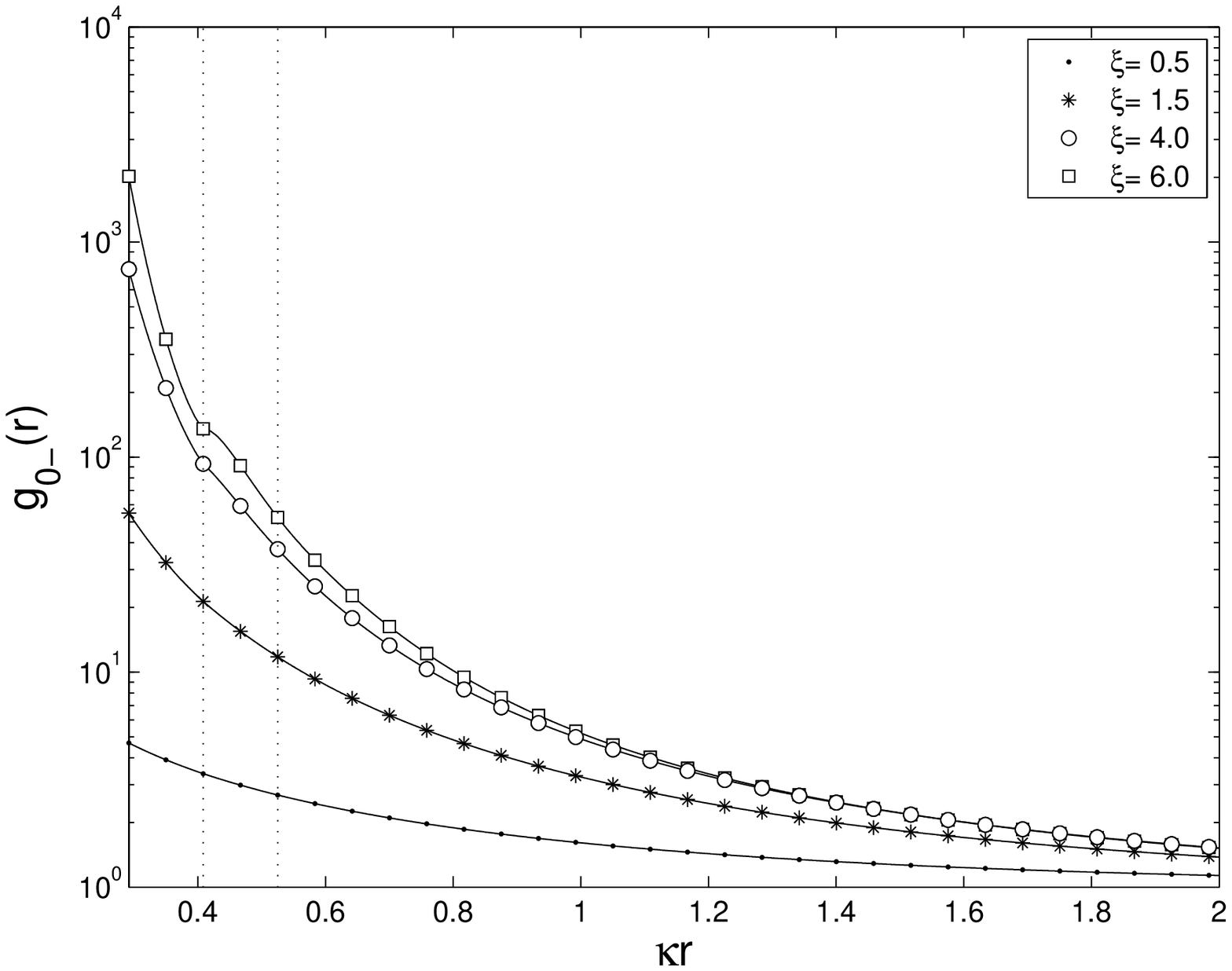} 
\begin{quotation}
\caption{Correlation functions $g_{0\pm}(r)$ for  1:2 (top) and 2:1 case (bottom) at  $n=0.005$M.  In the inset are shown the associated charge densities (dashed line is used for $\xi=-1.0$). Dotted lines marks $r=\sigma_c+2\sigma$ and $r=\sigma_c+4\sigma$. }  \label{colloida_gropm}
\end{quotation}
\end{center}
\end{figure}

Likewise, at high enough density (for instance, $n=0.005$M
($\kappa\sigma_c\approx 0.29$)), as $\xi$ increases the Manning radius
increases faster that its PB counterpart (dashed line in
figure~\ref{colloida_Rmann}) and eventually fuses with the inflection
point characterizing the salt. Again, when inflection points are close
enough, a peak in $g_{0+}(r)$ appears, i.e. a local increase of coions
density takes place, as it can be seen in the bottom frames of 
figure~\ref{colloida_gropm}. In these circumstances, the Manning
inflection point and the Debye length become comparable so that, the
charge renormalization is determined by a more correlated ionic cloud
giving place to a saturation value of $\xi_{ren}$ within the studied
range of $\xi$, as it is displayed in figure~\ref{colloida_Zren_a20}.

With respect to the 1:2 case, the differences between PB and HNC
results become more significant. For all the considered densities,
soon after the linear charge density is larger (in absolute value)
than the Manning condensation threshold ($\xi_m=-0.5$), the
renormalized charge decays quickly to zero and gets the positive sign
at $\xi \lesssim -1.3$ (see the inset of
figure~\ref{colloida_Zren_a20}). As well, there is no qualitative
similarity in the Manning radius evaluated from both of those
approaches, suggesting that the short-range correlations between
macroions and microions play a more predominant role in this
situation. Although in figure~\ref{colloida_gropm} the coion
densities for both 1:2 and 2:1 electrolytes exhibit a qualitatively
similar behavior as $|\xi|$ increases, we call the attention on the
difference between their orders of magnitude at $r=\sigma_c$: in the
2:1 case the first microion layer around the macroion corresponds only
to counterions whereas in the 1:2 case, this appears to be conformed
both of counter- and coions. This fact indicates a change in the
short range behavior of the mean force potential between the macroion
and the coions, as it was mentioned earlier.  In the 1:2 case,
increasing further $|\xi|$, the counterion fraction of the ionic cloud
around the macroion does not only compensates for its total charge but
even exceeds it, giving rise to a charge inversion phenomenon, as it
is displayed in the inset of figure~\ref{colloida_Zren_a20}.

\begin{figure}[!htbp] 
\begin{center}
\includegraphics[width=0.48\textwidth]{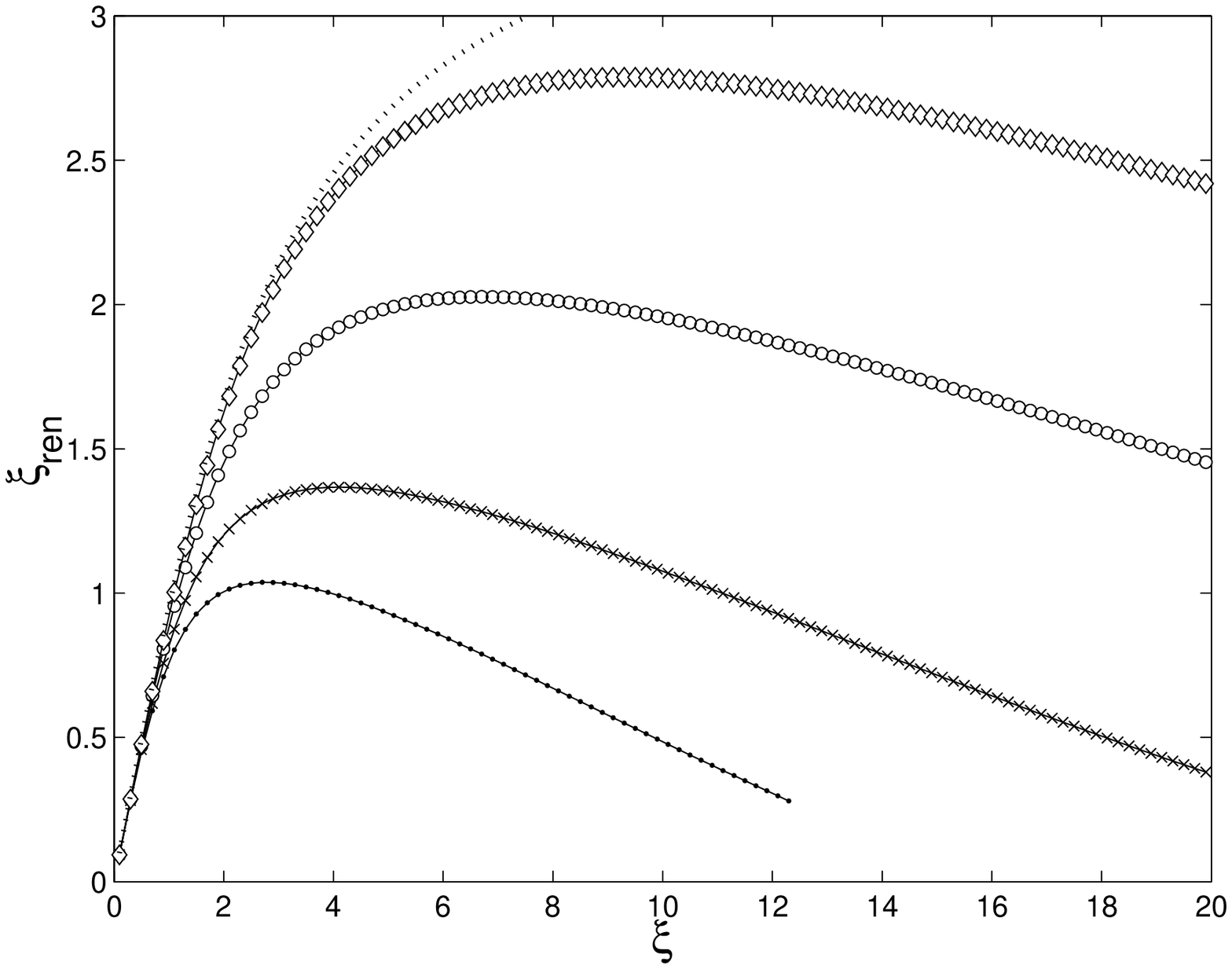}
\includegraphics[width=0.48\textwidth]{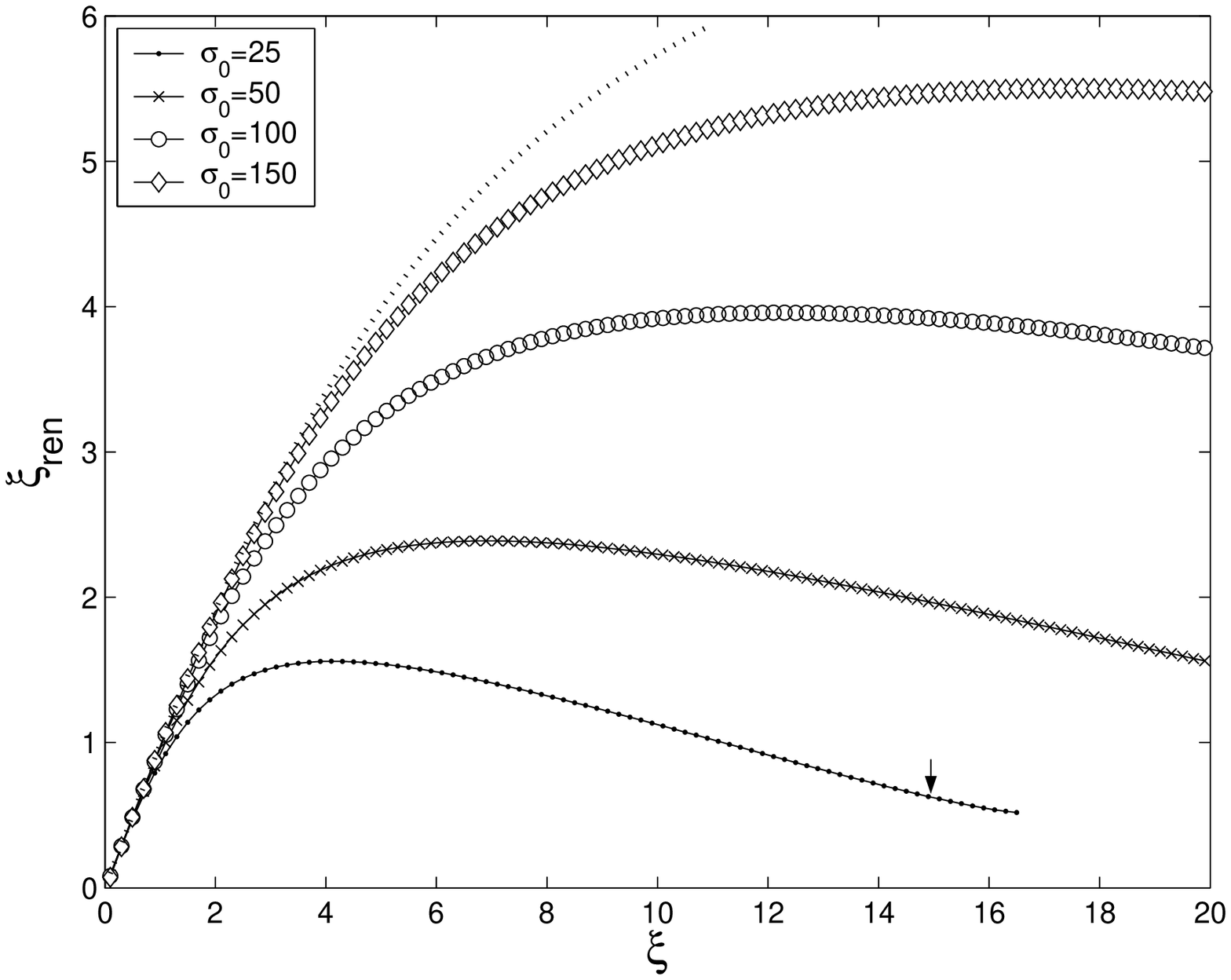}
\begin{quotation}
\caption{Renormalized charge in the symmetric case for different values of colloid radius ($[\sigma_0]=\An$) at $n=0.001$M (left) and $n=0.01$M (right). Dotted lines indicate the PB solutions from \cite{TellezTrizac} for $\sigma_0=150 \An$ ($\kappa\sigma_c\approx 1.6$ and $3.6$, respectively). Arrow indicates a value of $\xi$ from which  a less pronounced decrease is apparently found.}  \label{colloid_radios}
\end{quotation}
\end{center}
\end{figure}

\begin{figure}[p]
\begin{center}
\includegraphics[width=0.6\textwidth]{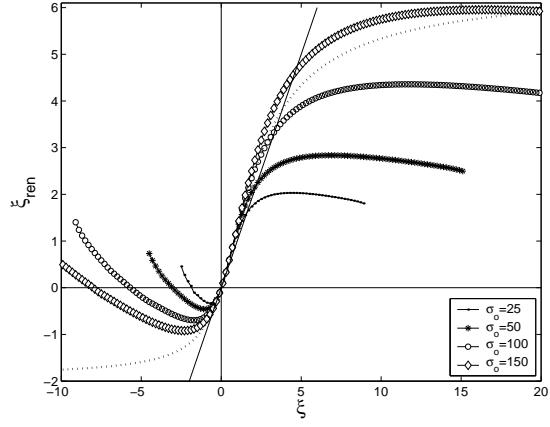}
\begin{quotation}
\caption{Renormalized charge in the asymmetric case for different values of colloid radius ($[\sigma_0]=\An$) at $n=0.001$M. Dotted line indicates the PB solution  for $\sigma_0=150\An$ ($\kappa\sigma_c\approx 1.6$) \cite{TellezTrizac}.}  \label{colloida_radios}
\end{quotation}
\end{center}
\end{figure}

\begin{figure}[!htbp] 
\begin{center}
\includegraphics[width=0.48\textwidth]{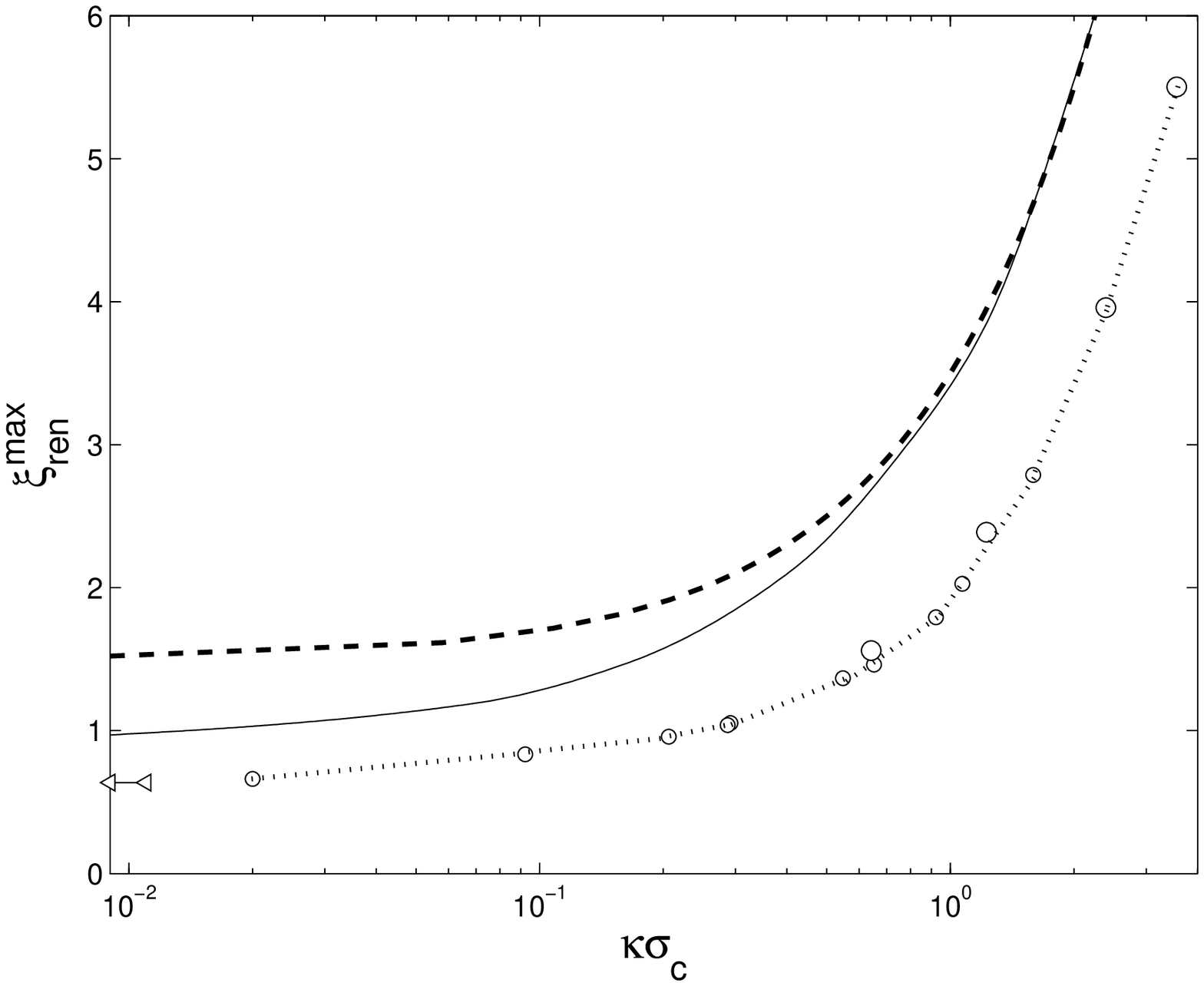}
\includegraphics[width=0.48\textwidth]{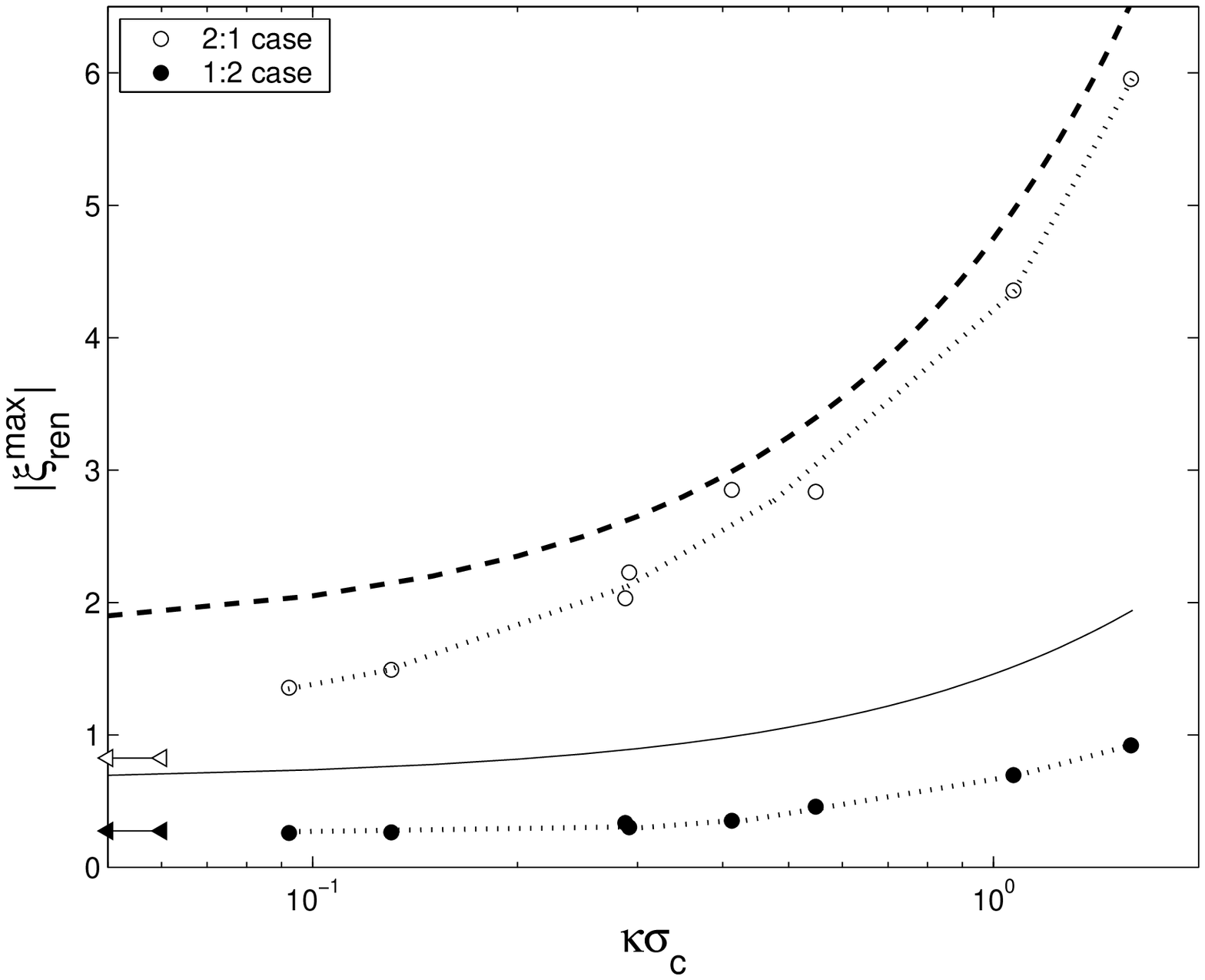}\vspace{-0.5cm}
\begin{quotation}
\caption{Maximum value of renormalized charge as a function of the
  reduced radius $\kappa\sigma_c$ for symmetric (left) and asymmetric
  (right) electrolyte. At the left, continuous and dashed lines are
  the theoretical PB upper bound for $Z_{ren}$ at saturation
  (analytical prediction from~\cite{TellezTrizac_PB} accurate for
  large $\sigma$) and the PB saturated charge evaluated numerically
  in~\cite{TellezTrizac_PB}, respectively. At the right, dashed and
  continuous lines are, respectively, the PB saturated charges for 2:1
  and 1:2 cases \cite{TellezTrizac}. Arrows indicates theoretical PB
  limit when $\kappa\sigma_c\to 0$. Dotted lines are intended as a
  guide to eye.}
  \label{colloid_Ximax}
\end{quotation}
\end{center}
\end{figure}

Also, the results show the existence of a maximum value
of the effective charge $\xi_{ren}^{max}$. In order to
describe the dependence of that maximum on the macroion radius
$\sigma_0$, we evaluate $\xi_{ren}$ at different values of $\sigma_0$
for both symmetric and asymmetric electrolytes. The outcome of these
calculations within the HNC convergence range are shown in
figures~\ref{colloid_radios} and~\ref{colloida_radios}. As it can be
seen, the qualitative behavior appears to be the same previously
discussed: the observation of the maximum of $\xi_{ren}(\xi)$ followed
by a monotonic decrease is systematically found in all studied
cases. In figure~\ref{colloid_Ximax}, the dependence of the maximum of
$\xi_{ren}$ on the reduced radius $\kappa\sigma_c$
($\sigma_c=\sigma_0+\sigma$) is displayed.  Although systematically
lesser, $\xi_{ren}^{max}$ qualitatively follows the same tendency (at
less within the considered range) that the saturation value from PB
theory ($\xi_{sat}^{PB})$ at $\kappa\sigma_c>1$ in both symmetric and
asymmetric electrolyte \cite{TellezTrizac,Aubouy}. We can associate
the $\xi_{ren}^{max}$ to the counterion condensation in the sense
that, around the bare charge $\xi^*$ at which the maximum is reached,
the renormalized charge is mainly fixed by the macroion-counterion
correlation. In this sense, since we use a 2D system, which
overestimates correlations between microions,
figures~\ref{colloid_Ximax} can be used to establish both upper (PB
theory, no microion correlations) and lower bound to the exact value
of $\xi_{ren}^{max}$ .

 Finally, coming back to the figure~\ref{colloid_radios} for
 $n=0.01$M and $\sigma_0=25\An$, a particular behavior of $\xi_{ren}$
 is found as $\xi\gtrsim 15$. At such conditions the decrease of
 $\xi_{ren}$ appears to be less pronounced that the one found for
 $8<\xi<15$. This fact would indicate the eventual appearance of a
 plateau region in $\xi_{ren}$ at large enough $\xi$; nevertheless,
 since the HNC calculation does not converge at $Z\gtrsim 16$, it is
 difficult to give a conclusive argument in this direction.

%%%%%%%%%%%%%%%%%%%%%%%%%%%%%%%%%%%%%%%%%%%%%%%%%%%%%%%%%%%%%%%%%
%%%%%%%%%%%%%%%%%%%%%%% CONCLUSIONS %%%%%%%%%%%%%%%%%%%%%%%%%%%%%
%%%%%%%%%%%%%%%%%%%%%%%%%%%%%%%%%%%%%%%%%%%%%%%%%%%%%%%%%%%%%%%%%

\section{Conclusions}
We have evaluated the renormalized charge in a 2D system formed by a
central charged particle (macroion) immersed in an electrolytic
solution. By solving numerically the Ornstein-Zernike equations within
the framework of the hypernetted chain approximation (HNC), we have
calculated the correlation functions between the macroion and the
microions ($g_{0\pm}(r)$). Under the assumption that the correlation
functions fall off exponentially, we found the renormalized charge by
fitting the asymptotic behavior of the correlation functions to the
one derived from the linearized PB theory (Debye-H\"uckel limit). The
adequacy of this procedure was proved by comparing the numerical
results to the analytical ones from exactly solved 2D models.

The 2D model could be applied to study a cylindrical macroion immersed
in an electrolyte formed by cylindrical co- and counterions, for
instance, charged stiff polymers. Moreover, since particles in the
model interact through a logarithmic potential, it would be expected
that it preserves aspects concerning the nature of the Coulomb
interaction, i.e., in a 3D model, the dependence of the renormalized
charge on the bare one would share the same qualitative
characteristics that those described here.

Recently, by using the theory of integrable systems, some analytical
results concerning the renormalized charge within the non-linearized
Poisson-Boltzmann (PB) theory have been reported
\cite{TellezTrizac_PB}. The results of this work show that the
renormalized charge is not explained qualitatively by the PB
theory. As the bare charge of the macroion increases sufficiently, the
PB theory predicts a constant value of the renormalized charge; on the
contrary, the HNC renormalized charge reaches a maximal value and then
monotonically decreases, both in symmetric and asymmetric
electrolytes. This behavior takes place because the HNC approach
appreciates the difference between the potential and the potential of
mean force, which is exactly what the Poisson-Boltzmann theory
neglects.

On the other hand, the structure of the ionic cloud around the
macroion, which provokes the charge renormalization, can be described
through either the density profiles $g_{0\pm} (r)$ or the inflection
points of the integrated charge density $P(r)$ (defined by
eq. (\ref{Charge_r})). For a weakly charged cylinder, the short
range behavior of the potential of mean force between macroion and
microions appears to correspond to the Coulomb interaction. Also,
there is one inflection point associated to the screening phenomena in
the bulk of the salt. Consequently, at that limit, the density
profiles suggest a diffuse cloud, which gives rise to a linear
screening that can be described by the Debye-H\"uckel approximation.

At low density, as the bare charge $\xi$ increases, the stronger
accumulation of counterions around the central particle results in a
change of the short-distance behavior of the potential of mean force
between macroion and coions, which has been interpreted as a precursor
of the (Manning) counterion condensation
\cite{Tellez_Guest}. Simultaneously, according to the inflection point
criterion, a new inflection point at $r_m<\kappa^{-1}$ can be
identified. This point can be associated to the Manning radius
provided that the bare charge is close to the Manning threshold; in
such situation the renormalized charge is mainly fixed by the
macroion-counterion correlation. Besides, since the maximum of the
renormalized charge ($\xi_{ren}^{max}$) exhibits the same
characteristics as the plateau of the PB theory, it can be assumed
that under these circumstances, the thermal energy of a counterion is
nearly compensated by the reversible work required to remove it from
the vicinity of the macroion.

On the other hand, by  studying $\xi_{ren}^{max}$ within an
intermediate range of reduced radius ($10^{-2}<\kappa\sigma_c\lesssim
1$),  we can conclude that $\xi_{ren}^{max}$ as a function of
$\kappa\sigma_c$ is systematically smaller and present the same
qualitative behavior than the saturation value of the PB theory. Since
in our model the microions interact through a logarithmic potential,
the role of microion correlations has been  overestimated from the
point of view of 3D systems. In this way,  our HNC results  could be  established  as lower bounds for the real $\xi_{ren}^{max}$. 

As $\xi$ increases further, $r_m$ becomes bigger until it fuses with
the inflection point characterizing the electrolyte. Under such
circumstances, most of the counterions are located at the cylinder
surface, whereas the rest of the screening charge, is spread in the
ionic cloud. Nevertheless, the merging of the inflection point is
related to the onset of a peak in the coions density profile. This
suggests that beyond the counterion layer, the ionic cloud is not a
diffuse layer which can be handled by means of the linearized theory,
as the two-state model claims, but a more complex structure is settled
by correlations between microions.

If we consider an asymmetric 1:2  electrolyte, the net charge in the 
vicinity   of the cylinder surface has opposite sign to that of 
the bare charge. This fact causes that the coions at some distance 
from the surface to be attracted  towards the macroion (giving rise 
to the mentioned peak) and that the counterions to be repelled  from 
the macroion, leading to a reversion  of the charge distribution in a 
region some distance from the cylinder. In this way, the evaluation of
 the renormalized charge from the asymptotic behavior of $g_{0\pm}(r)$ 
 allows to account for the charge inversion phenomenon  without additional
  assumptions, in contrast to  PB theory, in which the introduction of 
  the counterion binding conjecture becomes necessary.  

Therefore, the present study  suggest that the HNC approach is able to 
take into account, at least qualitatively, the behavior of correlations 
responsible for the counterion condensation process, as well as the charge 
inversion one. As well, it gives indications in order to obtain  a better
 description of the charge renormalization, because it  intrinsically  
 explains for  the structure of the electric double layer. So, a theoretical
  method to estimate more adequately  the  renormalized charge at high enough 
  $\xi$, must consider  a more detailed  description of the  electric double 
  layer structure by taking into account short range correlations between 
  macro- and microions  and between the last ones (for instance, density 
  functional  or dressed-ion theories).
 
Finally, a question remain to be solved: Does $\xi_{ren}$ always
continue to decrease at large $\xi$?  A unique of the outcomes
presented in this work appears to indicate the possibility of a
plateau region in $\xi_{ren}$ at large enough $\xi$; however, due to
the non convergence of HNC calculation at high $\xi$, we cannot
categorically state this conclusion.  As it was mentioned, HNC
approach is inadequate to treat the ion clustering effects, which
could be of extreme importance at such a strong coupling. A device to
overcome this difficult might be the introduction of a bridge
function, which takes indirectly into account correlations mediated by
two or more particles.
 
This work was partially supported by a ECOS Nord/COLCIENCIAS action
of French and Colombian cooperation.

%%%%%%%%%%%%%%%%%%%%%%%%%%%%%%%%%%%%%%%%%%%%%%%%%%%%%%%%%%%%%%%%%
%%%%%%%%%%%%%%%%%%%%%%% REFERENCES %%%%%%%%%%%%%%%%%%%%%%%%%%%%%%
%%%%%%%%%%%%%%%%%%%%%%%%%%%%%%%%%%%%%%%%%%%%%%%%%%%%%%%%%%%%%%%%%

%%%%%%%%%%%%%%%%%%%%%%%%%%%%%%%%%%%%%%%%%%%%%%%%%%%%%%%%%%%%%%%%%
%%%%%%%%%%%%%%%%%%%%%% FIGURES sTCP %%%%%%%%%%%%%%%%%%%%%%%%%%%%%
%%%%%%%%%%%%%%%%%%%%%%%%%%%%%%%%%%%%%%%%%%%%%%%%%%%%%%%%%%%%%%%%%
%%%%%%%%%%%%%%%    Guest-Microion  Correlations  %%%%%%% 1 %%%%%%
\newpage
%%%%%%%%%%%%%%%%%%%%%%%%%%%%%%%%%%%%%%%%%%%%%%%%%%%%%%%%%%%%%%%%%
%%%%%%%%%%%%%%%%   Microion-Microion  Correlations %%%%%%%%%%%%%%
%%%%%%%%%%%%%%%%%%%%%%%%%%%%%%%%%%%%%%%%%%%%%%%%%%%%%%%% 2 %%%%%%
\newpage
%%%%%%%%%%%%%%%%%%%%%%%%%%%%%%%%%%%%%%%%%%%%%%%%%%%%%%%%%%%%%%%%%
%%%%%%%%%%%%%%%  Linear fitted functions (sTCP) %%%%%%%%%%%%%%%%%
%%%%%%%%%%%%%%%%%%%%%%%%%%%%%%%%%%%%%%%%%%%%%%%%%%%%%%% 3 %%%%%%%
\newpage

%%%%%%%%%%%%%%%%%%%%%%%%%%%%%%%%%%%%%%%%%%%%%%%%%%%%%%%%%%%%%%%%%
%%%%%%%%%%%%%%%%%%   Renormalized charge (sTCP) %%%%%%%%%%%%%%%%%
%%%%%%%%%%%%%%%%%%%%%%%%%%%%%%%%%%%%%%%%%%%%%%%%%%%%%%%% 4 %%%%%%
\newpage

%%%%%%%%%%%%%%%%%%%%%%%%%%%%%%%%%%%%%%%%%%%%%%%%%%%%%%%%%%%%%%%%%
%%%%%%%%%%%%%%%%   Short-range behavior (sTCP) %%%%%%%%%%%%%%%%%%
%%%%%%%%%%%%%%%%%%%%%%%%%%%%%%%%%%%%%%%%%%%%%%%%%%%%%%%% 5 %%%%%%
\newpage
\clearpage
%%%%%%%%%%%%%%%%%%%%%%%%%%%%%%%%%%%%%%%%%%%%%%%%%%%%%%%%%%%%%%%%%
%%%%%%%%%%%%%%%%%%%%%%% Condensed charge %%%%%%%%%%%%%%%%%%%%%%%%
%%%%%%%%%%%%%%%%%%%%%%%%%%%%%%%%%%%%%%%%%%%%%%%%%%%%% 6 %%%%%%%%%
\clearpage
%%%%%%%%%%%%%%%%%%%%%%%%%%%%%%%%%%%%%%%%%%%%%%%%%%%%%%%%%%%%%%%%%
%%%%%%%%%%%%%%%%%%%%%%%% Manning Radius %%%%%%%%%%%%%%%%%%%%%%%%%
%%%%%%%%%%%%%%%%%%%%%%%%%%%%%%%%%%%%%%%%%%%%%%%%%%%%% 7 %%%%%%%%%
%%%%%%%%%%%%%%%%%%%%%%%%%%%%%%%%%%%%%%%%%%%%%%%%%%%%%%%%%%%%%%%%%
%%%%%%%%%%%%%%%%%%%%%% FIGURES aTCP %%%%%%%%%%%%%%%%%%%%%%%%%%%%%
%%%%%%%%%%%%%%%%%%%%%%%%%%%%%%%%%%%%%%%%%%%%%%%%%%%%%%%%%%%%%%%%%
%%%%%%%%%%%%%    Microion-Microion  Correlations  %%%%%%% 8 %%%%%
%%%%%%%%%%%%%%%%%%%%%%%%%%%%%%%%%%%%%%%%%%%%%%%%%%%%%%%%%%%%%%%%%
%%%%%%%%%%%%%%    Guest-Microion1  Correlations  1 %%%%%%%%%%%%%%
%%%%%%%%%%%%%%%%%%%%%%%%%%%%%%%%%%%%%%%%%%%%%%%%%%%%%%%%%% 9 %%%%
%%%%%%%%%%%%%%%%%%%%%%%%%%%%%%%%%%%%%%%%%%%%%%%%%%%%%%%%%%%%%%%%%
%%%%%%%%%%%%%%    Guest-Microion  Correlations  2 %%%%%%%%%%%%%%%
%%%%%%%%%%%%%%%%%%%%%%%%%%%%%%%%%%%%%%%%%%%%%%%%%%%%%% 10 %%%%%%%
%%%%%%%%%%%%%%%%%%%%%%%%%%%%%%%%%%%%%%%%%%%%%%%%%%%%%%%%%%%%%%%%%
%%%%%%%%%%%%%%%  Linear fitted functions (aTCP) %%%%%%%%%%%%%%%%%
%%%%%%%%%%%%%%%%%%%%%%%%%%%%%%%%%%%%%%%%%%%%%%%%%%%%%%% 11 %%%%%%
%%%%%%%%%%%%%%%%%%%%%%%%%%%%%%%%%%%%%%%%%%%%%%%%%%%%%%%%%%%%%%%%%
%%%%%%%%%%%%%%%%%%   Renormalized charge (aTCP) %%%%%%%%%%%%%%%%%
%%%%%%%%%%%%%%%%%%%%%%%%%%%%%%%%%%%%%%%%%%%%%%%%%%%%%% 12 %%%%%%%
%%%%%%%%%%%%%%%%%%%%%%%%%%%%%%%%%%%%%%%%%%%%%%%%%%%%%%%%%%%%%%%%%
%%%%%%%%%%%%%%%%% Short-range correlations (aTCP) %%%%%%%%%%%%%%%
%%%%%%%%%%%%%%%%%%%%%%%%%%%%%%%%%%%%%%%%%%%%%%%%%%%%%%% 13 %%%%%%
%%%%%%%%%%%%%%%%%%%%%%%%%%%%%%%%%%%%%%%%%%%%%%%%%%%%%%%%%%%%%%%%%
%%%%%%%%%%%%%%%%%%%%  Short-range factor (aTCP) %%%%%%%%%%%%%%%%%
%%%%%%%%%%%%%%%%%%%%%%%%%%%%%%%%%%%%%%%%%%%%%%%%%%%%%% 14 %%%%%%%
%%%%%%%%%%%%%%%%%%%%%%%%%%%%%%%%%%%%%%%%%%%%%%%%%%%%%%%%%%%%%%%%%
%%%%%%%%%%%%%%%%%%%%%   Manning Radius (aTCP) %%%%%%%%%%%%%%%%%%%
%%%%%%%%%%%%%%%%%%%%%%%%%%%%%%%%%%%%%%%%%%%%%%%%%%%%%%% 15 %%%%%%
%%%%%%%%%%%%%%%%%%%%%%%%%%%%%%%%%%%%%%%%%%%%%%%%%%%%%%%%%%%%%%%%%
%%%%%%%%%%%%%%%%%%   Charge distribution (aTCP) %%%%%%%%%%%%%%%%%
%%%%%%%%%%%%%%%%%%%%%%%%%%%%%%%%%%%%%%%%%%%%%%%%%%%%%%% 16 %%%%%%

%%%%%%%%%%%%%%%%%%%%%%%%%%%%%%%%%%%%%%%%%%%%%%%%%%%%%%%%%%%%%%%%%
%%%%%%%%%%%%%%%%%%%%%   Colloidal suspension  %%%%%%%%%%%%%%%%%%%
%%%%%%%%%%%%%%%%%%%  Renormalized charge (sTCP) %%%%%%%%%%%%%%%%%
%%%%%%%%%%%%%%%%%%%%%%%%%%%%%%%%%%%%%%%%%%%%%%%%%%%%%%% 17 %%%%%%
%%%%%%%%%%%%%%%%%%%%%%%%%%%%%%%%%%%%%%%%%%%%%%%%%%%%%%%%%%%%%%%%%
%%%%%%%%%%%%%%%%%%%%%  Manning radius (sTCP) %%%%%%%%%%%%%%%%%%%%
%%%%%%%%%%%%%%%%%%%%%%%%%%%%%%%%%%%%%%%%%%%%%%%%%%%%% 18 %%%%%%%%
%%%%%%%%%%%%%%%%%%%%%%%%%%%%%%%%%%%%%%%%%%%%%%%%%%%%%%%%%%%%%%%%%
%%%%%%%%%%%%%  Macroion-microions correlations (sTCP) %%%%%%%%%%%
%%%%%%%%%%%%%%%%%%%%%%%%%%%%%%%%%%%%%%%%%%%%%%%%%%%%%%%%% 19 %%%%
%%%%%%%%%%%%%%%%%%%%%%%%%%%%%%%%%%%%%%%%%%%%%%%%%%%%%%%%%%%%%%%%%
%%%%%%%%%%%%%%%%%%%  Renormalized charge (aTCP) %%%%%%%%%%%%%%%%%
%%%%%%%%%%%%%%%%%%%%%%%%%%%%%%%%%%%%%%%%%%%%%%%%%%%%%%%% 20 %%%%%
%%%%%%%%%%%%%%%%%%%%%%%%%%%%%%%%%%%%%%%%%%%%%%%%%%%%%%%%%%%%%%%%%
%%%%%%%%%%%%%%%%%%%%%  Manning radius (aTCP) %%%%%%%%%%%%%%%%%%%%
%%%%%%%%%%%%%%%%%%%%%%%%%%%%%%%%%%%%%%%%%%%%%%%%%%%%% 21 %%%%%%%%
%%%%%%%%%%%%%%%%%%%%%%%%%%%%%%%%%%%%%%%%%%%%%%%%%%%%%%%%%%%%%%%%%
%%%%%%%%%%%%%  Macroion-microions correlations (aTCP) %%%%%%%%%%%
%%%%%%%%%%%%%%%%%%%%%%%%%%%%%%%%%%%%%%%%%%%%%%%%%%%%%%%%% 22 %%%%
 \clearpage
%%%%%%%%%%%%%%%%%%%%%%%%%%%%%%%%%%%%%%%%%%%%%%%%%%%%%%%%%%%%%%%%%
%%%%%%%%%  Renormalized linear charge density 2(sTCP) %%%%%%%%%%%
%%%%%%%%%%%%%%%%%%%%%%%%%%%%%%%%%%%%%%%%%%%%%%%%%%%%%%%%%%% 23 %%
%%%%%%%%%%%%%%%%%%%%%%%%%%%%%%%%%%%%%%%%%%%%%%%%%%%%%%%%%%%%%%%%%
%%%%%%%%%  Renormalized linear charge density 2(aTCP) %%%%%%%%%%%
%%%%%%%%%%%%%%%%%%%%%%%%%%%%%%%%%%%%%%%%%%%%%%%%%%%%%%%% 24 %%%%%

%%%%%%%%%%%%%%%%%%%%%%%%%%%%%%%%%%%%%%%%%%%%%%%%%%%%%%%%%%%%%%%%%
%%%%%%%%%%  Maximal renormalized linear charge density %%%%%%%%%%
%%%%%%%%%%%%%%%%%%%%%%%%%%%%%%%%%%%%%%%%%%%%%%%%%%%%%%%%%% 25 %%%

\end{document}